\documentclass[a4paper,10pt]{article}
 
\usepackage{graphicx}
\usepackage{epstopdf}
\usepackage{latexsym}
\usepackage{amsmath,amssymb}

\usepackage{algorithm}
\usepackage{algpseudocode}

\usepackage{color}

\usepackage[nocompress]{cite}

\usepackage{enumitem}
\usepackage{url}

\usepackage{booktabs}

\usepackage{multirow}

\usepackage{mdframed}
\usepackage{mathtools}



\long\def\invis#1{}

\newcommand{\Err}{\mathrm{Err}}
\newcommand{\NA}{\text{-}}
\newcommand{\paircell}[2]{%
  \makebox[6.0em][c]{%
    (\makebox[2.5em][c]{\ensuremath{#1}},
     \makebox[2.5em][c]{\ensuremath{#2}})%
  }%
}

\usepackage{subcaption}

\title{Mixing Vector Model for Copolymer Inference via Mixed Integer Linear Programming
}
\author{
Jianshen Zhu$^{1}$ \and
Raveena Rai$^{2}$ \and
Taiyo Sohkawa$^{3}$ \and
Naveed Ahmed Azam$^{2,4}$ \and
Kazuya Haraguchi$^{3,}$\thanks{Current affiliation: Tokyo University of Marine Science and Technology.} \and
Liang Zhao$^4$ \and
Tatsuya Akutsu$^5$
}
\date{
 $^1$Department of Information Science and Technology, Tokyo University of Science, Noda 278-8510, Japan \\
$^2$Discrete Mathematics and Computational Intelligence Laboratory, Department of Mathematics, Quaid-i-Azam University, Islamabad 45320, Pakistan \\
$^3$Graduate School of Informatics, Kyoto University, Kyoto 606-8501, Japan \\
$^4$Graduate School of Advanced Integrated Studies in Human Survivability   (Shishu-Kan),  
 Kyoto University, Kyoto 606-8306, Japan  \\
 $^5$Bioinformatics Center,  Institute for Chemical Research, 
 Kyoto University, Uji 611-0011, Japan 
}

\newcommand{\filename}{2LMM\_copolymer\_mv\_v5}
\newcommand{\bbC}{\mathbb{C}}
\newcommand{\bbR}{\mathbb{R}}
\newcommand{\bbZ}{\mathbb{Z}}
\newcommand{\anC}{\langle \mathbb{C} \rangle}

\newcommand{\GC}{G_{\mathrm{C}}}
\newcommand{\VC}{V_{\mathrm{C}}}
\newcommand{\EC}{E_{\mathrm{C}}}

\newcommand{\Ez}{E_{(0/1)}}
\newcommand{\Ew}{E_{(\geq 1)}}
\newcommand{\Et}{E_{(\geq 2)}}
\newcommand{\Eew}{E_{(=1)}}

\newcommand{\ta}{{\tt a}}
\newcommand{\tb}{{\tt b}}

\newcommand{\tC}{{\tt C}}
\newcommand{\tO}{{\tt O}}
\newcommand{\tN}{{\tt N}}
\newcommand{\tS}{{\tt S}}
\newcommand{\tH}{{\tt H}}
\newcommand{\tP}{{\tt P}}

\newcommand{\VH}{V_{\tH}}

\newcommand{\calC}{\mathcal{C}}
\newcommand{\calG}{\mathcal{G}}
\newcommand{\calGmono}{\mathcal{G}_{\mathrm{mono}}}
\newcommand{\calM}{\mathcal{M}}
\newcommand{\calP}{\mathcal{P}}

\newcommand{\ylb}{\underline{y}^*}
\newcommand{\yub}{\overline{y}^*}

\newcommand{\val}{\mathrm{val}}
\newcommand{\h}{\mathrm{ht}}
\newcommand{\dg}{\mathrm{dg}}

\newcommand{\ex}{\mathrm{ex}}
\newcommand{\inte}{\mathrm{int}}
\newcommand{\lnk}{\mathrm{lnk}}

\newcommand{\Vleaf}{V_{\mathrm{leaf}}}
\newcommand{\Eleaf}{E_{\mathrm{leaf}}}
\newcommand{\Elnk}{E^\lnk}

\newcommand{\molinfer}{{\tt mol-infer}}

\newcommand{\etaxTB}{\eta_{\mathrm{xTB}}}

\begin{document} 
\maketitle

\begin{quote}  
{\bf Abstract}\\  
A novel two-phase molecule inference framework, \molinfer, has recently been developed to infer chemical graphs
with prescribed abstract structures and desired property values through mixed integer linear programming (MILP) under the two-layered model,
with guaranteed optimality and exactness relative to the given learned prediction function and structural constraints.
In this study, we extend this framework to copolymers by introducing a simple feature representation,
called the mixing vector (MV) model.
In the proposed model, a feature vector for a copolymer is represented as a convex combination of MILP-tractable
monomer descriptors weighted by the mixing ratio of the constituent monomers.
This representation does not require explicit sequence-class information and is therefore naturally compatible with 
the MILP-based inverse design.
Under this model, we construct prediction functions for a variety of copolymer property datasets with several machine-learning methods,
including artificial neural networks, reduced quadratic multiple linear regression, and random forests.
Experimental results show that the proposed representation achieves practically useful
predictive performance across multiple physicochemical property datasets;
in particular, the best test \(R^2\) score exceeds 0.7 for nine of the ten datasets
and exceeds 0.9 for six datasets.
We also formulate a multi-monomer inverse-design problem under the MV representation with a prescribed mixing ratio
and demonstrate that the resulting MILP instances remain tractable in practical computation time, 
even for three-monomer settings. 
We further perform an external consistency check by re-evaluating the inferred candidates
and comparing the re-computed property values
with those predicted by the learned model.
Overall, the experimental results demonstrate that the proposed framework provides a simple and tractable first step toward 
model-level exact inverse design of copolymers under the two-layered model.

\noindent 
{\bf Keywords: } Integer Programming, 
Chemoinformatics, Molecular Design,
Inverse QSAR/QSPR.

\end{quote}

\section{Introduction}\label{sec:introduction}

\subsection{Background}

\emph{Polymers} are macromolecular substances composed of a large number of covalently bonded repeating units (\emph{monomers}) and typically possess molar masses several orders of magnitude larger than those of ordinary small molecules.
In contrast to small molecules, polymer chains exhibit broad molecular-weight distributions and diverse architectures (linear, branched, cross-linked), and many of their macroscopic behaviors, such as viscoelasticity, glass transition, and melt rheology, emerge from collective, chain-level phenomena rather than from a single molecular entity~\cite{Flory:1953aa, Rodriguez:2014aa}.
Due to this structural versatility, both natural and synthetic polymers underpin a wide 
range of applications in both medical science and materials science~\cite{Connor:2017aa, Miccio:2020aa,Das:2023aa,Lim:2024aa,Li:2024aa,Trucillo:2024aa}.



From a structural viewpoint, polymers are often first classified according to the number of their repeat units into \emph{homopolymers}, which consist of a single monomer species, 
and \emph{copolymers}, which are formed from two or more distinct monomers~\cite{Das:2023aa}.
Copolymers can adopt a wide variety of architectures, including linear, comb-like, star-shaped, hyperbranched, and network topologies, 
and it is well established that such architectural differences can significantly affect the resulting
material properties~\cite{Kharchenko:2003aa, Gao:2004aa, Zhang:2020ab}.
For linear chains, copolymers can be further subdivided according to the sequence distribution of monomer units along the backbone into, for example, random (or statistical), alternating, block, gradient, and graft copolymers,
where such sequence distributions also have been found to affect the properties significantly~\cite{Meier:2019aa,Self:2022aa,Lefebvre:2004aa,Coldstream:2022aa}.
Within this classification, copolymers constitute an especially important class 
because combining chemically distinct monomers within a single macromolecule provides a powerful handle to tailor material performance beyond what is achievable with homopolymers. 
A prominent example is styrene-butadiene rubber (SBR), a type of synthetic copolymer obtained by
polymerization of styrene and 1,3-butadiene, which has been
widely used in tire tread compounds and impact-resistant elastomeric applications~\cite{Khan:2022aa,Sinclair:2019aa}.
In more advanced technologies, block copolymers and related architectures serve as versatile structure-directing agents for nanoporous membranes, filtration media, and templated nanostructured materials, as well as platforms for catalysis, nanodevices, and drug delivery~\cite{Zhai:2023aa,Olson:2008aa,Jackson:2010aa,Kumar:2021aa,Stoykovich:2007aa}.
Consequently, the design of copolymers with targeted properties has become a central theme in contemporary polymer and materials science.
However, relatively few studies have explicitly tackled representations that faithfully encode copolymer
composition, sequence distribution, and chain architecture~\cite{Zhao:2023aa, Patel:2022aa,Tao:2022aa,Vogel:2025aa}.
As a result, the development of novel and useful inverse-design frameworks for copolymers
remains a significant challenge.


One such approach involves \emph{inverse quantitative structure activity/property relationship} (inverse QSAR/QSPR)~\cite{Ikebata:2017aa, Miyao:2016aa, Rupakheti:2015aa},
which can be regarded as the inverse counterpart of the classical \emph{quantitative structure activity/property relationship} (QSAR/QSPR)~\cite{Cherkasov:2014aa, Skvortsova:1993aa}.
QSAR/QSPR focuses on predicting chemical activities or properties from given molecular structures~\cite{Cherkasov:2014aa}. 
A prediction function is usually constructed from existing structure-activity relation data, employing machine learning-based
methods, including artificial neural network (ANN)-based methods~\cite{Lo:2018aa, Tetko:2020aa}.
In contrast, inverse QSAR/QSPR aims to infer molecular structures that realize prescribed
activities or property values~\cite{Ikebata:2017aa, Miyao:2016aa, Rupakheti:2015aa}.
In most classical approaches,
molecular structures are represented as undirected graphs,  called \emph{chemical graphs},
 and are further encoded as vectors of real numbers (\emph{descriptors} or \emph{feature vectors}).
A typical workflow for inverse QSAR/QSPR involves first inferring feature vectors from given chemical activities and then reconstructing chemical graphs from these 
 feature vectors~\cite{Ikebata:2017aa, Miyao:2016aa, Rupakheti:2015aa, Shino:2025aa}. 

Recent advancements in deep learning, especially the development of
\emph{graph neural networks} (GNNs) and deep generative models,
have provided promising alternatives to traditional feature-based methods,
and state-of-the-art results on benchmark datasets such as QM9 have been
obtained~\cite{Gasteiger:2022aa,Zhang:2023ab}.
However, despite the remarkable progress made with GNNs and other
deep learning-based approaches in the forward QSAR/QSPR setting,
comparatively limited work has addressed inverse QSAR/QSPR in a form that
can provide mathematical guarantees on the inferred structures.
A central difficulty lies in guaranteeing two key properties of the generated
chemical structures:
\emph{optimality}---whether the obtained solution is optimal with respect to the
learned prediction function and the inverse-design formulation---and
\emph{exactness}---whether the solution corresponds to a valid chemical graph
satisfying the prescribed structural constraints.
While many deep learning-based generative models
(e.g.,~\cite{Shino:2025aa,Cai:2022aa,Bort:2022aa,Kaneko:2023aa})
aim to create chemically plausible molecules,
they are usually not designed to guarantee the optimality or exactness of the
inferred solutions mathematically, which can be problematic in practical
applications~\cite{Zhu:2023aa}.

A similar trend can be observed in polymer informatics, and more recently also in inverse design of copolymers.
For example, Vogel and Weber~\cite{Vogel:2025aa} proposed a graph-to-string variational autoencoder for copolymers that explicitly represents monomer stoichiometry 
and chain architecture, and demonstrated inverse design through optimization in latent space. 
Moreover, Yue~et~al.~\cite{Yue:2025aa} performed a benchmark study of several deep generative models for inverse polymer design, 
further highlighting the growing importance of such approaches in polymer discovery. 
However, these methods are likewise not designed to guarantee exactness of the inferred polymer structures or optimality relative to the learned prediction model.

\begin{figure}[t!]
\begin{center}
\includegraphics[width=.89\columnwidth]{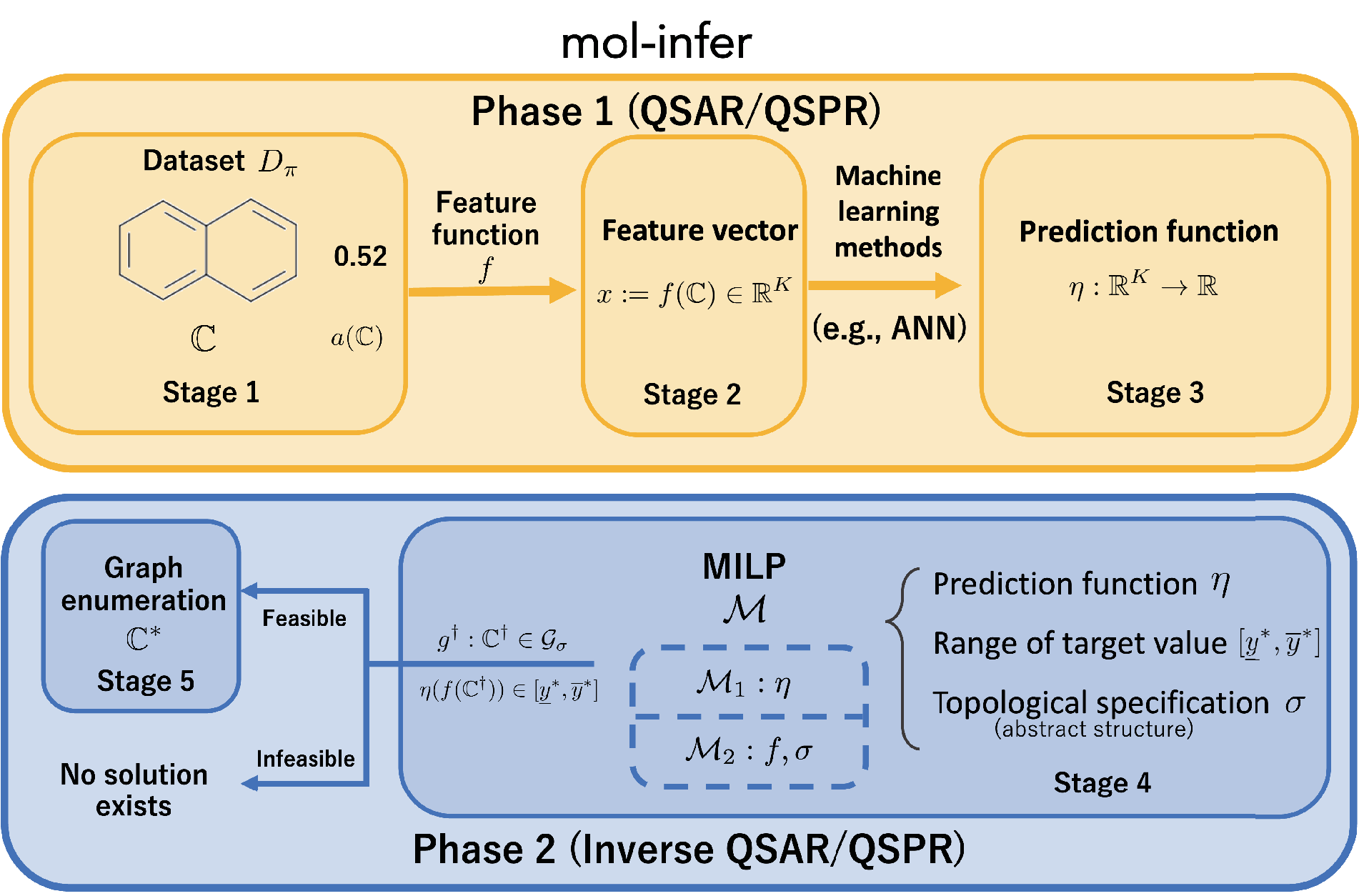}
\end{center}
\caption{Overview of the two-phase \molinfer\ framework.}
\label{fig:framework}
\end{figure}

\subsection{Overview of the \molinfer\ Framework}

To overcome these limitations, a framework called \molinfer~\cite{Azam:2020aa, Shi:2021aa}
has recently been proposed for inferring chemical compounds with prescribed abstract structures
and desired property values.
The central idea of this framework is to employ a \emph{mixed integer linear programming} (MILP) formulation
that simultaneously simulates the computational process of the underlying machine-learning model
and encodes the necessary and sufficient conditions for the existence of a valid chemical graph.
Consequently, \molinfer\ 
guarantees both the optimality and the exactness of the inferred solutions.
An overview of \molinfer\ is illustrated in Figure~\ref{fig:framework}.
Conceptually, the framework consists of two phases: 
Phase~1, the QSAR/QSPR phase, which consists of three stages, 
and Phase~2, the inverse QSAR/QSPR phase, which consists of two stages.
Phase~1 is the QSAR/QSPR phase, whose goal is to construct a prediction function $\eta$
relating chemical compounds to their observed property values.
Let $\calG$ denote the set of all possible chemical graphs.
In Stage~1, we collect a dataset $D_\pi \subseteq \calG$ consisting of chemical graphs $\bbC$
together with their observed values $a(\bbC)$.
In Stage~2, a feature function $f : \calG \to \bbR^K$ ($K$ is a positive integer) is defined
to convert each chemical graph into a $K$-dimensional real vector.
In Stage~3, a prediction function $\eta : \bbR^K \to \bbR$ is then learned by some
machine-learning method.
Phase~2 corresponds to the inverse QSAR/QSPR phase, where the aim is to infer chemical graphs
that exhibit specific property values.
Given a set of rules, called a topological specification $\sigma$, describing the desired
abstract structure of the inferred graphs, and a target range $[\ylb, \yub]$ of the property value,
Stage~4 seeks chemical graphs $\bbC^*$ satisfying $\sigma$ and
$\eta(f(\bbC^*)) \in [\ylb, \yub]$ by solving an MILP formulation $\mathcal{M}$ that encodes
\begin{itemize}
  \item[-] $\mathcal{M}_1$: the computation process of the prediction function $\eta$, and
  \item[-] $\mathcal{M}_2$: the computation process of the feature function $f$ together with the
    structural constraints for $\bbC \in \calG_\sigma$,
\end{itemize}
where $\calG_\sigma$ denotes the set of all chemical graphs satisfying $\sigma$.
In Stage~5, dynamic-programming-based graph enumeration algorithms~\cite{Ido:2025aa}
are applied to generate isomers of the inferred chemical graphs $\bbC^*$ obtained in Stage~4.
The original versions of this framework were developed only for restricted classes of chemical graphs,
such as trees~\cite{Zhang:2022aa, Azam:2021aa}, rank-1 graphs~\cite{Ito:2021aa},
and rank-2 graphs~\cite{Zhu:2020aa}.
The \emph{two-layered model} (2L-model)~\cite{Shi:2021aa} refines and generalizes \molinfer\ so that,
by providing an abstract structure as part of the input, one can in principle infer \emph{any} chemical graph,
and it has now become the standard setting for \molinfer.
One of the major advantages of \molinfer\ is that, when the MILP formulation $\mathcal{M}$ is infeasible,
it can rigorously certify that $\calG_\sigma$ does not contain any chemical graph with the specified property value,
whereas most existing inverse QSAR/QSPR models cannot provide such a guarantee.
The framework was first introduced for small molecules and has recently been extended to homopolymers
as well~\cite{Ido:2024aa, Zhu:2025ab}.
Various machine-learning models have been incorporated into \molinfer, including
artificial neural networks~\cite{Shi:2021aa}, linear regression~\cite{Zhu:2022ad, Zhu:2025aa},
decision trees~\cite{Tanaka:2021aa}, and graph neural networks~\cite{Zhu:2026aa}.
We refer the reader to Appendix~\ref{sec:frame_all} and the thesis~\cite{Zhu:2023aa} for a more comprehensive description of \molinfer.

\subsection{Contributions}
In this study, we extend the \molinfer\ framework to the copolymer setting by introducing a simple feature representation,
called the mixing vector (MV) model, and by showing that
this representation can be incorporated into the MILP formulation through linear constraints.
In the proposed model, the feature vector for a copolymer is represented by a weighted
combination of monomer-level descriptors, where the weights correspond to the mixing ratio of the constituent monomers.
This simple structure allows the representation to remain compatible with the MILP-based inverse-design phase of \molinfer.
We evaluate the proposed approach on multiple copolymer datasets covering several physicochemical properties, and demonstrate
both predictive performance in Phase~1 and computational effectiveness in Phase~2. Although the present representation does not explicitly
encode detailed sequence information, it provides a tractable framework for exact inverse design of copolymers and serves as
a foundation for more expressive future models.

The remainder of this paper is organized as follows.
Section~\ref{sec:copolymer} introduces the proposed extension of \molinfer\ to copolymers and formulates the MV model in both Phase~1 and Phase~2.
Section~\ref{sec:experiment} reports experimental results on predictive performance, computational effectiveness, and external consistency checks.
Section~\ref{sec:conclude} concludes the paper and discusses future directions.
%
%
%
All program codes and experimental results are available at {\url{https://github.com/ku-dml/mol-infer/tree/master/Copolymer}}.

\section{An Extension of the \molinfer\ Framework to Copolymers}\label{sec:copolymer}
This section outlines an extension of our \molinfer\ framework to the copolymer setting.
In Section~\ref{sec:copolymerclass}, we summarize basic notation for copolymers.
Phase~1 (QSAR/QSPR) is presented in Section~\ref{sec:mvmodel}, and
Phase~2 (inverse QSAR/QSPR) is presented in Section~\ref{sec:mvinverse}.

For a vector $x \in \bbR^K$, the $j$-th entry of $x$ is denoted by $x(j), j\in [1,K]$.

\subsection{Basic Notation for Copolymers}\label{sec:copolymerclass}

In this subsection, we introduce basic notation for copolymers used in this study.
We restrict our attention to so-called \emph{linear} copolymers, 
for which available experimental and computational data are currently 
much more abundant than for more complex
architectures such as comb, star, or network polymers. 
The term \emph{copolymer} is therefore used
to mean a linear copolymer unless explicitly stated otherwise throughout this paper.

\begin{figure}[t!]
  \centering
  \begin{subcaptiongroup}
    \begin{subfigure}[b]{0.3\textwidth}
      \centering
      \includegraphics[width=\textwidth]{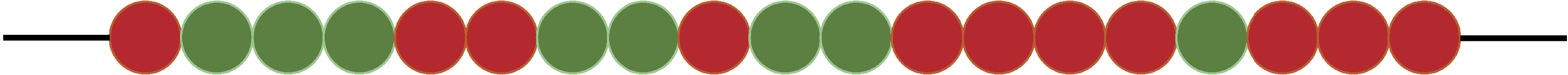}
      \caption{Random copolymer}
      \label{fig:copolymer-random}
    \end{subfigure}\hfill
    \begin{subfigure}[b]{0.3\textwidth}
      \centering
      \includegraphics[width=\textwidth]{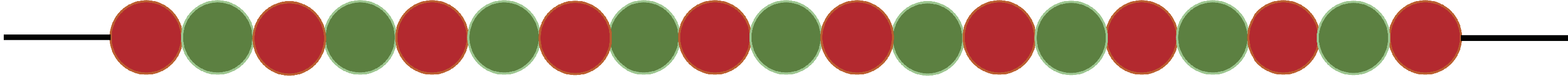}
      \caption{Alternating copolymer}
      \label{fig:copolymer-alternate}
    \end{subfigure}\hfill
    \begin{subfigure}[b]{0.3\textwidth}
      \centering
      \includegraphics[width=\textwidth]{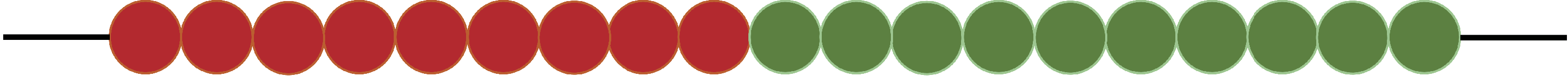}
      \caption{Block copolymer}
      \label{fig:copolymer-block}
    \end{subfigure}
  \end{subcaptiongroup}
  \caption{Examples of linear copolymers with different sequence distributions.}
  \label{fig:copolymer_classes}
\end{figure}

Even under the fixed linear architecture, 
copolymers differ in how monomer species are distributed along the backbone.
From the viewpoint of sequence distribution, 
common classes include random (or statistical), alternating, and block copolymers.
In a \emph{random copolymer}, the local monomer sequence follows a statistical rule and
does not exhibit long-range regularity. 
In an \emph{alternating copolymer}, different monomer units appear 
in an (almost) perfectly alternating fashion. 
In a \emph{block copolymer}, long contiguous segments (blocks) composed of a single monomer
species are covalently linked together along the chain.
Figure~\ref{fig:copolymer_classes} illustrates an example of each class mentioned above.
The importance of these sequence classes for tuning copolymer properties has been
highlighted in both classical sequence-distribution analyses and recent
copolymer-property studies~\cite{Meier:2019aa,Self:2022aa,Lefebvre:2004aa,Coldstream:2022aa}.
We refer to such sequence-distribution patterns as \emph{copolymer classes}.
In addition to these multi-monomer cases, 
a \emph{homopolymer} can be regarded as a special case of
a copolymer in which only one monomer species appears along the chain. 

\smallskip
We now formalize the notion of copolymers used in this paper.
Let $\calG$ be the set of all chemical graphs and let $\calGmono
\subseteq \calG$ denote the set of monomer units, typically determined
by the dataset under consideration.
We define a (linear) \emph{copolymer} to be a pair
\(
  P=(\mathbf{C},r),
\)
where
\begin{itemize}
  \item[-] $\mathbf{C}=(\bbC_1,\ldots,\bbC_m)\in \calGmono^m$ $(m\ge 1)$ is a finite
        indexed tuple of chemical graphs representing the monomer units that
        constitute the copolymer; and
  \item[-] $r : [1,m] \to \bbR_{>0}$ is a function satisfying
        $\sum_{i=1}^{m} r(i) = 1$, called the \emph{mixing ratio}, which represents the
        molar fraction of the $i$-th monomer in the copolymer.
\end{itemize}
Here, the ordering of $\mathbf{C}$ has no chemical meaning by itself; it is used only
to match each monomer $\bbC_i$ with its prescribed mixing ratio $r(i)$.
Let $\calP$ denote the set of all copolymers in this sense.
Under this definition, a homopolymer corresponds to the case \(m=1\) and
\(r(1)=1\).

Although sequence-distribution information can have a significant influence on
copolymer properties, the present study focuses on the composition-level information
contained in \(P=(\mathbf{C},r)\), namely the constituent monomer species and their mixing ratios.
This information is sufficient to define the copolymer representation introduced in
Section~\ref{sec:mvmodel}.
As will be demonstrated in Sections~\ref{sec:experiment_phase1} and~\ref{sec:experiment_phase2},
this choice leads to useful predictive performance in Phase~1 while preserving the
computational tractability of the MILP formulation in Phase~2.

\subsection{Mixing Vector Model: Phase 1}\label{sec:mvmodel}

\begin{figure}[t!]
  \centering
  \includegraphics[width=\textwidth]{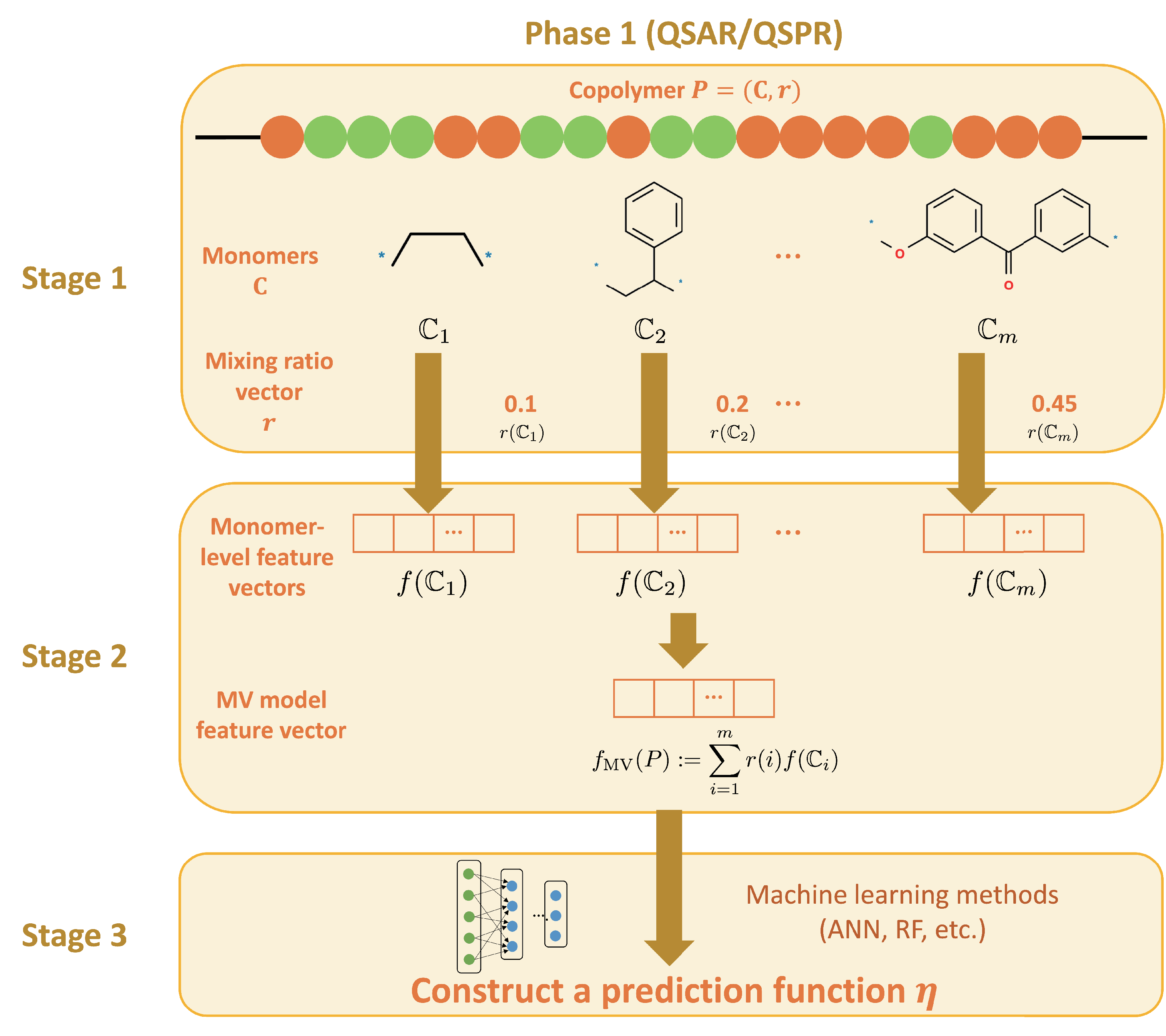}
  \caption{Illustration of Phase~1 (QSAR/QSPR) under the mixing vector model.}
  \label{fig:phase1_mv}
\end{figure}

In this subsection, we introduce the \emph{mixing vector (MV) model} as a feature representation for copolymers,
and explain how it integrates into Phase~1 of the \molinfer\ framework.
Figure~\ref{fig:phase1_mv} illustrates Phase~1 of the resulting modified framework for copolymers.

\paragraph{Stage 1}
As in \molinfer, for a target property $\pi$, we collect a dataset $D_\pi$ of copolymers $P$ together with their measured or simulated property values $a(P)$.

\paragraph{Stage 2}
Let $P=(\mathbf{C},r)\in\mathcal{P}$ be a copolymer as defined in
Section~\ref{sec:copolymerclass}, where
\(
  \mathbf{C}=(\bbC_1,\ldots,\bbC_m)
\)
is the indexed tuple of constituent monomers and $r$ is the mixing ratio.
We first compute monomer-level descriptors using the feature function
$f:\mathcal{G}_{\mathrm{mono}}\to\mathbb{R}^K$ proposed by Ido et al.~\cite{Ido:2024aa}
(see also Appendix~\ref{sec:descriptor}), where $K$ denotes the number of descriptors.
For each monomer graph $\bbC\in\mathcal{G}_{\mathrm{mono}}$, the vector $f(\bbC)$ consists solely of
graph-theoretic descriptors that are designed to be tractable with MILP formulations in the inverse-design phase.

The MV model maps a copolymer to a single vector in $\mathbb{R}^K$ by taking a convex combination of the
monomer descriptors according to the mixing ratio:
\[
  f_{\mathrm{MV}}(P)
  := \sum_{i=1}^{m} r(i)\, f(\bbC_i)\in \mathbb{R}^{K}.
\]
Here, $r(i)>0$ for all $i\in[1,m]$ and $\sum_{i=1}^{m} r(i)=1$.
A key advantage is that it does \emph{not} require any explicit sequence-distribution label.
Hence, the representation remains applicable even when the sequence-distribution type is missing,
uncertain, or intentionally ignored, as long as the indexed tuple of constituent monomers
$\mathbf{C}$ and the mixing ratio $r$ are available.

\paragraph{Stage~3}
The aim of Stage~3 is to construct a prediction function $\eta: \bbR^K \to \bbR$ from the feature vectors 
$f_{\mathrm{MV}} (P)$ obtained in Stage~2 and the property values $a(P)$.
The value $y:=\eta(f_{\mathrm{MV}}(P))$ will serve as the predicted value of the property $\pi$.

Due to the need for tractability in the MILP formulation in Phase~2
(the inverse design phase), we focus on model families that admit standard
MILP formulations or existing MILP encodings, such as multiple linear regression,
ReLU-based artificial neural networks, and tree-based models including decision
trees and random forests.
In practice, these choices are often sufficient to achieve strong predictive
performance while preserving end-to-end solvability.
The computational results are reported in Section~\ref{sec:experiment_phase1}.

\medskip
The simplicity of the MV model is an important feature of the present framework.
Since it aggregates monomer information through a weighted average, it provides a compact
descriptor that is directly expressible as linear constraints in MILP.
Moreover, the representation can be constructed from monomer composition and mixing ratio alone,
which are commonly available in many copolymer datasets.

Although different sequence arrangements with the same constituent monomers and mixing ratio may affect
property values~\cite{Meier:2019aa,Self:2022aa,Lefebvre:2004aa,Coldstream:2022aa},
the MV model deliberately focuses on this widely available composition-level information.
As will be shown in Sections~\ref{sec:experiment_phase1} and~\ref{sec:experiment_phase2},
this choice leads to useful predictive performance while preserving the computational tractability
of the inverse-design formulation.

\subsection{Mixing Vector Model: Phase 2}\label{sec:mvinverse}

\begin{figure}[t!]
  \centering
  \includegraphics[width=\textwidth]{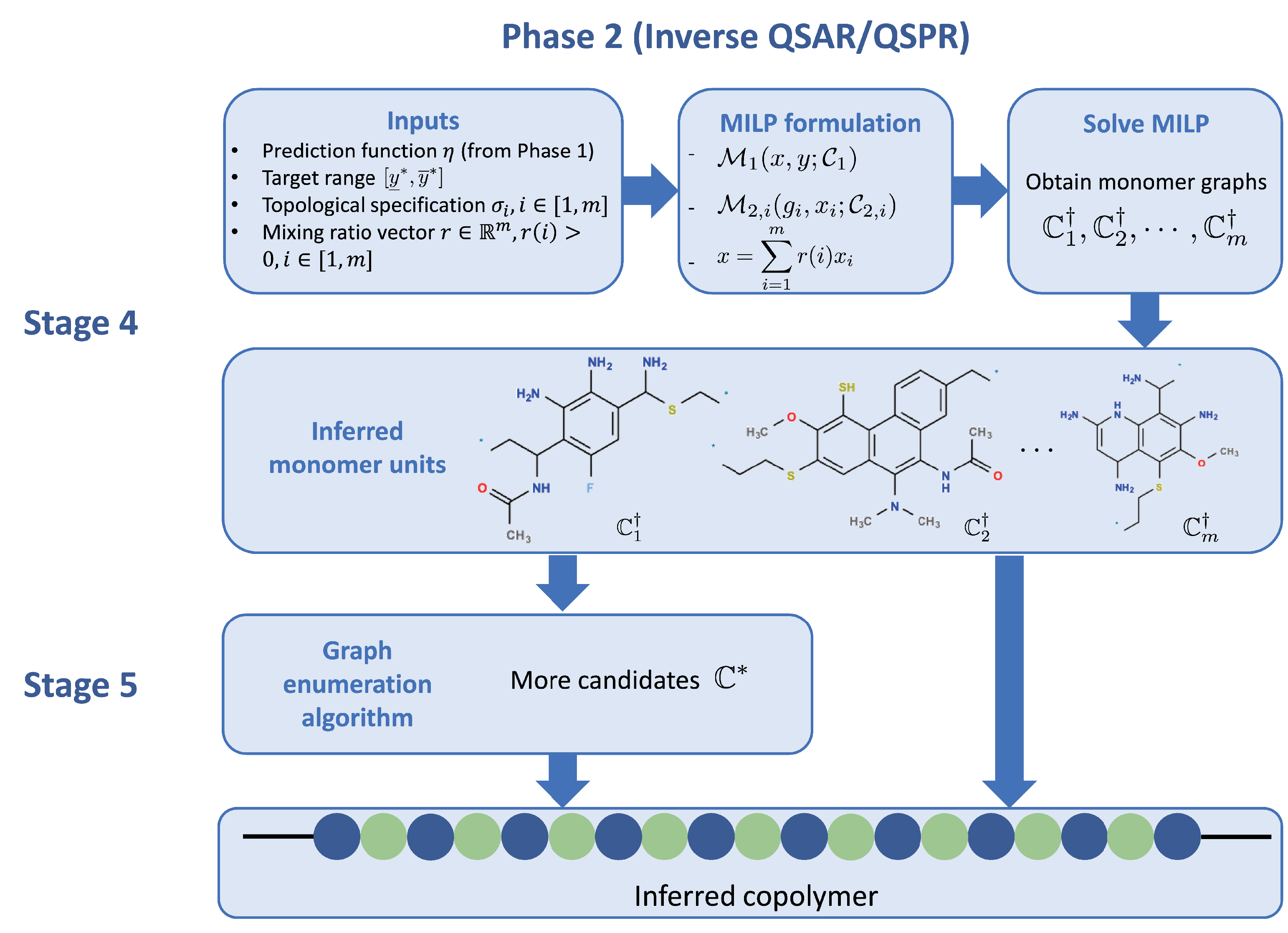}
  \caption{Illustration of Phase~2 (inverse QSAR/QSPR) under the mixing vector model. 
  The polymer chain shown in the lower part is schematic; the MV model specifies
the constituent monomers and their mixing ratio, but does not infer an explicit
sequence distribution.}
  \label{fig:phase2_mv}
\end{figure}

In this subsection, we formulate the Phase~2 (inverse QSAR/QSPR phase) of the extended framework
under the MV model introduced in Section~\ref{sec:mvmodel},
namely the Stages~4 and~5
in the original \molinfer\ framework, with Stage~4 modified so as to handle a copolymer described
by multiple monomers and their mixing ratio.
Figure~\ref{fig:phase2_mv} illustrates Phase~2 of the extended framework.

\paragraph{Stage~4}
Stage~4 is designed to infer the structures of constituent monomers under prescribed topological specifications 
and a prescribed mixing ratio vector. 

Let $\eta:\mathbb{R}^K\to\mathbb{R}$ be a prediction function constructed in Phase~1, Stage~3.
Given a target range $[\ylb,\yub]$, our goal is to infer
monomer structures that yield a copolymer whose predicted property value
$y=\eta(x)$ lies in $[\ylb,\yub]$,
where $x$ is the feature vector of the inferred copolymer under the MV model.

More precisely, we require a fixed positive integer $m$ that denotes the number of constituent
monomer positions in the copolymer representation, and a fixed mixing ratio vector
$r\in \bbR_{>0}^m$ satisfying $\sum_{i=1}^m r(i)=1$.
For each monomer position $i\in[1,m]$, we are also given a topological specification $\sigma_i$
that restricts the possible pattern of the inferred monomer graph (e.g., bounds on size, allowable atom/bond types,
and degree patterns; see Appendix~\ref{sec:specification} for a detailed description).
We denote by $\calG_{\mathrm{mono}}(\sigma_i)$ the set of chemical graphs satisfying $\sigma_i$.

The MILP formulation in Stage~4 in this extended framework 
simultaneously (i) simulates the computation of the prediction function $\eta$,
(ii) enforces the existence of valid monomer graphs $\bbC_i$ consistent with the specifications $\sigma_i$, and
(iii) couples the feature vectors of monomers under the MV model.
More precisely, the MILP formulation consists of the following constraints:
\begin{itemize}
  \item[-] $\calM_1(x,y;\calC_1)$: the computation process $y=\eta(x)$, where $\eta$ is the given prediction function;
  \item[-] $\calM_{2,i}(g_i,x_i;\calC_{2,i}), i\in[1,m]$: the constraints for $\bbC_i \in \calG_{\mathrm{mono}}(\sigma_i)$, together with
  the computation process of the feature function $x_i = f(\bbC_i)$ for each monomer position; and
  \item[-] the constraint $x = \sum_{i=1}^m r(i)x_i$ that simulates the MV model.
\end{itemize}
This constraint set is organized as a direct analogue of Stage~4 in \molinfer, with an additional constraint to calculate the feature
vector of the copolymer under the MV model.

A final feasibility constraint $\ylb \le y \le \yub$ is added to ensure that the predicted property value of the inferred copolymer 
falls in the target interval. If the MILP returns infeasible, it suggests that no copolymer exists in the given setting within the present model.
Additional practical requirements (e.g., bounds on size, counts of specific functional groups) can be incorporated as linear inequalities without
changing the MILP nature of the formulation.

The output of Stage~4 is an indexed tuple of monomer graphs
\(
  \mathbf{C}^{\dagger}
  :=
  (\bbC_1^\dagger,\bbC_2^\dagger,\ldots,\bbC_m^\dagger).
\)
A copolymer candidate is then obtained by setting
\(
  P^\dagger = (\mathbf{C}^{\dagger}, r).
\)

\paragraph{Stage~5}
After obtaining monomer solutions from Stage~4, we can apply the standard \molinfer\ enumeration step
to each monomer $\bbC_i^\dagger$ to generate alternative monomer graphs that share the same descriptor
vector $x_i^\dagger$ under $\sigma_i$, for example, using the dynamic-programming algorithm~\cite{Ido:2025aa}.
Combining such alternatives across $i\in[1,m]$ yields a family of copolymer candidates that preserve
the same MV feature vector $x$ and thus the same predicted property value $y$ under $\eta$.

This is an optional stage and is not the main object of the present study.
Its role is to provide additional structural candidates after Stage~4 has identified one feasible solution.
In particular, the main contribution of the present framework lies in the formulation of Stage~4,
namely the integration of the MV model into the inverse-design phase of \molinfer.

\smallskip
It should be noted that the MV model itself does not determine an explicit
sequence distribution along the polymer chain. Therefore, the output of 
the present framework should be interpreted as a composition-level copolymer
candidate, namely a collection of possible constituent monomers together with
the prescribed mixing ratio. When an explicit oligomer sequence is required for
a subsequent external computation, an additional modeling choice is needed.
For example, in Section~\ref{sec:experiment_phase2_compare}, we restrict attention to
binary alternating copolymers and instantiate each inferred monomer pair as
the two possible ordered alternating oligomers.

\section{Experimental Results}\label{sec:experiment}

To evaluate the effectiveness of our proposed inverse QSPR model for copolymers,
we collected several copolymer datasets from the literature.
We implemented the framework based on the MV model described in Section~\ref{sec:copolymer}.
All experiments were conducted on a MacBook Pro (Apple M3, 16 GB unified memory), 
and all MILP formulations were solved using CPLEX version 22.1.1~\cite{CPLEX} with a time limit of 3600 seconds.

\subsection{Phase 1: Predictive Performance}\label{sec:experiment_phase1}

We use multiple copolymer datasets collected from the literature to 
evaluate the predictive performance of our MV model.
Specifically, the following property datasets are used:
${}^{19}\mathrm{F}$ (${}^{19}\mathrm{F}$~NMR chemical shift) from Reis~et~al.~\cite{Reis:2021aa},
\textsc{EAvS} (EA vs SHE; electron affinity values on the standard hydrogen electrode scale), 
\textsc{IPvS} (IP vs SHE; ionization potential values on the standard hydrogen electrode scale), 
\textsc{OptG} (optical gap), and 
\textsc{OscS} (oscillator strength) from Bai~et~al.~\cite{Bai:2019aa}, 
\textsc{Tg} (glass transition temperature) from Brierley-Croft~et~al.~\cite{Brierley-Croft:2025aa}, 
\textsc{MelV} (melt viscosity) from Jain~et~al.~\cite{Jain:2025aa}, 
and \textsc{EA} (electron affinity), \textsc{IP} (ionization potential), and 
\textsc{ExE} (excitation energy) from Wilbraham~et~al.~\cite{Wilbraham:2019aa}. 
Following~\cite{Ido:2024aa}, each descriptor and the target property value were normalized
based on the range of values in each dataset.

Some datasets require additional preprocessing. 
For the dataset ${}^{19}\mathrm{F}$, we followed the preprocessing used by
Tao et al.~\cite{Tao:2022aa} and retained only the data points for which
a numerical positive ${}^{19}\mathrm{F}$ chemical-shift value is available.
The remaining entries are labeled as ``negative'' rather than being assigned
numerical chemical-shift values, and therefore cannot be used directly as
targets in the present regression setting.
For the dataset \textsc{MelV}, we used the base-10 logarithm of viscosity as the
target value, because viscosity values span several orders of magnitude.
Moreover, we added four additional descriptors: the base-10 logarithm of
\(M_w\), PDI, \(T\), and the base-10 logarithm of the shear rate. Here,
\(M_w\), PDI, and \(T\) denote molecular weight, polydispersity index, and
temperature, respectively. These variables are experimental conditions that
strongly affect melt viscosity. Data points with missing values in these
additional variables were removed so that all models were trained on a common
complete descriptor set.

The details of the datasets are summarized in Table~\ref{table:phase1a}.
We use the following notations:
\begin{itemize}
\item[-] $D_\pi$: the name of the dataset;
\item[-] Ref.: the source of the dataset;
\item[-] $\#$Copolymer: the number of copolymers (including homopolymers) $P$ in the dataset;
\item[-] Copolymer class: the classes of the copolymers contained in the dataset;
\item[-] $\underline{n}, \overline{n}$: the minimum and maximum numbers of 
non-hydrogen atoms of the repeating units in their monomer forms (see Appendix~\ref{sec:polymer} for a detailed description);
\item[-] $\underline{a}, \overline{a}$: the minimum and maximum property values $a(P)$; and
\item[-] $K$: the number of descriptors included in a feature vector for a copolymer $P$.
\end{itemize}

\begin{table}[t!]\caption{
Summary of the copolymer datasets.
}
\begin{center}\scalebox{0.99}{
\begin{tabular}{c c c c c c c} \hline
$D_\pi$ & Ref. & $\#$Copolymer & Copolymer class & $\underline{n}, \overline{n}$  & $\underline{a}, \overline{a}$ & $K$ \\ \hline
${}^{19}\mathrm{F}$  & \cite{Reis:2021aa} & 263  & random      & 8, 27  & 15, 111        & 48  \\ 
\textsc{EAvS} & \cite{Bai:2019aa}  & 6327 & alternating   & 5, 28  & -3.572, 0.281  & 159 \\
\textsc{IPvS} & \cite{Bai:2019aa}  & 6327 & alternating   & 5, 28  & -0.360, 2.737  & 159 \\
\textsc{OptG} & \cite{Bai:2019aa}  & 6327 & alternating   & 5, 28  & 0.288, 4.839   & 159 \\
\textsc{OscS} & \cite{Bai:2019aa}  & 6327 & alternating   & 5, 28  & 0.0, 14.088    & 159 \\
\textsc{Tg}   & \cite{Brierley-Croft:2025aa} & 146  & homopolymer, random & 7, 57  & 368, 551       & 51  \\
\textsc{MelV} & \cite{Jain:2025aa}  & 1920 & homopolymer, random & 2, 47  & -2.346, 12.730 & 124 \\
\textsc{EA}   & \cite{Wilbraham:2019aa} & 5000 & alternating   & 4, 31  & 0.697, 3.809   & 131 \\
\textsc{IP}   & \cite{Wilbraham:2019aa} & 5000 & alternating   & 4, 31  & 5.116, 7.793   & 131 \\
\textsc{ExE}  & \cite{Wilbraham:2019aa} & 5000 & alternating  & 4, 31  & 1.604, 3.979   & 131 \\ \hline
\end{tabular}
}
\end{center}\label{table:phase1a}
\end{table}

We then computed the feature vector for each copolymer following the mixing vector model
described in Section~\ref{sec:mvmodel}.
Using the resulting feature representation $f_{\mathrm{MV}}(P)$, we constructed a prediction function
$\eta$ with three learning algorithms:
an artificial neural network (ANN),
multiple linear regression with reduced quadratic descriptors (R-MLR)~\cite{Zhu:2025aa},
and random forest (RF).
For each dataset and each learning algorithm, we used the same evaluation protocol in order
to compare the predictive performance of the MV representation under a unified setting.
We first conducted a preliminary experiment to select hyperparameters, following the tuning
procedures described in~\cite{Zhu:2025aa,Azam:2021ab}. 
The selected hyperparameters were then fixed in the subsequent repeated cross-validation
experiments.
Specifically, we employed a repeated 5-fold cross-validation procedure.
For each repetition, the dataset $D_\pi$ was randomly partitioned into five disjoint subsets
$D_\pi^{(k)}$ $(k\in[1,5])$ of approximately equal size.
For each fold $k$, we trained a model $\eta^{(k)}$ on the training set
$D_\pi\setminus D_\pi^{(k)}$ and evaluated its performance on the held-out test set
$D_\pi^{(k)}$.
To reduce variability due to random splits, we repeated the entire 5-fold cross-validation
procedure ten times for each dataset and each learning method, resulting in 50 train/test
scores for each combination.
The reported values are aggregated from these 50 scores.

To evaluate the predictive performance of the constructed prediction functions, 
we use three standard evaluation metrics: the coefficient of determination \(R^2\), 
the mean absolute error (MAE), and the root mean squared error (RMSE).
Formally, let \(D\) be a dataset of copolymers.
For a prediction function \(\eta:\mathbb{R}^K\to\mathbb{R}\), define the sum of squared errors (SSE) on \(D\) by
\[
\mathrm{SSE}(\eta;D)
\triangleq
\sum_{P\in D}\bigl(a(P)-\eta(f_{\mathrm{MV}}(P))\bigr)^2.
\]
Let
\(
\bar{a}\triangleq \frac{1}{|D|}\sum_{P\in D} a(P)
\)
be the mean of the observed values, and define the total sum of squares (TSS) by
\[
\mathrm{TSS}(D)
\triangleq
\sum_{P\in D}\bigl(a(P)-\bar{a}\bigr)^2.
\]
The \emph{coefficient of determination} (the \(R^2\) score) is defined by
\[
R^2(\eta;D)
\triangleq
1-\frac{\mathrm{SSE}(\eta;D)}{\mathrm{TSS}(D)}.
\]
In addition, the \emph{mean absolute error} (MAE) and the \emph{root mean squared error} (RMSE) are defined by
\[
\mathrm{MAE}(\eta;D)
\triangleq
\frac{1}{|D|}\sum_{P\in D}\bigl|a(P)-\eta(f_{\mathrm{MV}}(P))\bigr|,
\qquad
\mathrm{RMSE}(\eta;D)
\triangleq
\sqrt{
\frac{1}{|D|}\sum_{P\in D}
\bigl(a(P)-\eta(f_{\mathrm{MV}}(P))\bigr)^2
}.
\]

The results of Phase~1 are summarized in Table~\ref{table:phase1b}. We use the following notations:
\begin{itemize}
\item[-] $D_\pi$: the name of the dataset;
\item[-] ML.: the machine-learning method;
\item[-] $R^2$, MAE, RMSE: the value of the coefficient of determination, the mean absolute error, and the root mean squared error, respectively;
\item[-] Train/Test: the median of train/test evaluation metrics ($R^2$, MAE, or RMSE) over all 50 trials in ten cross-validations; and
\item[-] boldface indicates the best test scores achieved among the three different machine-learning methods for each dataset.
\end{itemize}

\begin{table}[t!]
\centering
\caption{Results of predictive performance in Stage~3.}
\label{table:phase1b}
\scalebox{0.97}{
\begin{tabular}{c | c | c c | c c | c c}
\toprule
$D_\pi$ & ML & \multicolumn{2}{c}{$R^2$} & \multicolumn{2}{c}{MAE} & \multicolumn{2}{c}{RMSE} \\
\cmidrule(lr){3-4}\cmidrule(lr){5-6}\cmidrule(lr){7-8}
 &  & Train & Test & Train & Test & Train & Test \\
\midrule
 & ANN & 0.830 & 0.772 & 7.291 & 8.358 & 9.537 & 10.750 \\
${}^{19}\mathrm{F}$ & R-MLR & 0.761 & 0.731 & 8.702 & 8.878 & 11.265 & 11.629 \\
& RF & 0.913 & \textbf{0.787} & 4.902 & \textbf{7.944} & 6.771 & \textbf{10.379} \\ \hline

& ANN & 0.980 & \textbf{0.961} & 0.065 & \textbf{0.092} & 0.089 & \textbf{0.124} \\
\textsc{EAvS} & R-MLR & 0.951 & 0.947 & 0.107 & 0.111 & 0.140 & 0.145 \\
 & RF & 0.987 & 0.943 & 0.050 & 0.107 & 0.072 & 0.151 \\ \hline

& ANN & 0.980 & \textbf{0.956} & 0.045 & \textbf{0.070} & 0.062 & \textbf{0.092} \\
\textsc{IPvS} & R-MLR & 0.934 & 0.930 & 0.086 & 0.089 & 0.112 & 0.116 \\
 & RF & 0.981 & 0.906 & 0.043 & 0.095 & 0.060 & 0.134 \\ \hline

& ANN & 0.941 & \textbf{0.866} & 0.076 & 0.116 & 0.108 & \textbf{0.161} \\
\textsc{OptG} & R-MLR & 0.852 & 0.845 & 0.124 & 0.128 & 0.170 & 0.174 \\
 & RF & 0.945 & 0.857 & 0.072 & \textbf{0.113} & 0.105 & 0.168 \\ \hline

 & ANN & 0.696 & 0.591 & 0.911 & 1.073 & 1.223 & 1.419 \\
\textsc{OscS} & R-MLR & 0.651 & \textbf{0.633} & 0.998 & \textbf{1.028} & 1.312 & \textbf{1.349} \\
 & RF & 0.824 & 0.589 & 0.695 & 1.066 & 0.934 & 1.423 \\ \hline

& ANN & 0.871 & 0.813 & 9.032 & 10.666 & 11.836 & 14.076 \\
\textsc{Tg} & R-MLR & 0.943 & \textbf{0.924} & 5.432 & \textbf{6.372} & 7.843 & \textbf{8.996} \\
 & RF & 0.945 & 0.767 & 5.025 & 11.000 & 7.697 & 16.326 \\ \hline

& ANN & 0.980 & 0.944 & 0.199 & 0.273 & 0.295 & 0.495 \\
\textsc{MelV} & R-MLR & 0.961 & \textbf{0.953} & 0.259 & 0.282 & 0.418 & \textbf{0.452} \\
& RF & 0.992 & 0.949 & 0.087 & \textbf{0.221} & 0.186 & 0.480 \\ \hline

 & ANN & 0.980 & \textbf{0.943} & 0.057 & \textbf{0.097} & 0.085 & \textbf{0.143} \\
\textsc{EA} & R-MLR & 0.932 & 0.929 & 0.119 & 0.123 & 0.157 & 0.161 \\
& RF & 0.986 & 0.938 & 0.046 & 0.100 & 0.071 & 0.149 \\ \hline

 & ANN & 0.980 & \textbf{0.937} & 0.043 & \textbf{0.076} & 0.061 & \textbf{0.110} \\
\textsc{IP} & R-MLR & 0.927 & 0.922 & 0.090 & 0.093 & 0.118 & 0.122 \\
& RF & 0.984 & 0.923 & 0.038 & 0.084 & 0.055 & 0.121 \\ \hline

 & ANN & 0.960 & 0.847 & 0.054 & 0.112 & 0.078 & 0.153 \\
\textsc{ExE} & R-MLR & 0.816 & 0.807 & 0.131 & 0.135 & 0.168 & 0.172 \\
 & RF & 0.964 & \textbf{0.864} & 0.054 & \textbf{0.106} & 0.074 & \textbf{0.145} \\
\bottomrule
\end{tabular}}
\end{table}

We also compare our results with previously reported studies.
In this study, we adopted a unified repeated 5-fold cross-validation protocol
for all datasets and learning methods in order to evaluate the MV representation
under a common experimental setting. Since train/test splitting, preprocessing
procedures, descriptors, and cross-validation protocols are different from those
in the literature, these comparisons should be interpreted only as qualitative
references rather than strict head-to-head benchmarks.
Table~\ref{table:phase1_comparison} gives a summary of the datasets
for which directly related benchmark results are available.

\begin{table}[t!]
\caption{
Summary of qualitative comparisons with previously reported studies.
}
\label{table:phase1_comparison}
\begin{center}
\scalebox{0.95}{
\begin{tabular}{c l l}
\toprule
$D_\pi$ & \multicolumn{1}{c}{Previously reported results} & \multicolumn{1}{c}{Our results} \\
\midrule
\textsc{EA}
&
RMSE: \(<0.12\) eV~\cite{Wilbraham:2019aa}
&
RMSE: \(0.143\) eV (ANN)
\\
& RMSE: \(0.09\)--\(0.19\) eV~\cite{Tao:2022aa} & \\ \hline

\textsc{IP}
&
RMSE: \(<0.12\) eV~\cite{Wilbraham:2019aa}
&
RMSE: \(0.110\) eV (ANN)
\\
& RMSE: \(0.09\)--\(0.19\) eV~\cite{Tao:2022aa} & \\ \hline

\textsc{ExE}
&
RMSE: \(<0.12\) eV~\cite{Wilbraham:2019aa}
&
RMSE: \(0.145\) eV (RF)
\\ \hline

\textsc{Tg}
&
RMSE: \(8\) K (QSPR-GAP); \(5\)--\(12\) K (Lasso)~\cite{Brierley-Croft:2025aa}
&
RMSE: \(8.996\) K (R-MLR)
\\
\bottomrule
\end{tabular}
}
\end{center}
\end{table}

For the datasets \textsc{EA}, \textsc{IP}, and \textsc{ExE} of binary alternating copolymers,
Wilbraham~et~al.~\cite{Wilbraham:2019aa}
reported that a neural network model predicts EA, IP, and \textsc{ExE} simultaneously with an RMSE
of less than $0.12$ eV, while a linear regression baseline with the same fingerprint representation yields
an overall RMSE of $0.30$ eV.
In a subsequent comparative study, Tao~et~al.~\cite{Tao:2022aa} further examined machine-learning
strategies on the same Wilbraham dataset and reported RMSE values around $0.09$--$0.19$ eV, depending on
the model architecture.
Our results are broadly consistent with these earlier studies,
in the sense that ANN performs very strongly and yields competitive predictive performance.
In particular, our ANN model attains test RMSE values of $0.143$ eV for \textsc{EA} and $0.110$ eV for \textsc{IP}.
For \textsc{ExE}, ANN and RF yield test RMSE values of $0.153$ eV and $0.145$ eV, respectively.

For the dataset \textsc{Tg}, Brierley-Croft~et~al.~\cite{Brierley-Croft:2025aa} reported that their QSPR-GAP framework
achieves a median RMSE of about $8$ K, with the lasso-based model typically lying in the range of about
$5$--$12$ K and outperforming pure QSPR baselines.
Our best model on the corresponding \textsc{Tg} dataset is R-MLR, which gives a test RMSE of $8.996$ K,
together with a test $R^2$ of $0.924$.
This level of accuracy is therefore close to the scale reported in that work.

Taken together, these comparisons indicate that the predictive performance of
the proposed MV model is within a practically acceptable range and is 
comparable with the existing models.

Not all source papers provide directly comparable benchmarks for the present task.
For example, Jain~et~al.~\cite{Jain:2025aa} focused on extrapolative prediction for melt viscosity and used a different splitting strategy and evaluation metric,
while Reis~et~al.~\cite{Reis:2021aa} studied copolymer $^{19}$F MRI agents in terms of SNR rather than direct prediction of $^{19}$F chemical shift.
Likewise, Bai~et~al.~\cite{Bai:2019aa} used machine learning mainly to 
relate hydrogen-evolution performance to multiple computed descriptors rather than to predict each descriptor itself from copolymer structure.
Therefore, for these datasets, only a qualitative comparison is possible.
Nevertheless, the proposed mixing vector model achieves practically useful predictive performance across a broad range of copolymer datasets.
In particular, high test \(R^2\) values were obtained for datasets such as
\textsc{EAvS} (\(0.961\) by ANN), \textsc{IPvS} (\(0.956\) by ANN), \textsc{Tg} (\(0.924\) by R-MLR),
\textsc{MelV} (\(0.953\) by R-MLR), \textsc{EA} (\(0.943\) by ANN), and
\textsc{IP} (\(0.937\) by ANN),
suggesting that a weighted combination of monomer descriptors already captures a substantial part of the structure-property relationship in these settings.
On the other hand, the prediction accuracy for \textsc{OscS} was notably lower than for the other properties,
which may reflect the greater sensitivity of this property to electronic and sequence-dependent effects not explicitly represented in the MV model.

These results should be interpreted with the limitation of the MV representation
in mind. Since \(f_{\mathrm{MV}}\) depends only on the constituent monomers and
their mixing ratio, it is invariant under changes of sequence distribution. For
datasets consisting of a single copolymer class, such as alternating copolymers,
the sequence class is effectively fixed by the dataset, and the reported
performance should not be interpreted as evidence that the MV model can
distinguish random, alternating, and block architectures. Rather, the results
show that, within such fixed or weakly heterogeneous regimes, composition-level
information already provides a useful MILP-compatible baseline.

Overall, our computational results demonstrate the usefulness of the MV model as a simple, tractable, and reasonably
effective baseline representation for copolymer property prediction.

\subsection{Phase 2: Computational Effectiveness}\label{sec:experiment_phase2}

\begin{figure}[t!]
  \centering
  \begin{subcaptiongroup}
    \begin{subfigure}[b]{0.34\textwidth}
      \centering
      \includegraphics[width=\textwidth]{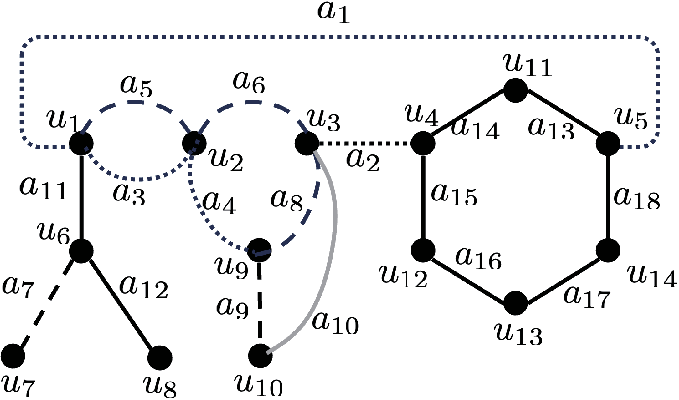}
      \caption{$I_a$}
      \label{fig:seed-graph-a}
    \end{subfigure}
    \hspace{0.02\textwidth}
    \begin{subfigure}[b]{0.36\textwidth}
      \centering
      \includegraphics[width=\textwidth]{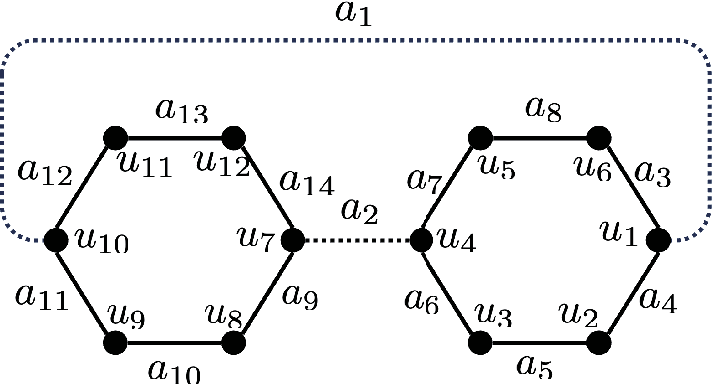}
      \caption{$I_b$}
      \label{fig:seed-graph-b}
    \end{subfigure}
    \hspace{0.02\textwidth}
    \begin{subfigure}[b]{0.20\textwidth}
      \centering
      \includegraphics[width=\textwidth]{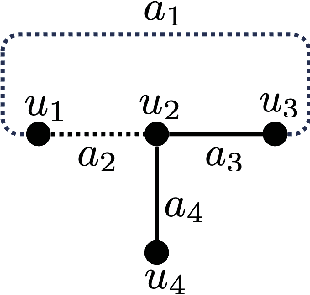}
      \caption{$I_c$}
      \label{fig:seed-graph-c}
    \end{subfigure}
  \end{subcaptiongroup}
  \caption{Illustrations of the seed graphs for the instances $I_a$, $I_b$, and $I_c$ used in Stage~4 of Phase~2, respectively.}
  \label{fig:seed-graph-abc}
\end{figure}

To evaluate the computational effectiveness of the proposed inverse-design model, 
we conducted numerical experiments for Phase~2 to infer copolymers with desired property values.
We used three types of seed graphs $I_a$, $I_b$, and $I_c$, as illustrated in Figure~\ref{fig:seed-graph-abc}.
See Appendix~\ref{sec:test_instances} for a detailed description of these instances and the associated topological specifications.
We would like to emphasize that
the experiment in this subsection should be understood primarily as an evaluation of 
the computational feasibility of Stage~4, rather than the ultimate chemical quality or synthetic accessibility of the inferred copolymers.

We executed Stage~4 on the following three datasets:
\textsc{IPvS} and \textsc{OptG} from Bai~et~al.~\cite{Bai:2019aa},  and
\textsc{Tg} from Brierley-Croft~et~al.~\cite{Brierley-Croft:2025aa}.
For \textsc{IPvS} and \textsc{OptG}, we used the prediction functions constructed by ANN,
and for \textsc{Tg}, we used the one constructed by R-MLR, following the experimental results in Stage~3 of Phase~1.
For \textsc{IPvS} and \textsc{OptG}, we assumed $m=2$ (i.e., we would like to infer a copolymer that consists of two monomers). 
We used two seed graphs from $I_a$, $I_b$, and $I_c$, and the mixing ratio vector $(0.5,0.5)$
to infer a copolymer with two constituent monomers.
For \textsc{Tg}, we assumed $m=3$ and used three seed graphs and the mixing ratio vector $(0.3,0.3,0.4)$
to infer a copolymer with three constituent monomers.
These settings are chosen only to provide representative multi-monomer inverse-design instances
for testing the computational behavior of the MILP formulation.

\begin{table}[t!]\caption{
Results of Stage~4 for the datasets \textsc{IPvS}, \textsc{OptG}, and \textsc{Tg}.
}
\begin{center}\scalebox{0.87}{
\begin{tabular}{c c c c c | c c c c c} \toprule
$D_\pi$ & ML & Instances & $r$ & $\ylb, \yub$ & \#v & \#c & I-time & $n$ & $\eta$ \\ \midrule
                   &          &                  &            & -0.1, 0.1 & 21183 & 21933 & 119.035 & 40, 31 & 0.090 \\
                   &          &                  &            & 0.1, 0.3 &  21183 & 21933 & 404.682 & 50, 24 & 0.241 \\
\textsc{IPvS} & ANN & $I_a, I_b$ & (0.5, 0.5) & 0.3, 0.5 &  21183 & 21933 & 242.039 & 50, 30 & 0.362 \\
                   &          &                  &            & 0.5, 0.7 &  21183 & 21933 & 111.630 & 48, 25 & 0.569 \\
                   &          &                  &            & 0.7, 0.9 &  21183 & 21933 & 878.439 & 50, 25 & 0.877 \\ \hline
                              &          &                 &             &  2.5, 2.7 & 20227 & 20825 & 214.444 & 50, 23 & 2.651 \\
                              &          &                 &               & 2.7, 2.9 &  20227 & 20825 & 147.496 & 50, 16 & 2.900 \\
\textsc{OptG} & ANN & $I_a, I_c$ & (0.5, 0.5) & 2.9, 3.1 &  20227 & 20825 & 152.234 & 50, 18 & 2.903\\
                               &          &                 &            & 3.1, 3.3 &  20227 & 20825 & 615.817 & 50, 18 & 3.228 \\
                               &          &                 &            & 3.3, 3.5 &  20227 & 20825 & 683.709 & 50, 23 & 3.439 \\ \hline
                              &          &                           &             &  400, 420 & 26893 & 27126 & 693.164 & 50, 31, 24 & 404.749 \\
                              &          &                            &               & 420, 440 &  26893 & 27126 & 496.320 & 50, 33, 25 & 439.146 \\
\textsc{Tg}                   & R-MLR & $I_a, I_b, I_c$ & (0.3, 0.3, 0.4) & 440, 460 &  26893 & 27126 & 148.047 & 50, 33, 17 & 458.998 \\
                               &          &                           &            & 460, 480 &  26893 & 27126 & 210.813 & 50, 25, 18 & 467.537 \\
                               &          &                           &            & 480, 500 &  26893 & 27126 & 3060.922 & 48, 35, 25 & 491.534 \\
\bottomrule
\end{tabular}
}
\end{center}\label{table:phase2a}
\end{table}

The computational results are shown in Table~\ref{table:phase2a}. We use the following notations:
\begin{itemize}
\item[-] $D_\pi$: the name of the dataset;
\item[-] ML: the machine-learning method used to construct the prediction function in Stage~3 of Phase~1;
\item[-] Instances: the instances corresponding to the constituent monomer positions
in the copolymer $P=(\mathbf{C},r)$ to be inferred;
\item[-] $r$: the given mixing ratio vector;
\item[-] $\ylb, \yub$: the target range $[\ylb, \yub]$ for the predicted property value;
\item[-] $\#v$ (respectively $\#c$): the number of variables (respectively, constraints) in the MILP;
\item[-] I-time: the time (in seconds) to solve the MILP;
\item[-] $n$: the numbers of non-hydrogen atoms of the inferred monomers, respectively; and
\item[-] $\eta$: the predicted property value of the inferred copolymer by the prediction function.
\end{itemize}

The main purpose of this subsection is to examine the computational feasibility of the proposed inverse-design model.
Table~\ref{table:phase2a} shows that the resulting MILP formulations involve roughly
$2\times 10^4$--$3\times 10^4$ variables and constraints, yet many target ranges were solved within
practical computation time; all cases were solved within one hour.
In particular, for the two-monomer instances arising from \textsc{IPvS} and \textsc{OptG},
the running time varies from about $10^2$ to $10^3$ seconds, and the inferred monomers typically contain
around $50$ and $20$ non-hydrogen atoms, respectively.
Likewise, for the three-monomer \textsc{Tg} instances, the running time remains mostly within the same order,
except for the most difficult target range, even though the inferred monomers typically contain around
$50$, $30$, and $25$ non-hydrogen atoms, respectively.
These results suggest that the extension preserves the tractability of the original \molinfer\ framework
even in the multi-monomer setting, and that the proposed MILP formulation remains practically solvable
even for relatively large monomer graphs.
Moreover, the predicted values of the inferred copolymers lie inside the prescribed target ranges,
indicating that the MILP formulation correctly simulates the learned prediction function with the MV coupling constraint.
This confirms that the proposed formulation preserves the key advantage of \molinfer, namely,
the ability to solve the inverse problem with guaranteed optimality under the given learned model and structural constraints.

\begin{figure}[t!]
  \centering
  \begin{subfigure}[t!]{\textwidth}
    \centering
    \begin{minipage}[t!]{0.35\textwidth}
      \centering
      \includegraphics[width=\textwidth]{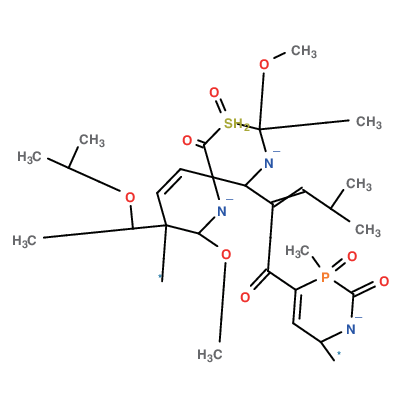}
    \end{minipage}
    \begin{minipage}[t!]{0.35\textwidth}
      \centering
      \includegraphics[width=\textwidth]{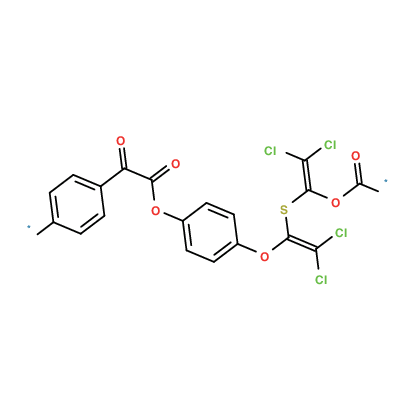}
    \end{minipage}
    \caption{}
    \label{fig:MILP_1_IP}
  \end{subfigure}
  \begin{subfigure}[t!]{\textwidth}
    \centering
    \begin{minipage}[t!]{0.35\textwidth}
      \centering
      \includegraphics[width=\textwidth]{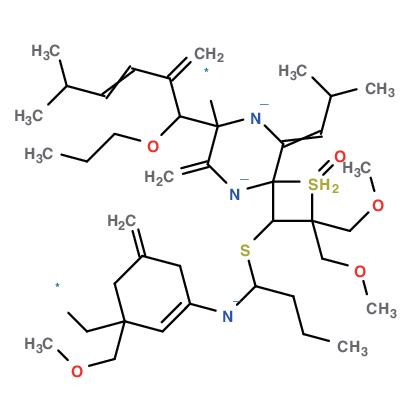}
    \end{minipage}
    \begin{minipage}[t!]{0.35\textwidth}
      \centering
      \includegraphics[width=\textwidth]{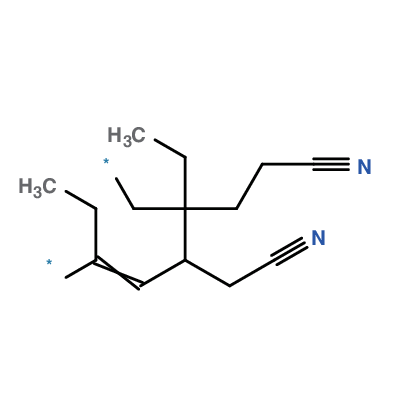}
    \end{minipage}
    \caption{}
    \label{fig:MILP_1_OG}
  \end{subfigure}
  \begin{subfigure}[t!]{\textwidth}
    \centering
    \begin{minipage}[t!]{0.32\textwidth}
      \centering
      \includegraphics[width=\textwidth]{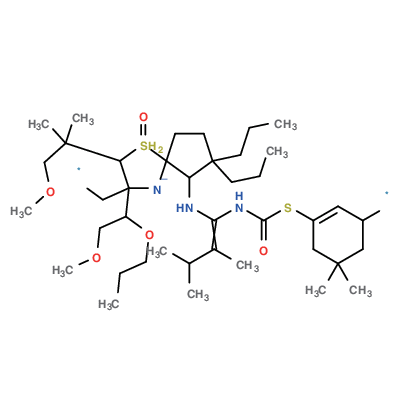}
    \end{minipage}
    \begin{minipage}[t!]{0.32\textwidth}
      \centering
      \includegraphics[width=\textwidth]{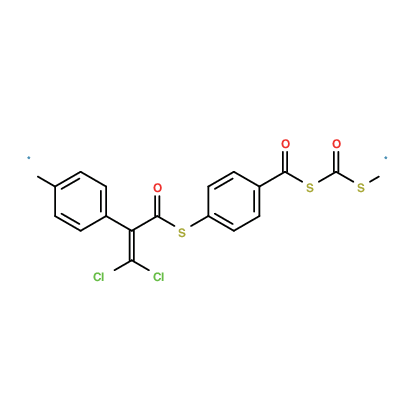}
    \end{minipage}
    \begin{minipage}[t!]{0.30\textwidth}
      \centering
      \includegraphics[width=\textwidth]{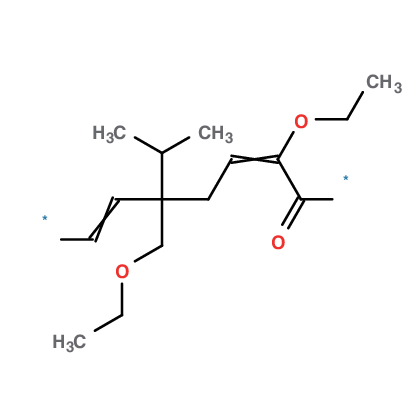}
    \end{minipage}
    \caption{}
    \label{fig:MILP_1_Tg}
  \end{subfigure}
  \caption{Examples of constituent monomers inferred in Stage~4. 
  (a) Monomer pair inferred for \textsc{IPvS} with target range $[\ylb,\yub]=[0.3,0.5]$. 
  (b) Monomer pair inferred for \textsc{OptG} with target range $[\ylb,\yub]=[2.9,3.1]$. 
  (c) Three monomers inferred for \textsc{Tg} with target range $[\ylb,\yub]=[460,480]$.}
  \label{fig:phase2_MILP_1}
\end{figure}

Examples of constituent monomers of the inferred copolymers are shown in Figure~\ref{fig:phase2_MILP_1}.

\subsection{A Comparison Study for External Consistency}\label{sec:experiment_phase2_compare}

In this subsection, we conduct an additional comparison study to examine the
self-consistency of the copolymers inferred in Stage~4.
We focus on the \textsc{EA} and \textsc{IP} datasets of binary alternating
copolymers from Wilbraham~et~al.~\cite{Wilbraham:2019aa}, since that work
provides a concrete computational workflow based on extended tight-binding
(xTB) methods.

The purpose of this comparison is not to claim that the inferred copolymers
experimentally realize the prescribed \textsc{EA}/\textsc{IP} values.
Rather, our aim is to check whether the candidates obtained by the
MILP-based inverse-design procedure are consistent with the same computational
pipeline used to construct the training data.
For this reason, we use the same monomer pairs as those in the dataset of
Wilbraham~et~al.~\cite{Wilbraham:2019aa}, but newly compute their
\textsc{EA}/\textsc{IP} values using our own implementation of the xTB-based
pipeline under the same overall computational setting.
The resulting re-computed dataset is then used both for constructing the
prediction functions and for the subsequent post hoc evaluation of the inferred
copolymers.

More concretely, the comparison proceeds as follows.
First, we re-compute the \textsc{EA}/\textsc{IP} values of the original
Wilbraham monomer pairs using the xTB-based pipeline.
Second, we train prediction functions on these re-computed datasets and use
them in Stage~4 to infer monomer pairs whose predicted values lie in prescribed
target ranges.
Finally, the inferred monomer pairs are converted into explicit alternating
oligomer structures and evaluated again by the same xTB-based pipeline.
The values obtained in this final evaluation are then compared with the values
predicted by the learned model.

This choice should be understood as a practical device for checking the
inverse-design framework within a unified computational setting.
Using the same xTB-based protocol for dataset generation and post hoc
evaluation avoids mixing different sources of computational error.
Since the accurate computation of \textsc{EA}/\textsc{IP} values for conjugated
copolymers is itself a nontrivial problem~\cite{Su:2019aa}, the present
comparison should be interpreted as a consistency check within a fixed
computational framework, rather than as a claim of chemically exact property
determination.

For the re-computation, we used essentially the same computational setting as in~\cite{Wilbraham:2019aa}.
In brief, geometries were optimized with the GFN-xTB family of methods~\cite{Grimme:2017aa},
and \textsc{EA} and \textsc{IP} were evaluated using IPEA-xTB~\cite{Asgeirsson:2017aa},
a variant of GFN-xTB.
To approximate the environment of a polymer chain in an amorphous polymeric solid,
we used the generalized Born surface area (GBSA) solvation model with the default benzene parameters distributed with \texttt{xtb}~\cite{Grimme-xtb}.
In the computation, polymers were modeled as oligomers containing eight monomer units in total.
Following this setting, our implementation constructs an alternating oligomer consisting of eight monomer units
from each monomer pair, optimizes its geometry with GFN2-xTB under GBSA (benzene),
and computes vertical \textsc{EA}/\textsc{IP} by \texttt{xtb --vipea} in GFN1 mode under the same GBSA setting.

\begin{figure}[t!]
  \centering
  \begin{subcaptiongroup}
    \begin{subfigure}[b]{0.26\textwidth}
      \centering
      \includegraphics[width=\textwidth]{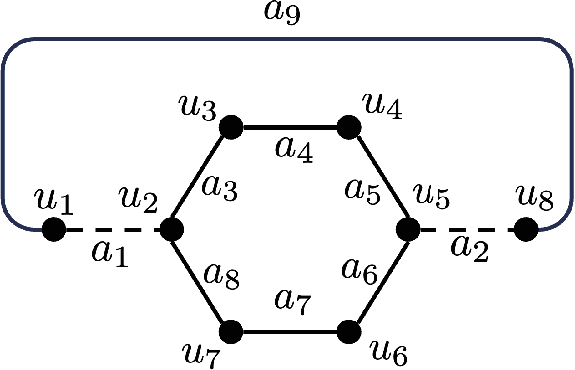}
      \caption{$I_{d1}$}
      \label{fig:seed-graph-d1}
    \end{subfigure}
    \hspace{0.02\textwidth}
    \begin{subfigure}[b]{0.26\textwidth}
      \centering
      \includegraphics[width=\textwidth]{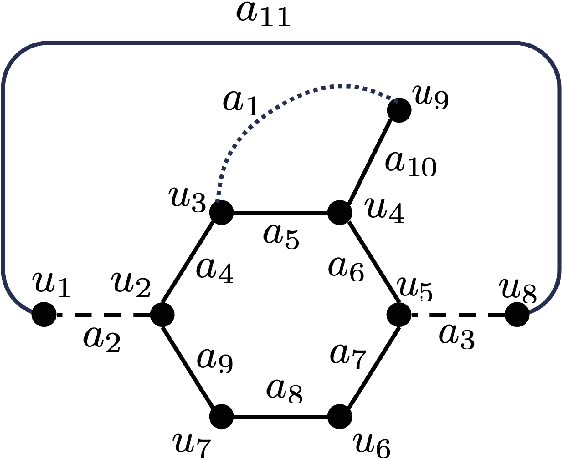}
      \caption{$I_{d2}$}
      \label{fig:seed-graph-d2}
    \end{subfigure}
    \hspace{0.02\textwidth}
    \begin{subfigure}[b]{0.38\textwidth}
      \centering
      \includegraphics[width=\textwidth]{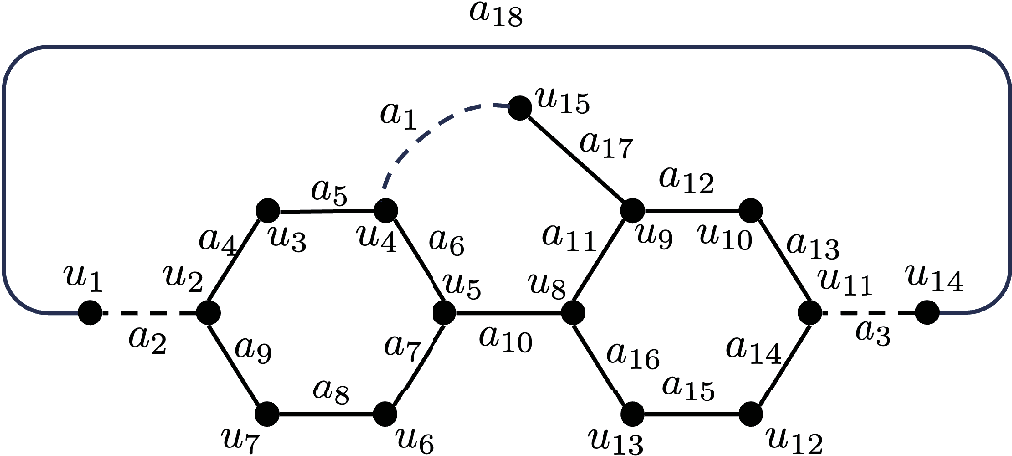}
      \caption{$I_{d3}$}
      \label{fig:seed-graph-d3}
    \end{subfigure}
  \end{subcaptiongroup}
  \caption{Illustrations of the seed graphs for the instances $I_{d1}$, $I_{d2}$, and $I_{d3}$ used in the comparison study, respectively.}
  \label{fig:seed-graph-d}
\end{figure}

Within this fixed computational framework, we formulate Stage~4 instances using the re-computed \textsc{EA} and \textsc{IP} datasets 
obtained from the same monomer pairs as in~\cite{Wilbraham:2019aa}.
For this purpose, we employ the three instances $I_{d1}, I_{d2}, I_{d3}$ illustrated in Figure~\ref{fig:seed-graph-d}, 
which were designed to mimic representative monomer types appearing in those datasets.
See Appendix~\ref{sec:test_instances} for a detailed description of these instances.

For each pair among $I_{d1}, I_{d2}, I_{d3}$, we solved Stage~4 with mixing ratio vector $(0.5,0.5)$.
Because a binary alternating copolymer admits two possible ordered connections of the two monomers, 
we evaluated both corresponding alternating oligomers in the subsequent xTB-based re-computation step.
Accordingly, the column $\etaxTB$ in Tables~\ref{table:phase-wilbraham-EA} and~\ref{table:phase-wilbraham-IP}
contains a pair of values whenever both ordered connections were successfully computed.
We also note that the prediction functions constructed from the re-computed datasets are less accurate than those obtained in Phase~1 from the original datasets.
In particular, the test RMSE values of the R-MLR models on the re-computed datasets are
\(0.269\) eV for \textsc{EA} and \(0.249\) eV for \textsc{IP}.
For comparison, the corresponding R-MLR models gave test RMSE values of
\(0.161\) eV for \textsc{EA} and \(0.122\) eV for \textsc{IP}, and the best
test RMSE values were \(0.143\) eV for \textsc{EA} and \(0.110\) eV for \textsc{IP},
both obtained by ANN; see Table~\ref{table:phase1b}.

\begin{table}[t!]\caption{
Results of Stage~4 for \textsc{EA} in the comparison study.
}
\begin{center}\scalebox{0.78}{
\begin{tabular}{c c c c | c c c c | c | c | c} \toprule
$D_\pi$ & Instances & $r$ & $\ylb, \yub$ & \#v & \#c & I-time & $\eta$ & $\etaxTB$ & $|\eta-\etaxTB|$ & $|\eta-\etaxTB|/\etaxTB$ \\ \midrule
 &  &  & 3.1, 3.3 & 15153 & 13146 & 25.562 & 3.242 & \paircell{\Err}{2.911} & \paircell{\NA}{0.331} & \paircell{\NA}{0.114} \\
 &  &  & 3.3, 3.5 & 15153 & 13146 & 2.102 & 3.417 & \paircell{3.165}{3.133} & \paircell{0.252}{0.284} & \paircell{0.080}{0.091} \\
 &  &  & 3.5, 3.7 & 15153 & 13146 & 4.162 & 3.740 & \paircell{4.191}{4.208} & \paircell{0.451}{0.468} & \paircell{0.108}{0.111} \\
 &  &  & 3.7, 3.9 & 15153 & 13146 & 2.066 & 3.871 & \paircell{\Err}{3.844} & \paircell{\NA}{0.027} & \paircell{\NA}{0.007} \\
\textsc{EA} & $I_{d1}, I_{d2}$ & (0.5, 0.5) & 3.9, 4.1 & 15153 & 13146 & 12.954 & 4.008 & \paircell{2.629}{2.707} & \paircell{1.379}{1.301} & \paircell{0.525}{0.481} \\
 &  &  & 4.1, 4.3 & 15153 & 13146 & 5.182 & 4.281 & \paircell{4.061}{4.092} & \paircell{0.220}{0.189} & \paircell{0.054}{0.046} \\
 &  &  & 4.3, 4.5 & 15153 & 13146 & 10.256 & 4.365 & \paircell{4.445}{4.391} & \paircell{0.080}{0.026} & \paircell{0.018}{0.006} \\
 &  &  & 4.5, 4.7 & 15153 & 13146 & 6.466 & 4.526 & \paircell{2.408}{2.406} & \paircell{2.118}{2.120} & \paircell{0.880}{0.881} \\
 &  &  & 4.7, 4.9 & 15153 & 13146 & 8.683 & 4.789 & \paircell{3.540}{3.604} & \paircell{1.249}{1.185} & \paircell{0.353}{0.329} \\
 &  &  & 4.9, 5.1 & 15153 & 13146 & 2.298 & 4.971 & \paircell{4.102}{4.005} & \paircell{0.869}{0.966} & \paircell{0.212}{0.241} \\ \hline

 &  &  & 3.1, 3.3 & 19292 & 17029 & 10.761 & 3.129 & \paircell{3.133}{\Err} & \paircell{0.004}{\NA} & \paircell{0.001}{\NA} \\
 &  &  & 3.3, 3.5 & 19292 & 17029 & 6.490 & 3.458 & \paircell{4.318}{\Err} & \paircell{0.860}{\NA} & \paircell{0.199}{\NA} \\
 &  &  & 3.5, 3.7 & 19292 & 17029 & 8.673 & 3.699 & \paircell{3.301}{3.389} & \paircell{0.398}{0.310} & \paircell{0.121}{0.091} \\
 &  &  & 3.7, 3.9 & 19292 & 17029 & 5.801 & 3.933 & \paircell{\Err}{\Err} & \paircell{\NA}{\NA} & \paircell{\NA}{\NA} \\
\textsc{EA} & $I_{d1}, I_{d3}$ & (0.5, 0.5) & 3.9, 4.1 & 19292 & 17029 & 3.460 & 4.105 & \paircell{4.092}{4.012} & \paircell{0.013}{0.093} & \paircell{0.003}{0.023} \\
 &  &  & 4.1, 4.3 & 19292 & 17029 & 9.684 & 4.345 & \paircell{3.408}{3.435} & \paircell{0.937}{0.910} & \paircell{0.275}{0.265} \\
 &  &  & 4.3, 4.5 & 19292 & 17029 & 5.216 & 4.329 & \paircell{3.582}{\Err} & \paircell{0.747}{\NA} & \paircell{0.209}{\NA} \\
 &  &  & 4.5, 4.7 & 19292 & 17029 & 6.476 & 4.631 & \paircell{4.212}{4.303} & \paircell{0.419}{0.328} & \paircell{0.100}{0.076} \\
 &  &  & 4.7, 4.9 & 19292 & 17029 & 11.852 & 4.708 & \paircell{\Err}{3.563} & \paircell{\NA}{1.145} & \paircell{\NA}{0.321} \\
 &  &  & 4.9, 5.1 & 19292 & 17029 & 6.357 & 4.936 & \paircell{\Err}{3.922} & \paircell{\NA}{1.014} & \paircell{\NA}{0.259} \\ \hline

 &  &  & 3.1, 3.3 & 19835 & 17587 & 10.531 & 3.349 & \paircell{3.313}{3.346} & \paircell{0.036}{0.003} & \paircell{0.011}{0.001} \\
 &  &  & 3.3, 3.5 & 19835 & 17587 & 9.891 & 3.461 & \paircell{3.398}{\Err} & \paircell{0.063}{\NA} & \paircell{0.019}{\NA} \\
 &  &  & 3.5, 3.7 & 19835 & 17587 & 4.487 & 3.622 & \paircell{3.519}{3.463} & \paircell{0.103}{0.159} & \paircell{0.029}{0.046} \\
 &  &  & 3.7, 3.9 & 19835 & 17587 & 7.029 & 3.869 & \paircell{\Err}{3.950} & \paircell{\NA}{0.081} & \paircell{\NA}{0.021} \\
\textsc{EA} & $I_{d2}, I_{d3}$ & (0.5, 0.5) & 3.9, 4.1 & 19835 & 17587 & 2.835 & 4.109 & \paircell{4.575}{4.534} & \paircell{0.466}{0.425} & \paircell{0.102}{0.094} \\
 &  &  & 4.1, 4.3 & 19835 & 17587 & 7.302 & 4.307 & \paircell{\Err}{\Err} & \paircell{\NA}{\NA} & \paircell{\NA}{\NA} \\
 &  &  & 4.3, 4.5 & 19835 & 17587 & 2.376 & 4.478 & \paircell{\Err}{4.714} & \paircell{\NA}{0.236} & \paircell{\NA}{0.050} \\
 &  &  & 4.5, 4.7 & 19835 & 17587 & 2.752 & 4.656 & \paircell{4.009}{3.784} & \paircell{0.647}{0.872} & \paircell{0.161}{0.230} \\
 &  &  & 4.7, 4.9 & 19835 & 17587 & 6.104 & 4.918 & \paircell{4.312}{4.334} & \paircell{0.606}{0.584} & \paircell{0.141}{0.135} \\
 &  &  & 4.9, 5.1 & 19835 & 17587 & 40.670 & 5.041 & \paircell{3.538}{3.567} & \paircell{1.503}{1.474} & \paircell{0.425}{0.413} \\
\bottomrule
\end{tabular}
}
\end{center}\label{table:phase-wilbraham-EA}
\end{table}

\begin{table}[t!]\caption{
Results of Stage~4 for \textsc{IP} in the comparison study.
}
\begin{center}\scalebox{0.78}{
\begin{tabular}{c c c c | c c c c | c | c | c} \toprule
$D_\pi$ & Instances & $r$ & $\ylb, \yub$ & \#v & \#c & I-time & $\eta$ & $\etaxTB$ & $|\eta-\etaxTB|$ & $|\eta-\etaxTB|/\etaxTB$ \\ \midrule
 &  &  & 4.8, 5.0 & 15321 & 13410 & 42.860 & 4.996 & \paircell{5.161}{5.107} & \paircell{0.165}{0.111} & \paircell{0.032}{0.022} \\
 &  &  & 5.0, 5.2 & 15321 & 13410 & 3.639 & 5.100 & \paircell{6.992}{6.727} & \paircell{1.892}{1.627} & \paircell{0.271}{0.242} \\
 &  &  & 5.2, 5.4 & 15321 & 13410 & 3.854 & 5.338 & \paircell{5.471}{\Err} & \paircell{0.133}{\NA} & \paircell{0.024}{\NA} \\
 &  &  & 5.4, 5.6 & 15321 & 13410 & 3.231 & 5.593 & \paircell{5.542}{5.471} & \paircell{0.051}{0.122} & \paircell{0.009}{0.022} \\
\textsc{IP} & $I_{d1}, I_{d2}$ & (0.5, 0.5) & 5.6, 5.8 & 15321 & 13410 & 3.842 & 5.799 & \paircell{5.934}{5.888} & \paircell{0.135}{0.089} & \paircell{0.023}{0.015} \\
 &  &  & 5.8, 6.0 & 15321 & 13410 & 6.992 & 5.942 & \paircell{5.936}{5.955} & \paircell{0.006}{0.013} & \paircell{0.001}{0.002} \\
 &  &  & 6.0, 6.2 & 15321 & 13410 & 4.463 & 6.104 & \paircell{6.200}{6.293} & \paircell{0.096}{0.189} & \paircell{0.015}{0.030} \\
 &  &  & 6.2, 6.4 & 15321 & 13410 & 3.010 & 6.221 & \paircell{5.847}{5.883} & \paircell{0.374}{0.338} & \paircell{0.064}{0.057} \\
 &  &  & 6.4, 6.6 & 15321 & 13410 & 3.974 & 6.469 & \paircell{6.455}{6.458} & \paircell{0.014}{0.011} & \paircell{0.002}{0.002} \\
 &  &  & 6.6, 6.8 & 15321 & 13410 & 5.965 & 6.633 & \paircell{6.259}{6.354} & \paircell{0.374}{0.279} & \paircell{0.060}{0.044} \\ \hline

 &  &  & 4.8, 5.0 & 19460 & 17293 & 14.211 & 5.012 & \paircell{5.690}{5.808} & \paircell{0.678}{0.796} & \paircell{0.119}{0.137} \\
 &  &  & 5.0, 5.2 & 19460 & 17293 & 2.650 & 5.139 & \paircell{5.958}{6.007} & \paircell{0.819}{0.868} & \paircell{0.137}{0.145} \\
 &  &  & 5.2, 5.4 & 19460 & 17293 & 7.118 & 5.202 & \paircell{6.638}{6.567} & \paircell{1.436}{1.365} & \paircell{0.216}{0.208} \\
 &  &  & 5.4, 5.6 & 19460 & 17293 & 2.638 & 5.405 & \paircell{6.016}{\Err} & \paircell{0.611}{\NA} & \paircell{0.102}{\NA} \\
\textsc{IP} & $I_{d1}, I_{d3}$ & (0.5, 0.5) & 5.6, 5.8 & 19460 & 17293 & 3.994 & 5.809 & \paircell{\Err}{\Err} & \paircell{\NA}{\NA} & \paircell{\NA}{\NA} \\
 &  &  & 5.8, 6.0 & 19460 & 17293 & 2.462 & 5.881 & \paircell{6.072}{6.018} & \paircell{0.191}{0.137} & \paircell{0.031}{0.023} \\
 &  &  & 6.0, 6.2 & 19460 & 17293 & 3.385 & 6.044 & \paircell{\Err}{6.094} & \paircell{\NA}{0.050} & \paircell{\NA}{0.008} \\
 &  &  & 6.2, 6.4 & 19460 & 17293 & 3.254 & 6.324 & \paircell{\Err}{\Err} & \paircell{\NA}{\NA} & \paircell{\NA}{\NA} \\
 &  &  & 6.4, 6.6 & 19460 & 17293 & 2.279 & 6.417 & \paircell{\Err}{6.522} & \paircell{\NA}{0.105} & \paircell{\NA}{0.016} \\
 &  &  & 6.6, 6.8 & 19460 & 17293 & 3.904 & 6.823 & \paircell{6.467}{6.450} & \paircell{0.356}{0.373} & \paircell{0.055}{0.058} \\ \hline

 &  &  & 4.8, 5.0 & 20003 & 17851 & 36.320 & 4.811 & \paircell{5.380}{5.333} & \paircell{0.569}{0.522} & \paircell{0.106}{0.098} \\
 &  &  & 5.0, 5.2 & 20003 & 17851 & 64.500 & 4.972 & \paircell{5.646}{5.603} & \paircell{0.674}{0.631} & \paircell{0.119}{0.113} \\
 &  &  & 5.2, 5.4 & 20003 & 17851 & 2.893 & 5.354 & \paircell{5.525}{5.498} & \paircell{0.171}{0.144} & \paircell{0.031}{0.026} \\
 &  &  & 5.4, 5.6 & 20003 & 17851 & 2.345 & 5.563 & \paircell{5.892}{\Err} & \paircell{0.329}{\NA} & \paircell{0.056}{\NA} \\
\textsc{IP} & $I_{d2}, I_{d3}$ & (0.5, 0.5) & 5.6, 5.8 & 20003 & 17851 & 3.125 & 5.758 & \paircell{5.767}{5.656} & \paircell{0.009}{0.102} & \paircell{0.002}{0.018} \\
 &  &  & 5.8, 6.0 & 20003 & 17851 & 2.279 & 5.874 & \paircell{\Err}{\Err} & \paircell{\NA}{\NA} & \paircell{\NA}{\NA} \\
 &  &  & 6.0, 6.2 & 20003 & 17851 & 3.043 & 6.008 & \paircell{5.910}{5.870} & \paircell{0.098}{0.138} & \paircell{0.017}{0.024} \\
 &  &  & 6.2, 6.4 & 20003 & 17851 & 2.384 & 6.322 & \paircell{6.430}{\Err} & \paircell{0.108}{\NA} & \paircell{0.017}{\NA} \\
 &  &  & 6.4, 6.6 & 20003 & 17851 & 3.222 & 6.525 & \paircell{5.927}{5.926} & \paircell{0.598}{0.599} & \paircell{0.101}{0.101} \\
 &  &  & 6.6, 6.8 & 20003 & 17851 & 3.362 & 6.621 & \paircell{5.416}{5.517} & \paircell{1.205}{1.104} & \paircell{0.222}{0.200} \\
\bottomrule
\end{tabular}
}
\end{center}\label{table:phase-wilbraham-IP}
\end{table}

For both re-computed datasets, we used R-MLR as the prediction function.
The computational results for the MILP and the subsequent external check of the inferred copolymers
are summarized in Tables~\ref{table:phase-wilbraham-EA} and~\ref{table:phase-wilbraham-IP}.
We use the following notations:
\begin{itemize}
\item[-] $D_\pi$: the name of the dataset;
\item[-] Instances: the instances corresponding to the constituent monomer positions
in the copolymer $P=(\mathbf{C},r)$ to be inferred;
\item[-] $r$: the given mixing ratio vector;
\item[-] $\ylb, \yub$: the target range $[\ylb, \yub]$ for the predicted property value;
\item[-] $\#v$ (respectively $\#c$): the number of variables (respectively, constraints) in the MILP;
\item[-] I-time: the time (in seconds) to solve the MILP;
\item[-] $\eta$: the predicted property value of the inferred copolymer by the prediction function;
\item[-] \(\etaxTB\): the value re-computed by our xTB-based implementation.
Because two ordered alternating oligomers are evaluated for each inferred monomer pair,
this column contains a pair of values whenever both computations are successful.
The symbol ``Err'' indicates that geometry generation or a subsequent quantum-chemical
calculation failed; 
\item[-] $|\eta-\etaxTB|$: the absolute difference between the property value predicted by $\eta$ and 
 the re-computed property value (where ``-'' indicates that the value is unavailable); and
\item[-] \(|\eta-\etaxTB|/\etaxTB\): the relative difference between the property value predicted by $\eta$
and the re-computed property value (where
``-'' indicates that the value is unavailable).
\end{itemize}

We would like to emphasize again that, in this subsection, 
the aim is to examine whether the inverse-designed candidates can be mapped back, under a reasonable implementation of the xTB pipeline, 
to values that are broadly consistent with the target ranges imposed in Stage~4.
Tables~\ref{table:phase-wilbraham-EA} and~\ref{table:phase-wilbraham-IP} show that the MILP itself is solved very quickly in this comparison study,
typically within a few seconds to a few tens of seconds.
As for the agreement between $\eta$ and $\etaxTB$, the discrepancy is often relatively small for target ranges in the middle of the data range,
whereas larger deviations are more frequently observed near the upper end of the target ranges.
For example, in Table~\ref{table:phase-wilbraham-IP}, the interval $[5.8,6.0]$ for the pair $(I_{d1},I_{d2})$ gives $\eta=5.942$ and $\etaxTB=(5.936,5.955)$,
corresponding to absolute differences
\((0.006,0.013)\) and relative differences \((0.001,0.002)\),
while several high-\textsc{EA} and high-\textsc{IP} intervals exhibit substantially larger discrepancies.

We also acknowledge that some entries in the $\etaxTB$ column are marked as ``Err''.
These correspond to cases where geometry generation or a subsequent xTB calculation failed.
Among the 60 ordered-oligomer evaluations for \textsc{EA}, 46 were successfully evaluated by
the xTB-based pipeline and 14 resulted in errors. For \textsc{IP}, 48 out of 60 evaluations
were successful and 12 resulted in errors.

These failures indicate that satisfying the MILP-encoded structural constraints and the
learned prediction function does not necessarily guarantee that the resulting candidate is
stable under subsequent geometry generation or quantum-chemical evaluation. In particular,
some of the inferred molecules appear to contain chemically unstable or unusual substructures.
The error cases and large discrepancies therefore illustrate the distinction between
optimization with respect to the learned predictor $\eta$ and validation with respect to an
external computational property-evaluation pipeline. This is a common issue in molecular
inference studies and highlights the need to incorporate additional chemical knowledge,
stability constraints, or post-screening procedures into the inference
procedure~\cite{Cheng:2021aa}. Addressing this issue is left as future work.

Overall, these results suggest that the proposed inverse-design framework is capable of finding candidates that are not only optimal for the learned model,
but also often reasonably consistent with the independently re-computed xTB-based values, especially in the central part of the property range.

\begin{figure}[t!]
  \centering
  \begin{subfigure}[t!]{\textwidth}
    \centering
    \begin{minipage}[t!]{0.35\textwidth}
      \centering
      \includegraphics[width=\textwidth]{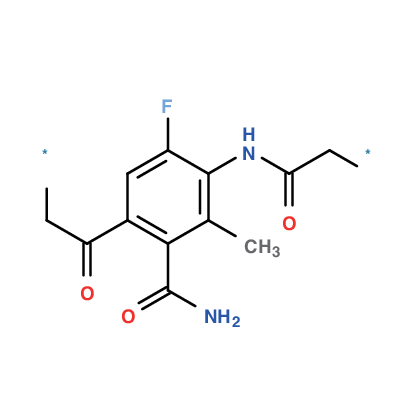}
    \end{minipage}
    \begin{minipage}[t!]{0.35\textwidth}
      \centering
      \includegraphics[width=\textwidth]{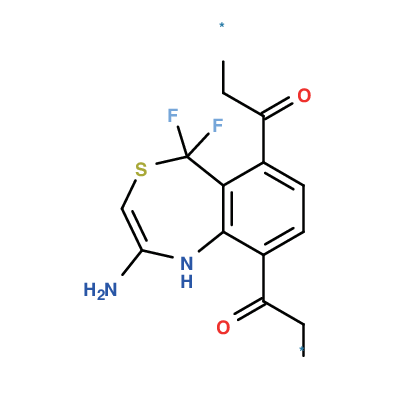}
    \end{minipage}
    \caption{}
    \label{fig:MILP_EA_d1d2}
  \end{subfigure}
  \begin{subfigure}[t!]{\textwidth}
    \centering
    \begin{minipage}[t!]{0.35\textwidth}
      \centering
      \includegraphics[width=\textwidth]{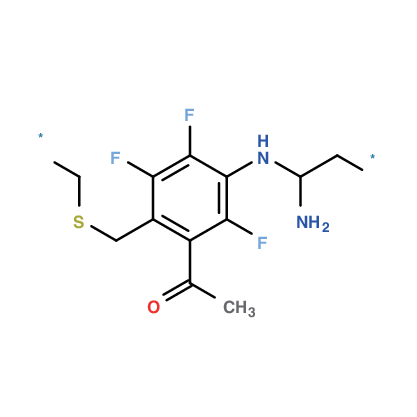}
    \end{minipage}
    \begin{minipage}[t!]{0.35\textwidth}
      \centering
      \includegraphics[width=\textwidth]{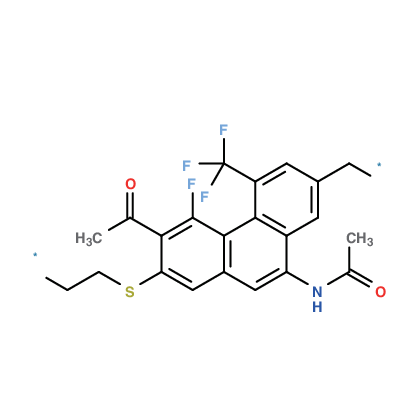}
    \end{minipage}
    \caption{}
    \label{fig:MILP_EA_d1d3}
  \end{subfigure}
 \begin{subfigure}[t!]{\textwidth}
    \centering
    \begin{minipage}[t!]{0.35\textwidth}
      \centering
      \includegraphics[width=\textwidth]{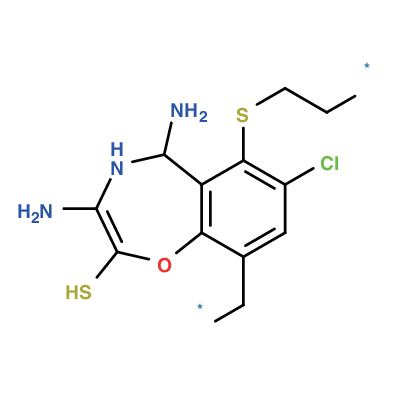}
    \end{minipage}
    \begin{minipage}[t!]{0.35\textwidth}
      \centering
      \includegraphics[width=\textwidth]{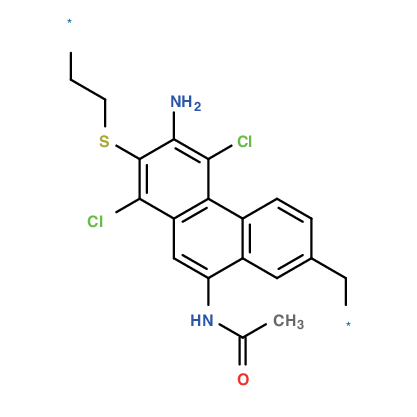}
    \end{minipage}
    \caption{}
    \label{fig:MILP_EA_d2d3}
  \end{subfigure}
  \caption{Examples of inferred monomers for the \textsc{EA} dataset. 
  (a) Instances $I_{d1}, I_{d2}$ with target range $[\ylb,\yub]=[4.3,4.5]$. 
  (b) Instances $I_{d1}, I_{d3}$ with target range $[\ylb,\yub]=[3.9,4.1]$. 
  (c) Instances $I_{d2}, I_{d3}$ with target range $[\ylb,\yub]=[3.1,3.3]$. }
  \label{fig:phase2_MILP_EA}
\end{figure}

\begin{figure}[t!]
  \centering
  \begin{subfigure}[t!]{\textwidth}
    \centering
    \begin{minipage}[t!]{0.35\textwidth}
      \centering
      \includegraphics[width=\textwidth]{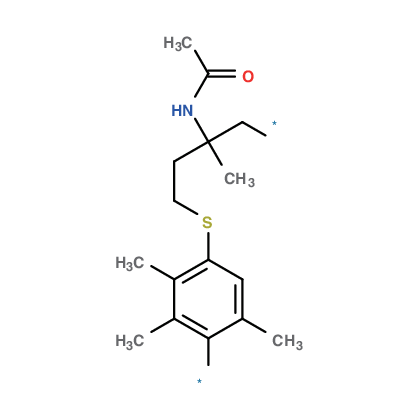}
    \end{minipage}
    \begin{minipage}[t!]{0.35\textwidth}
      \centering
      \includegraphics[width=\textwidth]{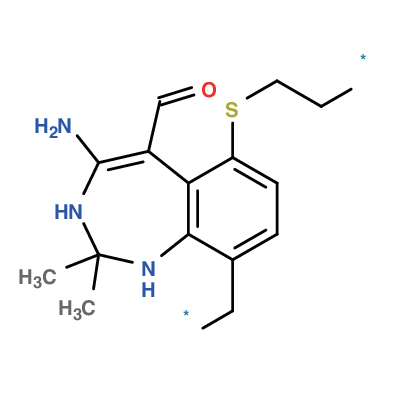}
    \end{minipage}
    \caption{}
    \label{fig:MILP_IP_d1d2}
  \end{subfigure}
  \begin{subfigure}[t!]{\textwidth}
    \centering
    \begin{minipage}[t!]{0.35\textwidth}
      \centering
      \includegraphics[width=\textwidth]{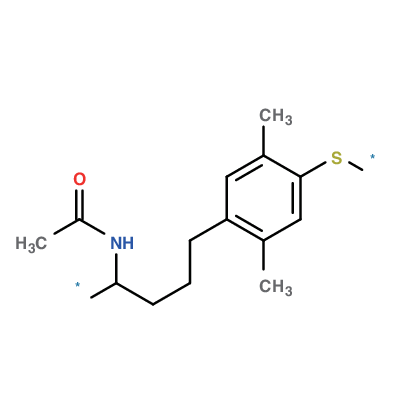}
    \end{minipage}
    \begin{minipage}[t!]{0.35\textwidth}
      \centering
      \includegraphics[width=\textwidth]{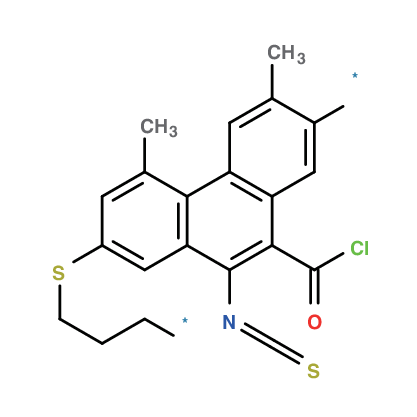}
    \end{minipage}
    \caption{}
    \label{fig:MILP_IP_d1d3}
  \end{subfigure}
 \begin{subfigure}[t!]{\textwidth}
    \centering
    \begin{minipage}[t!]{0.35\textwidth}
      \centering
      \includegraphics[width=\textwidth]{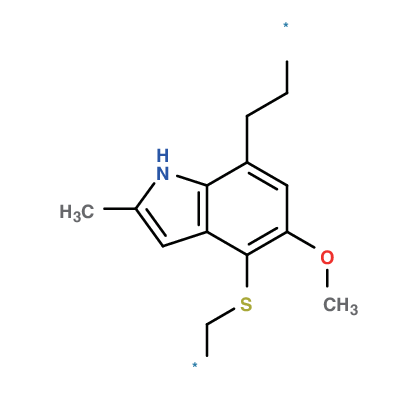}
    \end{minipage}
    \begin{minipage}[t!]{0.35\textwidth}
      \centering
      \includegraphics[width=\textwidth]{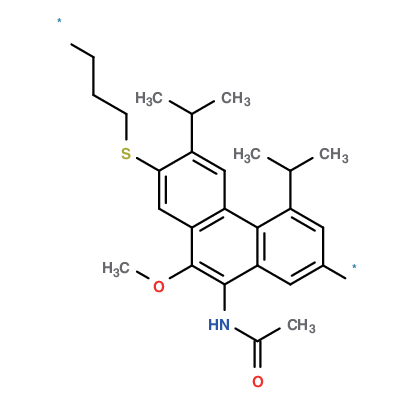}
    \end{minipage}
    \caption{}
    \label{fig:MILP_IP_d2d3}
  \end{subfigure}
  \caption{Examples of inferred monomers for the \textsc{IP} dataset. 
  (a) Instances $I_{d1}, I_{d2}$ with target range $[\ylb,\yub]=[5.8,6.0]$. 
  (b) Instances $I_{d1}, I_{d3}$ with target range $[\ylb,\yub]=[6.0,6.2]$. 
  (c) Instances $I_{d2}, I_{d3}$ with target range $[\ylb,\yub]=[5.6,5.8]$. }
  \label{fig:phase2_MILP_IP}
\end{figure}

Figures~\ref{fig:phase2_MILP_EA} and~\ref{fig:phase2_MILP_IP} show some inferred monomers for \textsc{EA} and \textsc{IP}, respectively.
A broader chemical assessment of such candidates, 
including synthetic accessibility or further post-screening by more accurate methods, 
is beyond the scope of the present subsection.

\section{Concluding Remarks}\label{sec:conclude}

In this study, we proposed an extension of the \molinfer\ framework to copolymers by introducing the mixing vector (MV) model. 
The key idea is to represent the feature vector for a copolymer by a weighted combination of monomer-level descriptors according to its mixing ratio, 
thereby obtaining a feature vector that remains compatible with the MILP-based inverse-design framework. 
This simple construction makes it possible to carry over the main advantage of \molinfer\ to the copolymer setting, 
namely, the ability to solve the inverse problem with guaranteed optimality relative to the learned model and the imposed structural constraints.

A notable feature of the present study is that the proposed representation was evaluated on a broad collection of copolymer property datasets gathered from the literature.
These datasets cover different target properties, copolymer classes, dataset sizes, and descriptor dimensions,
and therefore provide a useful test bed for examining the practical applicability of the MV model.
The computational experiments demonstrate that the proposed approach works at two levels. 
In Phase~1, the MV model provides practically useful predictive performance across a broad range of copolymer property datasets. 
In particular, the results show that even a composition-level representation based only on monomer descriptors and mixing ratios
can capture a substantial part of the structure-property relationship in many settings.
This suggests that monomer composition and mixing ratio already provide strong information for copolymer property prediction.
In Phase~2, the resulting multi-monomer MILP formulations remain computationally tractable in representative settings, 
showing that the original \molinfer\ framework can be extended to copolymers without losing its essential computational effectiveness. 
The additional comparison study on the re-computed EA/IP datasets further suggests that the inferred candidates 
are often reasonably consistent with the external re-evaluation, 
especially in the medium part of the property range.


The present MV model is intentionally designed as a simple composition-level representation.
From the viewpoint of the present study, this provides an important baseline:
the experiments show that meaningful predictive and inverse-design performance can already be obtained without explicitly encoding detailed sequence-distribution information.
At the same time, sequence-distribution information is expected to provide additional structural information,
and incorporating it into the representation may further improve predictive accuracy and the chemical relevance of inferred candidates.

There are several directions for future work.
One important direction is to develop richer copolymer representations that incorporate sequence information while preserving as much MILP compatibility as possible.
Recent machine-learning-based inverse-design studies have shown that higher-order structural information, 
such as monomer stoichiometry and chain architecture, can be incorporated into copolymer design models 
and may be useful for exploring broader polymer design spaces~\cite{Vogel:2025aa}.
Another is to strengthen the chemical realism of inferred candidates by integrating 
additional chemical constraints or stability-aware criteria into the inverse-design phase.
It is also important to extend the present framework beyond the linear copolymer setting and to investigate a wider range of polymer properties and architectures.
We hope that the present work serves as a useful first step toward more expressive and chemically reliable exact inverse-design frameworks for copolymers.

\section*{Acknowledgements}
This work was partially supported by JSPS KAKENHI Grant Numbers 22H00532, 25K24374, and 26KF0066. 

\bibliographystyle{abbrv}
\bibliography{./chemgraph.bib}

@inproceedings{Zhu:2026aa,
	address = {New York, NY, USA},
	author = {Zhu, Jianshen and Azam, Naveed Ahmed and Haraguchi, Kazuya and Zhao, Liang and Akutsu, Tatsuya},
	booktitle = {Proceedings of the 2025 9th International Conference on Computational Biology and Bioinformatics},
	doi = {10.1145/3789938.3789939},
	isbn = {9798400720697},
	numpages = {7},
	pages = {1--7},
	publisher = {Association for Computing Machinery},
	series = {ICCBB '25},
	title = {Combining Graph Neural Networks and Mixed Integer Linear Programming for Molecular Inference under the Two-Layered Model},
	url = {https://doi.org/10.1145/3789938.3789939},
	year = {2026}}

@article{Yue:2025aa,
	author = {Yue, Tianle and Tao, Lei and Varshney, Vikas and Li, Ying},
	doi = {10.1039/D4DD00395K},
	issue = {4},
	journal = {Digital Discovery},
	pages = {910-926},
	publisher = {RSC},
	title = {Benchmarking study of deep generative models for inverse polymer design},
	url = {http://dx.doi.org/10.1039/D4DD00395K},
	volume = {4},
	year = {2025}}

@article{Vogel:2025aa,
	author = {Vogel, Gabriel and Weber, Jana M.},
	doi = {10.1039/d4sc05900j},
	issn = {2041-6539},
	journal = {Chemical Science},
	number = {3},
	pages = {1161--1178},
	publisher = {Royal Society of Chemistry (RSC)},
	title = {Inverse design of copolymers including stoichiometry and chain architecture},
	url = {http://dx.doi.org/10.1039/d4sc05900j},
	volume = {16},
	year = {2025}}

@article{Su:2019aa,
	author = {Su, Neil Qiang and Xu, Xin},
	date = {2019/06/06},
	doi = {10.1021/acs.jpclett.9b01052},
	journal = {The Journal of Physical Chemistry Letters},
	journal1 = {The Journal of Physical Chemistry Letters},
	journal2 = {J. Phys. Chem. Lett.},
	month = {06},
	number = {11},
	pages = {2692--2699},
	publisher = {American Chemical Society},
	title = {Insights into Direct Methods for Predictions of Ionization Potential and Electron Affinity in Density Functional Theory},
	type = {doi: 10.1021/acs.jpclett.9b01052},
	url = {https://doi.org/10.1021/acs.jpclett.9b01052},
	volume = {10},
	year = {2019},
	year1 = {2019}}

@article{Jain:2025aa,
	author = {Jain, Ayush and Gurnani, Rishi and Rajan, Arunkumar and Qi, H. Jerry and Ramprasad, Rampi},
	date = {2025/02/20},
	doi = {10.1038/s41524-025-01532-6},
	id = {Jain2025},
	isbn = {2057-3960},
	journal = {npj Computational Materials},
	number = {1},
	pages = {42},
	title = {A physics-enforced neural network to predict polymer melt viscosity},
	url = {https://doi.org/10.1038/s41524-025-01532-6},
	volume = {11},
	year = {2025}}

@article{Brierley-Croft:2025aa,
	author = {Brierley-Croft, Sebastian and Olmsted, Peter D. and Hine, Peter J. and Mandle, Richard J. and Chaplin, Adam and Grasmeder, John and Mattsson, Johan},
	date = {2025/07/08},
	doi = {10.1021/acs.macromol.5c00178},
	isbn = {0024-9297},
	journal = {Macromolecules},
	journal1 = {Macromolecules},
	journal2 = {Macromolecules},
	month = {07},
	number = {13},
	pages = {6407--6417},
	publisher = {American Chemical Society},
	title = {Polymer Informatics Method for Fast and Accurate Prediction of the Glass Transition Temperature from Chemical Structure},
	type = {doi: 10.1021/acs.macromol.5c00178},
	url = {https://doi.org/10.1021/acs.macromol.5c00178},
	volume = {58},
	year = {2025},
	year1 = {2025}}

@article{Bai:2019aa,
	author = {Bai, Yang and Wilbraham, Liam and Slater, Benjamin J. and Zwijnenburg, Martijn A. and Sprick, Reiner Sebastian and Cooper, Andrew I.},
	date = {2019/06/05},
	doi = {10.1021/jacs.9b03591},
	isbn = {0002-7863},
	journal = {Journal of the American Chemical Society},
	journal1 = {Journal of the American Chemical Society},
	journal2 = {J. Am. Chem. Soc.},
	month = {06},
	number = {22},
	pages = {9063--9071},
	publisher = {American Chemical Society},
	title = {Accelerated Discovery of Organic Polymer Photocatalysts for Hydrogen Evolution from Water through the Integration of Experiment and Theory},
	type = {doi: 10.1021/jacs.9b03591},
	url = {https://doi.org/10.1021/jacs.9b03591},
	volume = {141},
	year = {2019},
	year1 = {2019}}

@article{Reis:2021aa,
	author = {Reis, Marcus and Gusev, Filipp and Taylor, Nicholas G. and Chung, Sang Hun and Verber, Matthew D. and Lee, Yueh Z. and Isayev, Olexandr and Leibfarth, Frank A.},
	date = {2021/10/27},
	doi = {10.1021/jacs.1c08181},
	isbn = {0002-7863},
	journal = {Journal of the American Chemical Society},
	journal1 = {Journal of the American Chemical Society},
	journal2 = {J. Am. Chem. Soc.},
	month = {10},
	number = {42},
	pages = {17677--17689},
	publisher = {American Chemical Society},
	title = {Machine-Learning-Guided Discovery of {19F MRI} Agents Enabled by Automated Copolymer Synthesis},
	type = {doi: 10.1021/jacs.1c08181},
	url = {https://doi.org/10.1021/jacs.1c08181},
	volume = {143},
	year = {2021},
	year1 = {2021}}

@online{Grimme-xtb,
	key = {Grimme-xtb},
	lastaccessed = {Jan 13, 2026},
	title = {{GitHub - grimme-lab/xtb: Semiempirical Extended Tight-Binding Program Package}},
	url = {https://github.com/grimme-lab/xtb},
	year = {2024}}

@online{CPLEX,
	key = {CPLEX},
	lastaccessed = {July 14, 2025},
	title = {{IBM ILOG CPLEX Optimization Studio}},
	url = {https://www.ibm.com/products/ilog-cplex-optimization-studio},
	year = {2025}}

@article{Asgeirsson:2017aa,
	author = {{\'A}sgeirsson, Vilhj{\'a}lmur and Bauer, Christoph A. and Grimme, Stefan},
	doi = {10.1039/C7SC00601B},
	issue = {7},
	journal = {Chemical Science},
	pages = {4879-4895},
	publisher = {The Royal Society of Chemistry},
	title = {Quantum chemical calculation of electron ionization mass spectra for general organic and inorganic molecules},
	url = {http://dx.doi.org/10.1039/C7SC00601B},
	volume = {8},
	year = {2017}}

@article{Grimme:2017aa,
	author = {Grimme, Stefan and Bannwarth, Christoph and Shushkov, Philip},
	date = {2017/05/09},
	doi = {10.1021/acs.jctc.7b00118},
	isbn = {1549-9618},
	journal = {Journal of Chemical Theory and Computation},
	journal1 = {Journal of Chemical Theory and Computation},
	journal2 = {J. Chem. Theory Comput.},
	month = {05},
	number = {5},
	pages = {1989--2009},
	publisher = {American Chemical Society},
	title = {A Robust and Accurate Tight-Binding Quantum Chemical Method for Structures, Vibrational Frequencies, and Noncovalent Interactions of Large Molecular Systems Parametrized for All spd-Block Elements (Z = 1--86)},
	type = {doi: 10.1021/acs.jctc.7b00118},
	url = {https://doi.org/10.1021/acs.jctc.7b00118},
	volume = {13},
	year = {2017},
	year1 = {2017}}

@article{Wilbraham:2019aa,
	author = {Wilbraham, Liam and Sprick, Reiner Sebastian and Jelfs, Kim E. and Zwijnenburg, Martijn A.},
	doi = {10.1039/C8SC05710A},
	issue = {19},
	journal = {Chemical Science},
	pages = {4973-4984},
	publisher = {The Royal Society of Chemistry},
	title = {Mapping binary copolymer property space with neural networks},
	url = {http://dx.doi.org/10.1039/C8SC05710A},
	volume = {10},
	year = {2019}}

@article{Zhang:2020ab,
	article-number = {2488},
	author = {Zhang, Xuewei and Daigle, Jean-Christophe and Zaghib, Karim},
	doi = {10.3390/ma13112488},
	issn = {1996-1944},
	journal = {Materials},
	number = {11},
	pubmedid = {32486029},
	title = {Comprehensive Review of Polymer Architecture for All-Solid-State Lithium Rechargeable Batteries},
	url = {https://www.mdpi.com/1996-1944/13/11/2488},
	volume = {13},
	year = {2020}}

@article{Gao:2004aa,
	author = {C Gao and D Yan},
	doi = {https://doi.org/10.1016/j.progpolymsci.2003.12.002},
	issn = {0079-6700},
	journal = {Progress in Polymer Science},
	number = {3},
	pages = {183-275},
	title = {Hyperbranched polymers: from synthesis to applications},
	url = {https://www.sciencedirect.com/science/article/pii/S0079670003001345},
	volume = {29},
	year = {2004}}

@article{Kharchenko:2003aa,
	author = {Kharchenko, Semen B. and Kannan, Rangaramanujam M. and Cernohous, Jeff J. and Venkataramani, Shivshankar},
	date = {2003/01/01},
	doi = {10.1021/ma0256486},
	isbn = {0024-9297},
	journal = {Macromolecules},
	journal1 = {Macromolecules},
	journal2 = {Macromolecules},
	month = {01},
	number = {2},
	pages = {399--406},
	publisher = {American Chemical Society},
	title = {Role of Architecture on the Conformation, Rheology, and Orientation Behavior of Linear, Star, and Hyperbranched Polymer Melts. 1. Synthesis and Molecular Characterization},
	type = {doi: 10.1021/ma0256486},
	url = {https://doi.org/10.1021/ma0256486},
	volume = {36},
	year = {2003},
	year1 = {2003}}

@article{Zhu:2025ab,
	author = {Zhu, Jianshen and Takekida, Mao and Azam, Naveed Ahmed and Haraguchi, Kazuya and Zhao, Liang and Akutsu, Tatsuya},
	date = {2025/10/14},
	doi = {10.1021/acsomega.5c02319},
	journal = {ACS Omega},
	journal1 = {ACS Omega},
	journal2 = {ACS Omega},
	month = {10},
	number = {40},
	pages = {46467--46481},
	publisher = {American Chemical Society},
	title = {Toward Environment-Sensitive Molecular Inference via Mixed Integer Linear Programming},
	type = {doi: 10.1021/acsomega.5c02319},
	url = {https://doi.org/10.1021/acsomega.5c02319},
	volume = {10},
	year = {2025},
	year1 = {2025}}

@article{Tao:2022aa,
	author = {Lei Tao and John Byrnes and Vikas Varshney and Ying Li},
	doi = {https://doi.org/10.1016/j.isci.2022.104585},
	issn = {2589-0042},
	journal = {iScience},
	number = {7},
	pages = {104585},
	title = {Machine learning strategies for the structure-property relationship of copolymers},
	url = {https://www.sciencedirect.com/science/article/pii/S2589004222008574},
	volume = {25},
	year = {2022}}

@article{Patel:2022aa,
	author = {Patel, Roshan A. and Borca, Carlos H. and Webb, Michael A.},
	doi = {10.1039/D1ME00160D},
	issue = {6},
	journal = {Mol. Syst. Des. Eng.},
	pages = {661-676},
	publisher = {The Royal Society of Chemistry},
	title = {Featurization strategies for polymer sequence or composition design by machine learning},
	url = {http://dx.doi.org/10.1039/D1ME00160D},
	volume = {7},
	year = {2022}}

@article{Zhao:2023aa,
	author = {Zhao, Yuankai and Mulder, Roger J. and Houshyar, Shadi and Le, Tu C.},
	doi = {10.1039/D3PY00395G},
	issue = {29},
	journal = {Polymer Chemistry},
	pages = {3325-3346},
	publisher = {The Royal Society of Chemistry},
	title = {A review on the application of molecular descriptors and machine learning in polymer design},
	url = {http://dx.doi.org/10.1039/D3PY00395G},
	volume = {14},
	year = {2023}}

@article{Stoykovich:2007aa,
	author = {Stoykovich, Mark P. and Kang, Huiman and Daoulas, Kostas Ch. and Liu, Guoliang and Liu, Chi-Chun and de Pablo, Juan J. and M{\"u}ller, Marcus and Nealey, Paul F.},
	date = {2007/10/31},
	doi = {10.1021/nn700164p},
	isbn = {1936-0851},
	journal = {ACS Nano},
	journal1 = {ACS Nano},
	journal2 = {ACS Nano},
	month = {10},
	number = {3},
	pages = {168--175},
	publisher = {American Chemical Society},
	title = {Directed Self-Assembly of Block Copolymers for Nanolithography: Fabrication of Isolated Features and Essential Integrated Circuit Geometries},
	type = {doi: 10.1021/nn700164p},
	url = {https://doi.org/10.1021/nn700164p},
	volume = {1},
	year = {2007},
	year1 = {2007}}

@article{Kumar:2021aa,
	article-number = {318},
	author = {Kumar, Labeesh and Singh, Sajan and Horechyy, Andriy and Fery, Andreas and Nandan, Bhanu},
	doi = {10.3390/membranes11050318},
	issn = {2077-0375},
	journal = {Membranes},
	number = {5},
	pubmedid = {33925335},
	title = {Block Copolymer Template-Directed Catalytic Systems: Recent Progress and Perspectives},
	url = {https://www.mdpi.com/2077-0375/11/5/318},
	volume = {11},
	year = {2021}}

@article{Jackson:2010aa,
	author = {Jackson, Elizabeth A. and Hillmyer, Marc A.},
	date = {2010/07/27},
	doi = {10.1021/nn1014006},
	isbn = {1936-0851},
	journal = {ACS Nano},
	journal1 = {ACS Nano},
	journal2 = {ACS Nano},
	month = {07},
	number = {7},
	pages = {3548--3553},
	publisher = {American Chemical Society},
	title = {Nanoporous Membranes Derived from Block Copolymers: From Drug Delivery to Water Filtration},
	type = {doi: 10.1021/nn1014006},
	url = {https://doi.org/10.1021/nn1014006},
	volume = {4},
	year = {2010},
	year1 = {2010}}

@article{Olson:2008aa,
	author = {Olson, David A. and Chen, Liang and Hillmyer, Marc A.},
	date = {2008/02/01},
	doi = {10.1021/cm702239k},
	isbn = {0897-4756},
	journal = {Chemistry of Materials},
	journal1 = {Chemistry of Materials},
	journal2 = {Chem. Mater.},
	month = {02},
	number = {3},
	pages = {869--890},
	publisher = {American Chemical Society},
	title = {Templating Nanoporous Polymers with Ordered Block Copolymers},
	type = {doi: 10.1021/cm702239k},
	url = {https://doi.org/10.1021/cm702239k},
	volume = {20},
	year = {2008},
	year1 = {2008}}

@article{Zhai:2023aa,
	author = {Yi Zhai and Chao Li and Longcheng Gao},
	doi = {https://doi.org/10.1016/j.giant.2023.100183},
	issn = {2666-5425},
	journal = {Giant},
	pages = {100183},
	title = {Degradable block copolymer-derived nanoporous membranes and their applications},
	url = {https://www.sciencedirect.com/science/article/pii/S2666542523000450},
	volume = {16},
	year = {2023}}

@article{Sinclair:2019aa,
	author = {Sinclair, Alex and Zhou, Xiaoyi and Tangpong, Siwakorn and Bajwa, Dilpreet S. and Quadir, Mohiuddin and Jiang, Long},
	date = {2019/08/20},
	doi = {10.1021/acsomega.9b01313},
	journal = {ACS Omega},
	journal1 = {ACS Omega},
	journal2 = {ACS Omega},
	month = {08},
	number = {8},
	pages = {13189--13199},
	publisher = {American Chemical Society},
	title = {High-Performance Styrene-Butadiene Rubber Nanocomposites Reinforced by Surface-Modified Cellulose Nanofibers},
	type = {doi: 10.1021/acsomega.9b01313},
	url = {https://doi.org/10.1021/acsomega.9b01313},
	volume = {4},
	year = {2019},
	year1 = {2019}}

@article{Khan:2022aa,
	article-number = {161},
	author = {Khan, Anish and Kian, Lau Kia and Jawaid, Mohammad and Khan, Aftab Aslam Parwaz and Alotaibi, Maha Moteb and Asiri, Abdullah M. and Marwani, Hadi M.},
	doi = {10.3390/gels8030161},
	issn = {2310-2861},
	journal = {Gels},
	number = {3},
	pubmedid = {35323274},
	title = {Preparation of Styrene-Butadiene Rubber {(SBR)} Composite Incorporated with Collagen-Functionalized Graphene Oxide for Green Tire Application},
	url = {https://www.mdpi.com/2310-2861/8/3/161},
	volume = {8},
	year = {2022}}

@article{Coldstream:2022aa,
	author = {Coldstream, Jonathan G. and Camp, Philip J. and Phillips, Daniel J. and Dowding, Peter J.},
	doi = {10.1039/D2SM00741J},
	issue = {35},
	journal = {Soft Matter},
	pages = {6538-6549},
	publisher = {The Royal Society of Chemistry},
	title = {Gradient copolymers versus block copolymers: self-assembly in solution and surface adsorption},
	url = {http://dx.doi.org/10.1039/D2SM00741J},
	volume = {18},
	year = {2022}}

@article{Lefebvre:2004aa,
	author = {Lefebvre, Michelle D. and Olvera de la Cruz, Monica and Shull, Kenneth R.},
	date = {2004/02/01},
	doi = {10.1021/ma035141a},
	isbn = {0024-9297},
	journal = {Macromolecules},
	journal1 = {Macromolecules},
	journal2 = {Macromolecules},
	month = {02},
	number = {3},
	pages = {1118--1123},
	publisher = {American Chemical Society},
	title = {Phase Segregation in Gradient Copolymer Melts},
	type = {doi: 10.1021/ma035141a},
	url = {https://doi.org/10.1021/ma035141a},
	volume = {37},
	year = {2004},
	year1 = {2004}}

@article{Self:2022aa,
	author = {Self, Jeffrey L. and Zervoudakis, Aristotle J. and Peng, Xiayu and Lenart, William R. and Macosko, Christopher W. and Ellison, Christopher J.},
	date = {2022/02/28},
	doi = {10.1021/jacsau.1c00500},
	journal = {JACS Au},
	journal1 = {JACS Au},
	journal2 = {JACS Au},
	month = {02},
	number = {2},
	pages = {310--321},
	publisher = {American Chemical Society},
	title = {Linear, Graft, and Beyond: Multiblock Copolymers as Next-Generation Compatibilizers},
	type = {doi: 10.1021/jacsau.1c00500},
	url = {https://doi.org/10.1021/jacsau.1c00500},
	volume = {2},
	year = {2022},
	year1 = {2022}}

@article{Meier:2019aa,
	author = {Meier, Michael A. R. and Barner-Kowollik, Christopher},
	doi = {https://doi.org/10.1002/adma.201806027},
	eprint = {https://advanced.onlinelibrary.wiley.com/doi/pdf/10.1002/adma.201806027},
	journal = {Advanced Materials},
	number = {26},
	pages = {1806027},
	title = {A New Class of Materials: Sequence-Defined Macromolecules and Their Emerging Applications},
	url = {https://advanced.onlinelibrary.wiley.com/doi/abs/10.1002/adma.201806027},
	volume = {31},
	year = {2019}}

@article{Trucillo:2024aa,
	article-number = {456},
	author = {Trucillo, Paolo},
	doi = {10.3390/ma17020456},
	issn = {1996-1944},
	journal = {Materials},
	number = {2},
	pubmedid = {38255624},
	title = {Biomaterials for Drug Delivery and Human Applications},
	url = {https://www.mdpi.com/1996-1944/17/2/456},
	volume = {17},
	year = {2024}}

@article{Li:2024aa,
	author = {Li, Xiaohan and Huang, Jiateng and Chen, Yawen and Zhu, Feiyu and Wang, Yepeng and Wei, Wei and Feng, Yakai},
	date = {2024/12/27},
	doi = {10.1021/acsapm.4c03086},
	journal = {ACS Applied Polymer Materials},
	journal1 = {ACS Applied Polymer Materials},
	journal2 = {ACS Appl. Polym. Mater.},
	month = {12},
	number = {24},
	pages = {14948--14969},
	publisher = {American Chemical Society},
	title = {Polymer-Based Electronic Packaging Molding Compounds, Specifically Thermal Performance Improvement: An Overview},
	type = {doi: 10.1021/acsapm.4c03086},
	url = {https://doi.org/10.1021/acsapm.4c03086},
	volume = {6},
	year = {2024},
	year1 = {2024}}

@article{Lim:2024aa,
	article-number = {3698},
	author = {Lim, Ju Won},
	doi = {10.3390/ma17153698},
	issn = {1996-1944},
	journal = {Materials},
	number = {15},
	pubmedid = {39124361},
	title = {Polymer Materials for Optoelectronics and Energy Applications},
	url = {https://www.mdpi.com/1996-1944/17/15/3698},
	volume = {17},
	year = {2024}}

@article{Das:2023aa,
	author = {Das, Abinash and Ringu, Togam and Ghosh, Sampad and Pramanik, Nabakumar},
	date = {2023/07/01},
	doi = {10.1007/s00289-022-04443-4},
	id = {Das2023},
	isbn = {1436-2449},
	journal = {Polymer Bulletin},
	number = {7},
	pages = {7247--7312},
	title = {A comprehensive review on recent advances in preparation, physicochemical characterization, and bioengineering applications of biopolymers},
	url = {https://doi.org/10.1007/s00289-022-04443-4},
	volume = {80},
	year = {2023}}

@book{Rodriguez:2014aa,
	author = {Rodriguez, Ferdinand and Cohen, Claude and Ober, Christopher K and Archer, Lynden},
	publisher = {CRC press},
	title = {Principles of polymer systems},
	year = {2014}}

@book{Flory:1953aa,
	author = {Flory, Paul J},
	publisher = {Cornell university press},
	title = {Principles of polymer chemistry},
	year = {1953}}

@article{Ido:2025aa,
	author = {Ido, Ryota and Azam, Naveed Ahmed and Zhu, Jianshen and Nagamochi, Hiroshi and Akutsu, Tatsuya},
	date = {2025/07/01},
	doi = {10.1038/s41598-025-05976-0},
	id = {Ido2025},
	isbn = {2045-2322},
	journal = {Scientific Reports},
	number = {1},
	pages = {22214},
	title = {A dynamic programming algorithm for generating chemical isomers based on frequency vectors},
	url = {https://doi.org/10.1038/s41598-025-05976-0},
	volume = {15},
	year = {2025}}

@article{Cheng:2021aa,
	author = {Cheng, Yu and Gong, Yongshun and Liu, Yuansheng and Song, Bosheng and Zou, Quan},
	doi = {10.1093/bib/bbab344},
	eprint = {https://academic.oup.com/bib/article-pdf/22/6/bbab344/41089800/bbab344.pdf},
	issn = {1477-4054},
	journal = {Briefings in Bioinformatics},
	month = {08},
	number = {6},
	pages = {bbab344},
	title = {Molecular design in drug discovery: a comprehensive review of deep generative models},
	url = {https://doi.org/10.1093/bib/bbab344},
	volume = {22},
	year = {2021}}

@article{Kaneko:2023aa,
	author = {Kaneko, Hiromasa},
	date = {2023/06/20},
	doi = {10.1021/acsomega.3c01332},
	journal = {ACS Omega},
	journal1 = {ACS Omega},
	journal2 = {ACS Omega},
	month = {06},
	number = {24},
	pages = {21781--21786},
	publisher = {American Chemical Society},
	title = {Molecular Descriptors, Structure Generation, and Inverse {QSAR/QSPR} Based on {SELFIES}},
	type = {doi: 10.1021/acsomega.3c01332},
	url = {https://doi.org/10.1021/acsomega.3c01332},
	volume = {8},
	year = {2023},
	year1 = {2023}}

@article{Bort:2022aa,
	author = {Bort, William and Mazitov, Daniyar and Horvath, Dragos and Bonachera, Fanny and Lin, Arkadii and Marcou, Gilles and Baskin, Igor and Madzhidov, Timur and Varnek, Alexandre},
	date = {2022/11/28},
	doi = {10.1021/acs.jcim.2c01086},
	isbn = {1549-9596},
	journal = {Journal of Chemical Information and Modeling},
	journal1 = {Journal of Chemical Information and Modeling},
	journal2 = {J. Chem. Inf. Model.},
	month = {11},
	number = {22},
	pages = {5471--5484},
	publisher = {American Chemical Society},
	title = {Inverse {QSAR}: Reversing Descriptor-Driven Prediction Pipeline Using Attention-Based Conditional Variational Autoencoder},
	type = {doi: 10.1021/acs.jcim.2c01086},
	url = {https://doi.org/10.1021/acs.jcim.2c01086},
	volume = {62},
	year = {2022},
	year1 = {2022}}

@article{Cai:2022aa,
	author = {Cai, Hanxuan and Zhang, Huimin and Zhao, Duancheng and Wu, Jingxing and Wang, Ling},
	doi = {10.1093/bib/bbac408},
	eprint = {https://academic.oup.com/bib/article-pdf/23/6/bbac408/47144410/bbac408.pdf},
	issn = {1477-4054},
	journal = {Briefings in Bioinformatics},
	month = {09},
	number = {6},
	pages = {bbac408},
	title = {{FP-GNN}: a versatile deep learning architecture for enhanced molecular property prediction},
	url = {https://doi.org/10.1093/bib/bbac408},
	volume = {23},
	year = {2022}}

@article{Ido:2024aa,
	author = {Ido, Ryota and Cao, Shengjuan and Zhu, Jianshen and Azam, Naveed Ahmed and Haraguchi, Kazuya and Zhao, Liang and Nagamochi, Hiroshi and Akutsu, Tatsuya},
	doi = {10.1109/TCBB.2024.3447780},
	journal = {IEEE/ACM Transactions on Computational Biology and Bioinformatics},
	number = {6},
	pages = {1623-1632},
	title = {A Method for Inferring Polymers Based on Linear Regression and Integer Programming},
	volume = {21},
	year = {2024}}

@article{Zhu:2025aa,
	author = {Zhu, Jianshen and Azam, Naveed Ahmed and Cao, Shengjuan and Ido, Ryota and Haraguchi, Kazuya and Zhao, Liang and Nagamochi, Hiroshi and Akutsu, Tatsuya},
	doi = {10.3389/fgene.2024.1483490},
	issn = {1664-8021},
	journal = {Frontiers in Genetics},
	pages = {1483490},
	title = {Quadratic descriptors and reduction methods in a two-layered model for compound inference},
	url = {https://www.frontiersin.org/journals/genetics/articles/10.3389/fgene.2024.1483490},
	volume = {15},
	year = {2025}}

@article{Shino:2025aa,
	author = {Shino, Yuto and Kaneko, Hiromasa},
	doi = {https://doi.org/10.1002/minf.202400227},
	eprint = {https://onlinelibrary.wiley.com/doi/pdf/10.1002/minf.202400227},
	journal = {Molecular Informatics},
	number = {1},
	pages = {e202400227},
	title = {Improving Molecular Design with Direct Inverse Analysis of {QSAR/QSPR} Model},
	url = {https://onlinelibrary.wiley.com/doi/abs/10.1002/minf.202400227},
	volume = {44},
	year = {2025}}

@article{Zhang:2023ab,
	author = {Zhang, Shuo and Liu, Yang and Xie, Lei},
	date = {2023/11/06},
	doi = {10.1038/s41598-023-46382-8},
	id = {Zhang2023},
	isbn = {2045-2322},
	journal = {Scientific Reports},
	number = {1},
	pages = {19171},
	title = {A universal framework for accurate and efficient geometric deep learning of molecular systems},
	url = {https://doi.org/10.1038/s41598-023-46382-8},
	volume = {13},
	year = {2023}}

@misc{Gasteiger:2022aa,
	archiveprefix = {arXiv},
	author = {Johannes Gasteiger and Shankari Giri and Johannes T. Margraf and Stephan G{\"u}nnemann},
	eprint = {2011.14115},
	howpublished = {arXiv:2011.14115},
	primaryclass = {cs.LG},
	title = {Fast and Uncertainty-Aware Directional Message Passing for Non-Equilibrium Molecules},
	year = {2022}}

@article{Zhu:2022ad,
	author = {Zhu, Jianshen and Azam, Naveed Ahmed and Haraguchi, Kazuya and Zhao, Liang and Nagamochi, Hiroshi and Akutsu, Tatsuya},
	journal = {Frontiers in Bioscience-Landmark},
	number = {6},
	pages = {188},
	publisher = {IMR Press},
	title = {An Inverse {QSAR} Method Based on Linear Regression and Integer Programming},
	volume = {27},
	year = {2022}}

@phdthesis{Zhu:2023aa,
	author = {Zhu, Jianshen},
	month = {9},
	school = {Kyoto University},
	title = {Novel Methods for Chemical Compound Inference Based on Machine Learning and Mixed Integer Linear Programming},
	url = {http://hdl.handle.net/2433/285872},
	year = {2023}}

@article{Miccio:2020aa,
	author = {Luis A. Miccio and Gustavo A. Schwartz},
	doi = {https://doi.org/10.1016/j.polymer.2020.122341},
	issn = {0032-3861},
	journal = {Polymer},
	pages = {122341},
	title = {From chemical structure to quantitative polymer properties prediction through convolutional neural networks},
	url = {https://www.sciencedirect.com/science/article/pii/S0032386120301786},
	volume = {193},
	year = {2020}}

@article{Connor:2017aa,
	author = {Connor, Eric F. and Lees, Inez and Maclean, Derek},
	doi = {https://doi.org/10.1002/pola.28703},
	eprint = {https://onlinelibrary.wiley.com/doi/pdf/10.1002/pola.28703},
	journal = {Journal of Polymer Science Part A: Polymer Chemistry},
	number = {18},
	pages = {3146-3157},
	title = {Polymers as drugs---Advances in therapeutic applications of polymer binding agents},
	url = {https://onlinelibrary.wiley.com/doi/abs/10.1002/pola.28703},
	volume = {55},
	year = {2017}}

@article{Cherkasov:2014aa,
	author = {Cherkasov, Artem and Muratov, Eugene N and Fourches, Denis and Varnek, Alexandre and Baskin, Igor I and Cronin, Mark and Dearden, John and Gramatica, Paola and Martin, Yvonne C and Todeschini, Roberto and others},
	journal = {Journal of Medicinal Chemistry},
	number = {12},
	pages = {4977--5010},
	publisher = {ACS Publications},
	title = {{QSAR} modeling: where have you been? Where are you going to?},
	volume = {57},
	year = {2014}}

@article{Lo:2018aa,
	author = {Lo, Yu-Chen and Rensi, Stefano E and Torng, Wen and Altman, Russ B},
	journal = {Drug Discovery Today},
	number = {8},
	pages = {1538--1546},
	publisher = {Elsevier},
	title = {Machine learning in chemoinformatics and drug discovery},
	volume = {23},
	year = {2018}}

@article{Tetko:2020aa,
	author = {Tetko, Igor V and Engkvist, Ola},
	journal = {Journal of Cheminformatics},
	pages = {1--3},
	publisher = {Springer},
	title = {From Big Data to Artificial Intelligence: chemoinformatics meets new challenges},
	volume = {12},
	year = {2020}}

@article{Miyao:2016aa,
	author = {Miyao, Tomoyuki and Kaneko, Hiromasa and Funatsu, Kimito},
	journal = {Journal of Chemical Information and Modeling},
	number = {2},
	pages = {286--299},
	publisher = {ACS Publications},
	title = {Inverse {QSPR/QSAR} analysis for chemical structure generation (from y to x)},
	volume = {56},
	year = {2016}}

@article{Ikebata:2017aa,
	author = {Ikebata, Hisaki and Hongo, Kenta and Isomura, Tetsu and Maezono, Ryo and Yoshida, Ryo},
	journal = {Journal of Computer-aided Molecular Design},
	pages = {379--391},
	publisher = {Springer},
	title = {Bayesian molecular design with a chemical language model},
	volume = {31},
	year = {2017}}

@article{Rupakheti:2015aa,
	author = {Rupakheti, Chetan and Virshup, Aaron and Yang, Weitao and Beratan, David N},
	journal = {Journal of Chemical Information and Modeling},
	number = {3},
	pages = {529--537},
	publisher = {ACS Publications},
	title = {Strategy to discover diverse optimal molecules in the small molecule universe},
	volume = {55},
	year = {2015}}

@article{Skvortsova:1993aa,
	author = {Skvortsova, Mariya I and Baskin, Igor I and Slovokhotova, Olga L and Palyulin, Vladimir A and Zefirov, Nikolai S},
	journal = {Journal of Chemical Information and Computer Sciences},
	number = {4},
	pages = {630--634},
	publisher = {ACS Publications},
	title = {Inverse problem in {QSAR/QSPR} studies for the case of topological indexes characterizing molecular shape ({Kier} indices)},
	volume = {33},
	year = {1993}}

@inproceedings{Akutsu:2019aa,
	author = {Akutsu, Tatsuya and Nagamochi, Hiroshi},
	booktitle = {Proceedings of the 2nd International Conference on Information Science and Systems},
	pages = {215--220},
	title = {A mixed integer linear programming formulation to artificial neural networks},
	year = {2019}}

@conference{Azam:2020aa,
	author = {Azam, Naveed Ahmed and Chiewvanichakorn, Rachaya and Zhang, Fan and Shurbevski, Aleksandar and Nagamochi, Hiroshi and Akutsu, Tatsuya},
	booktitle = {Proceedings of the 13th International Joint Conference on Biomedical Engineering Systems and Technologies - BIOINFORMATICS},
	organization = {INSTICC},
	pages = {101-108},
	publisher = {SciTePress},
	title = {A Novel Method for the Inverse {QSAR/QSPR} based on Artificial Neural Networks and Mixed Integer Linear Programming with Guaranteed Admissibility},
	year = {2020}}

@article{Zhang:2022aa,
	author = {Zhang, Fan and Zhu, Jianshen and Chiewvanichakorn, Rachaya and Shurbevski, Aleksandar and Nagamochi, Hiroshi and Akutsu, Tatsuya},
	journal = {Applied Intelligence},
	number = {15},
	pages = {17058--17072},
	publisher = {Springer US New York},
	title = {A new approach to the design of acyclic chemical compounds using skeleton trees and integer linear programming},
	volume = {52},
	year = {2022}}

@incollection{Ito:2021aa,
	author = {Ito, Ren and Azam, Naveed Ahmed and Wang, Chenxi and Shurbevski, Aleksandar and Nagamochi, Hiroshi and Akutsu, Tatsuya},
	booktitle = {Advances in Computer Vision and Computational Biology: Proceedings from IPCV'20, HIMS'20, BIOCOMP'20, and BIOENG'20},
	pages = {641--655},
	publisher = {Springer},
	title = {A novel method for the inverse {QSAR/QSPR} to monocyclic chemical compounds based on artificial neural networks and integer programming},
	year = {2021}}

@article{Zhu:2020aa,
	author = {Zhu, Jianshen and Wang, Chenxi and Shurbevski, Aleksandar and Nagamochi, Hiroshi and Akutsu, Tatsuya},
	journal = {Algorithms},
	number = {5},
	pages = {124},
	publisher = {MDPI},
	title = {A novel method for inference of chemical compounds of cycle index two with desired properties based on artificial neural networks and integer programming},
	volume = {13},
	year = {2020}}

@article{Azam:2021aa,
	author = {Azam, Naveed Ahmed and Zhu, Jianshen and Sun, Yanming and Shi, Yu and Shurbevski, Aleksandar and Zhao, Liang and Nagamochi, Hiroshi and Akutsu, Tatsuya},
	journal = {Algorithms for Molecular Biology},
	pages = {1--39},
	publisher = {Springer},
	title = {A novel method for inference of acyclic chemical compounds with bounded branch-height based on artificial neural networks and integer programming},
	volume = {16},
	year = {2021}}

@article{Shi:2021aa,
	author = {Shi, Yu and Zhu, Jianshen and Azam, Naveed Ahmed and Haraguchi, Kazuya and Zhao, Liang and Nagamochi, Hiroshi and Akutsu, Tatsuya},
	journal = {International Journal of Molecular Sciences},
	number = {6},
	pages = {2847},
	publisher = {MDPI},
	title = {An inverse {QSAR} method based on a two-layered model and integer programming},
	volume = {22},
	year = {2021}}

@inproceedings{Tanaka:2021aa,
	author = {Tanaka, Kouki and Zhu, Jianshen and Azam, Naveed Ahmed and Haraguchi, Kazuya and Zhao, Liang and Nagamochi, Hiroshi and Akutsu, Tatsuya},
	booktitle = {Intelligent Computing Theories and Application: 17th International Conference, ICIC 2021, Shenzhen, China, August 12--15, 2021, Proceedings, Part II},
	organization = {Springer},
	pages = {628--644},
	title = {An inverse {QSAR} method based on decision tree and integer programming},
	year = {2021}}

@inproceedings{Azam:2021ab,
	author = {Azam, Naveed Ahmed and Zhu, Jianshen and Haraguchi, Kazuya and Zhao, Liang and Nagamochi, Hiroshi and Akutsu, Tatsuya},
	booktitle = {2021 IEEE International Conference on Bioinformatics and Biomedicine (BIBM)},
	organization = {IEEE},
	pages = {360--363},
	title = {Molecular design based on artificial neural networks, integer programming and grid neighbor search},
	year = {2021}}
 
\appendix
\newpage

\renewcommand\thefigure{A\arabic{figure}}    
\renewcommand{\thetable}{A\arabic{table}}

\section*{Appendix}

\newcommand{\cnt}{\mathrm{cnt}}  

\newcommand{\nlnk}{\mathrm{n}^\mathrm{lnk}}  

\newcommand{\SAE}{\mathrm{SAE}}  
\newcommand{\MAE}{\mathrm{MAE}}  

\newcommand{\Tfr}{\mathcal{T}_\mathrm{fr}}  
 
\newcommand{\lf}{\mathrm{lf}}  

\newcommand{\newone}{\marginpar{\bf NEW!}}
\newcommand{\CHeck}{\marginpar{\bf Check!}}

\newcommand{\eledegC}{\mathrm{eledeg}_\mathrm{C}}   
\newcommand{\eledegT}{\mathrm{eledeg}_\mathrm{T}}   
\newcommand{\eledegF}{\mathrm{eledeg}_\mathrm{F}}   
\newcommand{\eledegX}{\mathrm{eledeg}_\mathrm{X}}   

\newcommand{\vion}{\mathrm{v}_\mathrm{ion}}  

\newcommand{\ttH}{{\tt H}}  
\newcommand{\ttC}{{\tt C}}  
\newcommand{\ttO}{{\tt O}}  
\newcommand{\ttN}{{\tt N}}  
\newcommand{\ttP}{{\tt P}}  
\newcommand{\ttF}{{\tt F}}  
\newcommand{\ttCl}{{\tt Cl}}  
\newcommand{\ttS}{{\tt S}}  
\newcommand{\ttSi}{{\tt Si}}

\newcommand{\oH}{\overline{{\tt H}}}  

\newcommand{\Z}{\mathbb{Z}}  
\newcommand{\Sbb}{\mathbb{S}}  
\newcommand{\C}{\mathbb{C}}  
\newcommand{\Co}{\mathbb{C}}  
\newcommand{\fr}{\mathrm{fr}}    
\newcommand{\st}{\mathrm{st}}   
\newcommand{\anpsi}{\langle \psi \rangle}  

\newcommand{\cv}{\mathrm{cv}}  

\newcommand{\R}{\mathbb{R}} 
\newcommand{\RKw}{\mathbb{R}^{K+1}} 

\newcommand{\Rtcv}{\mathrm{R}^2_\mathrm{cv}}  
\newcommand{\Rtmax}{\mathrm{R}^2_\mathrm{max}}  

\newcommand{\bs}{\mathrm{bias}}  

\newcommand{\deghyd}{\deg^\mathrm{hyd}} 
\newcommand{\hydg}{\mathrm{hydg}} 
\newcommand{\delhyd}{\delta_\mathrm{hyd}} 
\newcommand{\delhydC}{\delta_\mathrm{hyd}^\mathrm{C}} 
\newcommand{\delhydT}{\delta_\mathrm{hyd}^\mathrm{T}} 
\newcommand{\delhydF}{\delta_\mathrm{hyd}^\mathrm{F}} 
\newcommand{\delhydX}{\delta_\mathrm{hyd}^\mathrm{X}} 

\newcommand{\thftr}{\theta_\mathrm{ftr}} 
\newcommand{\thC}{\theta^\mathrm{C}} 
\newcommand{\thChd}{\theta_\mathrm{head}^\mathrm{C}} 
\newcommand{\thCtl}{\theta_\mathrm{tail}^\mathrm{C}} 
\newcommand{\thT}{\theta^\mathrm{T}} 
\newcommand{\thF}{\theta^\mathrm{F}} 
\newcommand{\thX}{\theta^\mathrm{X}} 
\newcommand{\thCT}{\theta^\mathrm{CT}} 
\newcommand{\thTC}{\theta^\mathrm{TC}} 
\newcommand{\thCF}{\theta^\mathrm{CF}} 
\newcommand{\thTF}{\theta^\mathrm{TF}} 
\newcommand{\thCTT}{\theta_\mathrm{T}^\mathrm{CT}} 
\newcommand{\thTCT}{\theta_\mathrm{T}^\mathrm{TC}} 
\newcommand{\thCTC}{\theta_\mathrm{C}^\mathrm{CT}} 
\newcommand{\thTCC}{\theta_\mathrm{C}^\mathrm{TC}} 
\newcommand{\thCFF}{\theta_\mathrm{F}^\mathrm{CF}} 
\newcommand{\thCFC}{\theta_\mathrm{C}^\mathrm{CF}} 
\newcommand{\thTFF}{\theta_\mathrm{F}^\mathrm{TF}} 
\newcommand{\thTFT}{\theta_\mathrm{T}^\mathrm{TF}} 

\newcommand{\FrL}{\mathcal{F}_\Lambda} 
\newcommand{\FrC}{\mathcal{F}^\mathrm{C}} 
\newcommand{\FrT}{\mathcal{F}^\mathrm{T}} 
\newcommand{\FrF}{\mathcal{F}^\mathrm{F}} 
\newcommand{\FrX}{\mathcal{F}^\mathrm{X}}  

\newcommand{\dcp}{\mathrm{dcp}} 
\newcommand{\std}{\mathrm{std}} 
\newcommand{\avg}{\mathrm{avg}} 

\newcommand{\tauac}{\tau_\mathrm{ac}} 

\newcommand{\sint}{\sigma_\mathrm{int}} 
\newcommand{\sce}{\sigma_\mathrm{ce}}  

\newcommand{\sco}{\sigma_\mathrm{co}} 
\newcommand{\snc}{\sigma_\mathrm{nc}} 
\newcommand{\sab}{\sigma_\mathrm{\alpha \beta}} 
\newcommand{\sch}{\sigma_\mathrm{ch}} 
\newcommand{\sC}{\sigma^\mathrm{C}} 
\newcommand{\sT}{\sigma^\mathrm{T}} 
\newcommand{\sF}{\sigma^\mathrm{F}} 
\newcommand{\sX}{\sigma^\mathrm{X}} 

\newcommand{\idfT}{\mathrm{id}_\mathrm{first}^\mathrm{T}} 
\newcommand{\idlT}{\mathrm{id}_\mathrm{last}^\mathrm{T}} 

\newcommand{\mul}{\mathrm{mul}}

\newcommand{\NC}{N_\mathrm{C}}
\newcommand{\Nz}{N_{(0/1)}}
\newcommand{\Nw}{N_{(\geq 1)}}
\newcommand{\Nt}{N_{(\geq 2)}}
\newcommand{\New}{N_{(=1)}}

\newcommand{\Ilnk}{I_\mathrm{lnk}} 
\newcommand{\Iz}{I_{(0/1)}}
\newcommand{\Iw}{I_{(\geq 1)}}
\newcommand{\It}{I_{(\geq 2)}}
\newcommand{\Iew}{I_{(=1)}}
\newcommand{\IVCs}{I_\mathrm{VC*}} 
\newcommand{\Iprv}{I_\mathrm{prv}} 
\newcommand{\tIprv}{\widetilde{I}_\mathrm{prv}} 
\newcommand{\tTree}{\widetilde{T}} 

\newcommand{\Gac}{\Gamma_\mathrm{ac}} 
\newcommand{\Gacs}{\Gamma_\mathrm{ac,<}} 
\newcommand{\Gace}{\Gamma_\mathrm{ac,=}} 
\newcommand{\Gacl}{\Gamma_\mathrm{ac,>}} 

\newcommand{\GacX}{{\Gamma}_\mathrm{ac}^\mathrm{X}}  
\newcommand{\GacC}{{\Gamma}_\mathrm{ac}^\mathrm{C}}  
\newcommand{\GacT}{{\Gamma}_\mathrm{ac}^\mathrm{T}} 
\newcommand{\GacF}{{\Gamma}_\mathrm{ac}^\mathrm{F}}  
\newcommand{\GacCT}{{\Gamma}_\mathrm{ac}^\mathrm{CT}}  
\newcommand{\GacTC}{{\Gamma}_\mathrm{ac}^\mathrm{TC}}  
\newcommand{\GacCF}{{\Gamma}_\mathrm{ac}^\mathrm{CF}}  
\newcommand{\GacTF}{{\Gamma}_\mathrm{ac}^\mathrm{TF}}  

\newcommand{\tGacex}{\widetilde{\Gamma}_\mathrm{ac}^\mathrm{ex}}  
\newcommand{\tGacX}{\widetilde{\Gamma}_\mathrm{ac}^\mathrm{X}}  
\newcommand{\tGacC}{\widetilde{\Gamma}_\mathrm{ac}^\mathrm{C}}  
\newcommand{\tGacT}{\widetilde{\Gamma}_\mathrm{ac}^\mathrm{T}} 
\newcommand{\tGacF}{\widetilde{\Gamma}_\mathrm{ac}^\mathrm{F}}  
\newcommand{\tGacCT}{\widetilde{\Gamma}_\mathrm{ac}^\mathrm{CT}}  
\newcommand{\tGacTC}{\widetilde{\Gamma}_\mathrm{ac}^\mathrm{TC}}  
\newcommand{\tGacCF}{\widetilde{\Gamma}_\mathrm{ac}^\mathrm{CF}}  
\newcommand{\tGacTF}{\widetilde{\Gamma}_\mathrm{ac}^\mathrm{TF}}

\newcommand{\LdgX}{\Lambda_\mathrm{dg}^\mathrm{X}}   
\newcommand{\LdgC}{\Lambda_\mathrm{dg}^\mathrm{C}}   
\newcommand{\LdgT}{\Lambda_\mathrm{dg}^\mathrm{T}}   
\newcommand{\LdgF}{\Lambda_\mathrm{dg}^\mathrm{F}}     
\newcommand{\LdgCco}{\Lambda_\mathrm{dg}^\mathrm{C,co}}
\newcommand{\LdgCnc}{\Lambda_\mathrm{dg}^\mathrm{C,nc}}   
\newcommand{\LdgTnc}{\Lambda_\mathrm{dg}^\mathrm{T,nc}}   
\newcommand{\LdgFnc}{\Lambda_\mathrm{dg}^\mathrm{F,nc}}   
\newcommand{\LdgXnc}{\Lambda_\mathrm{dg}^\mathrm{X,nc}}  
\newcommand{\LdgXtyp}{\Lambda_\mathrm{dg}^\mathrm{X,t}}     

\newcommand{\tLdgX}{\widetilde{\Lambda}_\mathrm{dg}^\mathrm{X}}   
\newcommand{\tLdgC}{\widetilde{\Lambda}_\mathrm{dg}^\mathrm{C}}   
\newcommand{\tLdgT}{\widetilde{\Lambda}_\mathrm{dg}^\mathrm{T}}   
\newcommand{\tLdgF}{\widetilde{\Lambda}_\mathrm{dg}^\mathrm{F}}     
\newcommand{\tLdgCnc}{\widetilde{\Lambda}_\mathrm{dg}^\mathrm{C,nc}}   
\newcommand{\tLdgTnc}{\widetilde{\Lambda}_\mathrm{dg}^\mathrm{T,nc}}   
\newcommand{\tLdgFnc}{\widetilde{\Lambda}_\mathrm{dg}^\mathrm{F,nc}}   
\newcommand{\tLdgXnc}{\widetilde{\Lambda}_\mathrm{dg}^\mathrm{X,nc}}  
\newcommand{\tLdgXtyp}{\widetilde{\Lambda}_\mathrm{dg}^\mathrm{X,t}}     

\newcommand{\Lbd}{\Lambda_\mathrm{bd}}   
\newcommand{\LbdX}{\Lambda_\mathrm{bd}^\mathrm{X}}   
\newcommand{\LbdC}{\Lambda_\mathrm{bd}^\mathrm{C}}   
\newcommand{\LbdT}{\Lambda_\mathrm{bd}^\mathrm{T}}   
\newcommand{\LbdF}{\Lambda_\mathrm{bd}^\mathrm{F}}     
\newcommand{\LbdCco}{\Lambda_\mathrm{bd}^\mathrm{C,co}}
\newcommand{\LbdCnc}{\Lambda_\mathrm{bd}^\mathrm{C,nc}}   
\newcommand{\LbdTnc}{\Lambda_\mathrm{bd}^\mathrm{T,nc}}   
\newcommand{\LbdFnc}{\Lambda_\mathrm{bd}^\mathrm{F,nc}}   
\newcommand{\LbdXnc}{\Lambda_\mathrm{bd}^\mathrm{X,nc}}  
\newcommand{\LbdXtyp}{\Lambda_\mathrm{bd}^\mathrm{X,t}}    

\newcommand{\Lbdp}{\Lambda_\mathrm{bd+}}   
\newcommand{\LbdpX}{\Lambda_\mathrm{bd+}^\mathrm{X}}   
\newcommand{\LbdpC}{\Lambda_\mathrm{bd+}^\mathrm{C}}   
\newcommand{\LbdpT}{\Lambda_\mathrm{bd+}^\mathrm{T}}   
\newcommand{\LbdpF}{\Lambda_\mathrm{bd+}^\mathrm{F}}     
\newcommand{\LbdpCco}{\Lambda_\mathrm{bd+}^\mathrm{C,co}}
\newcommand{\LbdpCnc}{\Lambda_\mathrm{bd+}^\mathrm{C,nc}}   
\newcommand{\LbdpTnc}{\Lambda_\mathrm{bd+}^\mathrm{T,nc}}   
\newcommand{\LbdpFnc}{\Lambda_\mathrm{bd+}^\mathrm{F,nc}}   
\newcommand{\LbdpXnc}{\Lambda_\mathrm{bd+}^\mathrm{X,nc}}  
\newcommand{\LbdpXtyp}{\Lambda_\mathrm{bd+}^\mathrm{X,t}}     
 
\newcommand{\Lbdm}{\Lambda_\mathrm{bd-}}   
\newcommand{\LbdmX}{\Lambda_\mathrm{bd-}^\mathrm{X}}   
\newcommand{\LbdmC}{\Lambda_\mathrm{bd-}^\mathrm{C}}   
\newcommand{\LbdmT}{\Lambda_\mathrm{bd-}^\mathrm{T}}   
\newcommand{\LbdmF}{\Lambda_\mathrm{bd-}^\mathrm{F}}     
\newcommand{\LbdmCco}{\Lambda_\mathrm{bd-}^\mathrm{C,co}}
\newcommand{\LbdmCnc}{\Lambda_\mathrm{bd-}^\mathrm{C,nc}}   
\newcommand{\LbdmTnc}{\Lambda_\mathrm{bd-}^\mathrm{T,nc}}   
\newcommand{\LbdmFnc}{\Lambda_\mathrm{bd-}^\mathrm{F,nc}}   
\newcommand{\LbdmXnc}{\Lambda_\mathrm{bd-}^\mathrm{X,nc}}  
\newcommand{\LbdmXtyp}{\Lambda_\mathrm{bd-}^\mathrm{X,t}}    

\newcommand{\tLbdX}{\widetilde{\Lambda}_\mathrm{bd}^\mathrm{X}}   
\newcommand{\tLbdC}{\widetilde{\Lambda}_\mathrm{bd}^\mathrm{C}}   
\newcommand{\tLbdT}{\widetilde{\Lambda}_\mathrm{bd}^\mathrm{T}}   
\newcommand{\tLbdF}{\widetilde{\Lambda}_\mathrm{bd}^\mathrm{F}}     
\newcommand{\tLbdCnc}{\widetilde{\Lambda}_\mathrm{bd}^\mathrm{C,nc}}   
\newcommand{\tLbdTnc}{\widetilde{\Lambda}_\mathrm{bd}^\mathrm{T,nc}}   
\newcommand{\tLbdFnc}{\widetilde{\Lambda}_\mathrm{bd}^\mathrm{F,nc}}   
\newcommand{\tLbdXnc}{\widetilde{\Lambda}_\mathrm{bd}^\mathrm{X,nc}}  
\newcommand{\tLbdXtyp}{\widetilde{\Lambda}_\mathrm{bd}^\mathrm{X,t}}     

\newcommand{\tGecex}{\widetilde{\Gamma}_\mathrm{ec}^\mathrm{ex}}  
\newcommand{\tGecX}{\widetilde{\Gamma}_\mathrm{ec}^\mathrm{X}}  
\newcommand{\tGecC}{\widetilde{\Gamma}_\mathrm{ec}^\mathrm{C}}  
\newcommand{\tGecT}{\widetilde{\Gamma}_\mathrm{ec}^\mathrm{T}} 
\newcommand{\tGecF}{\widetilde{\Gamma}_\mathrm{ec}^\mathrm{F}}  
\newcommand{\tGecCT}{\widetilde{\Gamma}_\mathrm{ec}^\mathrm{CT}}  
\newcommand{\tGecTC}{\widetilde{\Gamma}_\mathrm{ec}^\mathrm{TC}}  
\newcommand{\tGecCF}{\widetilde{\Gamma}_\mathrm{ec}^\mathrm{CF}}  
\newcommand{\tGecTF}{\widetilde{\Gamma}_\mathrm{ec}^\mathrm{TF}}  

\newcommand{\rX}{r_\mathrm{X}}  
\newcommand{\rC}{r_\mathrm{C}}  
\newcommand{\rT}{r_\mathrm{T}} 
\newcommand{\rF}{r_\mathrm{F}}  
\newcommand{\rCT}{r_\mathrm{CT}}  
\newcommand{\rTC}{r_\mathrm{TC}}  
\newcommand{\rCF}{r_\mathrm{CF}}  
\newcommand{\rTF}{r_\mathrm{TF}}  

\newcommand{\typ}{\mathrm{t}}

\newcommand{\f}{f}
\newcommand{\w}{w}
\newcommand{\x}{x}
\newcommand{\y}{\pmb{y}}
\newcommand{\z}{\pmb{z}}
\newcommand{\s}{\pmb{s}}
\newcommand{\m}{\pmb{m}}
\newcommand{\1}{\pmb{1}} 
\newcommand{\0}{\pmb{0}}

\newcommand{\tbc}{{\tt bc}}
\newcommand{\Ldg}{\Lambda_{\mathrm{dg}}}

\newcommand{\qmax}{q_{\max}}
\newcommand{\mmax}{m_{\max}}
\newcommand{\mUB}{m_{\mathrm{UB}}}

\newcommand{\ints}{\mathrm{ints}}  
\newcommand{\fc}{\mathrm{fc}} 
\newcommand{\betar}{\beta_\mathrm{r}} 

\newcommand{\dmax}{d_{\max}}
\newcommand{\dia}{\mathrm{dia}}
\newcommand{\en}{\mathrm{end}}
\newcommand{\pair}{\mathrm{pair}}
\newcommand{\pred}{\mathrm{pred}}
\newcommand{\code}{\mathrm{code}}

\newcommand{\Pt}{\mathcal{P}}
\newcommand{\F}{\mathcal{F}}
\newcommand{\Sb}{\mathcal{S}}
\newcommand{\Cr}{\mathrm{Cr}}  
\newcommand{\M}{\mathcal{M}}
\newcommand{\FT}{\mathcal{FT}} 
\newcommand{\PnT}{\mathcal{PT}} 
\newcommand{\W}{\mathrm{W}}  
\newcommand{\T}{\mathcal{T}}   
\newcommand{\Ww}{\mathcal{W}}
\newcommand{\V}{\mathrm{V}}  
\newcommand{\Fv}{\mathrm{F}}  
\newcommand{\Nft}{N_\mathrm{ft}}  
\newcommand{\Nin}{N_\mathrm{in}}  
\newcommand{\Nr}{N_\mathrm{r}}  
\newcommand{\Nco}{N_\mathrm{co}}  
\newcommand{\Nnc}{N_\mathrm{nc}}  
\newcommand{\Nend}{N_\mathrm{end}}  
\newcommand{\Nsrc}{N_\mathrm{src}}  
\newcommand{\Nsink}{N_\mathrm{sink}}  
\newcommand{\Nbase}{N_\mathrm{base}}  
\newcommand{\Aco}{A_\mathrm{co}}  
\newcommand{\Anc}{A_\mathrm{nc}}

\newcommand{\nint}{\mathrm{n}^\mathrm{int}}  
\newcommand{\nscd}{\mathrm{n}^\mathrm{scd}}  

\newcommand{\co}{\mathrm{co}}
\newcommand{\nc}{\mathrm{nc}}
\newcommand{\cs}{\mathrm{cs}}
\newcommand{\chch}{\mathrm{ch}}
\newcommand{\lh}{\mathrm{lh}}
\newcommand{\ce}{\mathrm{ce}}

\newcommand{\na}{\mathrm{na}}
 
\newcommand{\naX}{\mathrm{na}_\mathrm{X}}
\newcommand{\naY}{\mathrm{na}_\mathrm{Y}}
\newcommand{\naXp}{\mathrm{na}_{\mathrm{X}(p)}}
\newcommand{\naC}{\mathrm{na}_\mathrm{C}}
\newcommand{\naT}{\mathrm{na}_\mathrm{T}}
\newcommand{\naF}{\mathrm{na}_\mathrm{F}}
\newcommand{\naCT}{\mathrm{na}_\mathrm{CT}}
\newcommand{\naTC}{\mathrm{na}_\mathrm{TC}}
\newcommand{\naTF}{\mathrm{na}_\mathrm{TF}}
\newcommand{\naCF}{\mathrm{na}_\mathrm{CF}}
 
\newcommand{\ecX}{\mathrm{ec}_\mathrm{X}}
\newcommand{\ecY}{\mathrm{ec}_\mathrm{Y}}
\newcommand{\ecXp}{\mathrm{ec}_{\mathrm{X}(p)}}
\newcommand{\ecCp}{\mathrm{ec}_{\mathrm{C}(p)}}
\newcommand{\ecTp}{\mathrm{ec}_{\mathrm{T}(p)}}
\newcommand{\ecFp}{\mathrm{ec}_{\mathrm{F}(p)}}
\newcommand{\ecXz}{\mathrm{ec}_{\mathrm{X}(0)}}
\newcommand{\ecXw}{\mathrm{ec}_{\mathrm{X}(1)}}
\newcommand{\ecXpp}{\mathrm{ec}_{\mathrm{X}(p+1)}}
\newcommand{\ecXr}{\mathrm{ec}_{\mathrm{X}(\rho)}}
\newcommand{\ecCr}{\mathrm{ec}_{\mathrm{C}(\rho)}}
\newcommand{\ecTr}{\mathrm{ec}_{\mathrm{T}(\rho)}}
\newcommand{\ecFr}{\mathrm{ec}_{\mathrm{F}(\rho)}}
\newcommand{\ecXrp}{\mathrm{ec}_{\mathrm{X}(\rho+1)}}
\newcommand{\ecC}{\mathrm{ec}_\mathrm{C}}
\newcommand{\ecT}{\mathrm{ec}_\mathrm{T}}
\newcommand{\ecF}{\mathrm{ec}_\mathrm{F}}
\newcommand{\ecCT}{\mathrm{ec}_\mathrm{CT}}
\newcommand{\ecTC}{\mathrm{ec}_\mathrm{TC}}
\newcommand{\ecTF}{\mathrm{ec}_\mathrm{TF}}
\newcommand{\ecCF}{\mathrm{ec}_\mathrm{CF}}
 
\newcommand{\acX}{\mathrm{ac}_\mathrm{X}}
\newcommand{\acY}{\mathrm{ac}_\mathrm{Y}}
\newcommand{\acXp}{\mathrm{ac}_{\mathrm{X}(p)}}
\newcommand{\acC}{\mathrm{ac}_\mathrm{C}}
\newcommand{\acT}{\mathrm{ac}_\mathrm{T}}
\newcommand{\acF}{\mathrm{ac}_\mathrm{F}}
\newcommand{\acCT}{\mathrm{ac}_\mathrm{CT}}
\newcommand{\acTC}{\mathrm{ac}_\mathrm{TC}}
\newcommand{\acTF}{\mathrm{ac}_\mathrm{TF}}
\newcommand{\acCF}{\mathrm{ac}_\mathrm{CF}}

\newcommand{\acqX}{\mathrm{ac}_{q,\mathrm{X}}}
\newcommand{\acqY}{\mathrm{ac}_{q,\mathrm{Y}}}
\newcommand{\acqXp}{\mathrm{ac}_{q,\mathrm{X}(p)}}
\newcommand{\acqC}{\mathrm{ac}_{q,\mathrm{C}}}
\newcommand{\acqT}{\mathrm{ac}_{q,\mathrm{T}}}
\newcommand{\acqF}{\mathrm{ac}_{q,\mathrm{F}}}
\newcommand{\acqCT}{\mathrm{ac}_{q,\mathrm{CT}}}
\newcommand{\acqTC}{\mathrm{ac}_{q,\mathrm{TC}}}
\newcommand{\acqTF}{\mathrm{ac}_{q,\mathrm{TF}}}
\newcommand{\acqCF}{\mathrm{ac}_{q,\mathrm{CF}}}

\newcommand{\GDc}{\Gamma_\mathrm{Dcarbon}}
\newcommand{\dDcX}{\delta_\mathrm{Dcarbon}^\mathrm{X}}
\newcommand{\dDcY}{\delta_\mathrm{Dcarbon}^\mathrm{Y}}
\newcommand{\dDcXp}{\delta_\mathrm{Dcarbon}^{\mathrm{X}(p)}}
\newcommand{\dDcC}{\delta_\mathrm{Dcarbon}^\mathrm{C}}
\newcommand{\dDcT}{\delta_\mathrm{Dcarbon}^\mathrm{T}}
\newcommand{\dDcF}{\delta_\mathrm{Dcarbon}^\mathrm{F}}

\newcommand{\bdX}{\mathrm{bd}_\mathrm{X}}
\newcommand{\bdY}{\mathrm{bd}_\mathrm{Y}}
\newcommand{\bdXp}{\mathrm{bd}_{\mathrm{X}(p)}}
\newcommand{\bdC}{\mathrm{bd}_\mathrm{C}}
\newcommand{\bdT}{\mathrm{bd}_\mathrm{T}}
\newcommand{\bdF}{\mathrm{bd}_\mathrm{F}}
\newcommand{\bdCT}{\mathrm{bd}_\mathrm{CT}}
\newcommand{\bdTC}{\mathrm{bd}_\mathrm{TC}}
\newcommand{\bdTF}{\mathrm{bd}_\mathrm{TF}}
\newcommand{\bdCF}{\mathrm{bd}_\mathrm{CF}}

\newcommand{\sd}{\mathrm{sd}} 
\newcommand{\sdX}{\mathrm{sd}_\mathrm{X}}
\newcommand{\sdY}{\mathrm{sd}_\mathrm{Y}}
\newcommand{\sdXw}{\mathrm{sd}_{\mathrm{X}(1)}}
\newcommand{\sdXt}{\mathrm{sd}_{\mathrm{X}(2)}}
\newcommand{\sdCw}{\mathrm{sd}_{\mathrm{C}(1)}}
\newcommand{\sdCt}{\mathrm{sd}_{\mathrm{C}(2)}}
\newcommand{\sdCp}{\mathrm{sd}_{\mathrm{C}(p)}}
\newcommand{\sdXp}{\mathrm{sd}_{\mathrm{X}(p)}}
\newcommand{\sdC}{\mathrm{sd}_\mathrm{C}}
\newcommand{\sdT}{\mathrm{sd}_\mathrm{T}}
\newcommand{\sdF}{\mathrm{sd}_\mathrm{F}}
\newcommand{\sdCT}{\mathrm{sd}_\mathrm{CT}}
\newcommand{\sdTC}{\mathrm{sd}_\mathrm{TC}}
\newcommand{\sdTF}{\mathrm{sd}_\mathrm{TF}}
\newcommand{\sdCF}{\mathrm{sd}_\mathrm{CF}}

\newcommand{\sbX}{\mathrm{sb}_\mathrm{X}}
\newcommand{\sbY}{\mathrm{sb}_\mathrm{Y}}
\newcommand{\sbXw}{\mathrm{sb}_{\mathrm{X}(1)}}
\newcommand{\sbXt}{\mathrm{sb}_{\mathrm{X}(2)}}
\newcommand{\sbCw}{\mathrm{sb}_{\mathrm{C}(1)}}
\newcommand{\sbCt}{\mathrm{sb}_{\mathrm{C}(2)}}
\newcommand{\sbCp}{\mathrm{sb}_{\mathrm{C}(p)}}
\newcommand{\sbXp}{\mathrm{sb}_{\mathrm{X}(p)}}
\newcommand{\sbC}{\mathrm{sb}_\mathrm{C}}
\newcommand{\sbT}{\mathrm{sb}_\mathrm{T}}
\newcommand{\sbF}{\mathrm{sb}_\mathrm{F}}
\newcommand{\sbCT}{\mathrm{sb}_\mathrm{CT}}
\newcommand{\sbTC}{\mathrm{sb}_\mathrm{TC}}
\newcommand{\sbTF}{\mathrm{sb}_\mathrm{TF}}
\newcommand{\sbCF}{\mathrm{sb}_\mathrm{CF}}

\newcommand{\ap}{\mathrm{ap}} 
\newcommand{\apX}{\mathrm{ap}_\mathrm{X}}
\newcommand{\apY}{\mathrm{ap}_\mathrm{Y}}
\newcommand{\apXp}{\mathrm{ap}_{\mathrm{X}(p)}}
\newcommand{\apC}{\mathrm{ap}_\mathrm{C}}
\newcommand{\apT}{\mathrm{ap}_\mathrm{T}}
\newcommand{\apF}{\mathrm{ap}_\mathrm{F}}
\newcommand{\apCT}{\mathrm{ap}_\mathrm{CT}}
\newcommand{\apTC}{\mathrm{ap}_\mathrm{TC}}
\newcommand{\apTF}{\mathrm{ap}_\mathrm{TF}}
\newcommand{\apCF}{\mathrm{ap}_\mathrm{CF}}

\newcommand{\inc}{\mathrm{inc}} 
\newcommand{\incX}{\mathrm{inc}_\mathrm{X}}
\newcommand{\incY}{\mathrm{inc}_\mathrm{Y}}
\newcommand{\incXp}{\mathrm{inc}_{\mathrm{X}(p)}}
\newcommand{\incC}{\mathrm{inc}_\mathrm{C}}
\newcommand{\incT}{\mathrm{inc}_\mathrm{T}}
\newcommand{\incF}{\mathrm{inc}_\mathrm{F}}
\newcommand{\incCT}{\mathrm{inc}_\mathrm{CT}}
\newcommand{\incTC}{\mathrm{inc}_\mathrm{TC}}
\newcommand{\incTF}{\mathrm{inc}_\mathrm{TF}}
\newcommand{\incCF}{\mathrm{inc}_\mathrm{CF}}

\newcommand{\ns}{\mathrm{ns}}
\newcommand{\nsX}{\mathrm{ns}_\mathrm{X}}
\newcommand{\nsY}{\mathrm{ns}_\mathrm{Y}}
\newcommand{\nsXp}{\mathrm{ns}_{\mathrm{X}(p)}}
\newcommand{\nsC}{\mathrm{ns}_\mathrm{C}}
\newcommand{\nsT}{\mathrm{ns}_\mathrm{T}}
\newcommand{\nsF}{\mathrm{ns}_\mathrm{F}}
\newcommand{\nsCT}{\mathrm{ns}_\mathrm{CT}}
\newcommand{\nsTC}{\mathrm{ns}_\mathrm{TC}}
\newcommand{\nsTF}{\mathrm{ns}_\mathrm{TF}}
\newcommand{\nsCF}{\mathrm{ns}_\mathrm{CF}}

\newcommand{\dgY}{\mathrm{dg}_\mathrm{Y}}
\newcommand{\dgXp}{\mathrm{dg}_{\mathrm{X}(p)}}
\newcommand{\dgX}{\mathrm{dg}_\mathrm{X}}
\newcommand{\dgC}{\mathrm{dg}_\mathrm{C}}
\newcommand{\dgT}{\mathrm{dg}_\mathrm{T}}
\newcommand{\dgF}{\mathrm{dg}_\mathrm{F}}
\newcommand{\dgCT}{\mathrm{dg}_\mathrm{CT}}
\newcommand{\dgTC}{\mathrm{dg}_\mathrm{TC}}
\newcommand{\dgTF}{\mathrm{dg}_\mathrm{TF}}
\newcommand{\dgCF}{\mathrm{dg}_\mathrm{CF}}
 
\newcommand{\tdgC}{\mathrm{\widetilde{dg}}_\mathrm{C}}
\newcommand{\tdgX}{\mathrm{\widetilde{dg}}_\mathrm{X}}
\newcommand{\tdgY}{\mathrm{\widetilde{dg}}_\mathrm{Y}}
\newcommand{\tdgXp}{\mathrm{\widetilde{dg}}_{\mathrm{X}(p)}}
\newcommand{\tdgT}{\mathrm{\widetilde{dg}}_\mathrm{T}}
\newcommand{\tdgF}{\mathrm{\widetilde{dg}}_\mathrm{F}} 
\newcommand{\tdgCT}{\mathrm{\widetilde{dg}}_\mathrm{CT}}
\newcommand{\tdgTC}{\mathrm{\widetilde{dg}}_\mathrm{TC}}
\newcommand{\tdgTF}{\mathrm{\widetilde{dg}}_\mathrm{TF}}
\newcommand{\tdgCF}{\mathrm{\widetilde{dg}}_\mathrm{CF}}

\newcommand{\ec}{\mathrm{ec}}
\newcommand{\ac}{\mathrm{ac}}
\newcommand{\ib}{\mathrm{ib}}
\newcommand{\bt}{\mathrm{bt}}
\newcommand{\bp}{\mathrm{bp}}
\newcommand{\bl}{\mathrm{bl}}
\newcommand{\ft}{\mathrm{ft}}
\newcommand{\bc}{\mathrm{bc}}
\newcommand{\bd}{\mathrm{bd}}
\newcommand{\bn}{\mathrm{bn}}
\newcommand{\br}{\mathrm{br}}
\newcommand{\bh}{\mathrm{bh}}
\newcommand{\Bc}{\mathrm{Bc}} 
\newcommand{\Bd}{\mathrm{Bd}} 
\newcommand{\Dg}{\mathrm{Dg}}

\newcommand{\UB}{\mathrm{UB}}
\newcommand{\LB}{\mathrm{LB}}

\newcommand{\LtoD}{\mathrm{\Lambda\cup \Gamma\cup \Dg}}
 
\newcommand{\inn}{\mathrm{in}} 

\newcommand{\VcoC}{V^\mathrm{co}_\mathrm{C}}
\newcommand{\VcoT}{V^\mathrm{co}_\mathrm{T}}
\newcommand{\VinF}{V^\mathrm{in}_\mathrm{F}}

\newcommand{\EcoC}{E^\mathrm{co}_\mathrm{C}}
\newcommand{\EcoT}{E^\mathrm{co}_\mathrm{T}}
\newcommand{\EinF}{E^\mathrm{in}_\mathrm{F}}

\newcommand{\Gfst}{G_\mathrm{fst}}
\newcommand{\Gsnd}{G_\mathrm{snd}}
\newcommand{\GS}{G_\mathrm{S}}
 
\newcommand{\mC}{m_\mathrm{C}}

\newcommand{\jC}{j^\mathrm{C}}
\newcommand{\jT}{j^\mathrm{T}} 
\newcommand{\jF}{j^\mathrm{F}} 
\newcommand{\jX}{j^\mathrm{X}} 
\newcommand{\hC}{h^\mathrm{C}}
\newcommand{\hT}{h^\mathrm{T}} 
\newcommand{\hF}{h^\mathrm{F}} 
\newcommand{\hX}{h^\mathrm{X}}

\newcommand{\PC}{P_\mathrm{C}}
\newcommand{\PT}{P_\mathrm{T}} 
\newcommand{\PF}{P_\mathrm{F}} 
 
\newcommand{\VF}{V_\mathrm{F}}
\newcommand{\VT}{V_\mathrm{T}}
\newcommand{\VS}{V_\mathrm{S}} 
\newcommand{\VX}{V_\mathrm{X}}

\newcommand{\ET}{E_\mathrm{T}}
\newcommand{\EB}{E_\mathrm{B}}
\newcommand{\ES}{E_\mathrm{S}}
\newcommand{\EF}{E_\mathrm{F}}
\newcommand{\ECT}{E_\mathrm{CT}}
\newcommand{\ETC}{E_\mathrm{TC}}
\newcommand{\ETF}{E_\mathrm{TF}}
\newcommand{\ECB}{E_\mathrm{CB}}
\newcommand{\ECF}{E_\mathrm{CF}}
\newcommand{\EFC}{E_\mathrm{FC}}
\newcommand{\EFB}{E_\mathrm{FB}}
\newcommand{\EBF}{E_\mathrm{BF}}
\newcommand{\EX}{E_\mathrm{X}}
\newcommand{\nT}{n_\mathrm{T}}
\newcommand{\nC}{n_\mathrm{C}} 
\newcommand{\nS}{n_\mathrm{S}} 
\newcommand{\nX}{n_\mathrm{X}}
\newcommand{\nB}{n_\mathrm{B}}
\newcommand{\nF}{n_\mathrm{F}}

\newcommand{\vT}{{v^\mathrm{T}}}
\newcommand{\vC}{{v^\mathrm{C}}} 
\newcommand{\vS}{{v^\mathrm{S}}}  
\newcommand{\vX}{{v^\mathrm{X}}}
\newcommand{\vY}{{v^\mathrm{Y}}}
\newcommand{\vB}{{v^\mathrm{B}}}
\newcommand{\vF}{{v^\mathrm{F}}}

\newcommand{\eF}{{e^\mathrm{F}}}
\newcommand{\eT}{{e^\mathrm{T}}}
\newcommand{\eC}{{e^\mathrm{C}}} 
\newcommand{\eS}{{e^\mathrm{S}}} 
\newcommand{\eX}{{e^\mathrm{X}}} 
\newcommand{\eY}{{e^\mathrm{Y}}}

\newcommand{\eCF}{{e^\mathrm{CF}}}
\newcommand{\eST}{{e^\mathrm{ST}}}
\newcommand{\eCT}{{e^\mathrm{CT}}}
\newcommand{\eTC}{{e^\mathrm{TC}}}
\newcommand{\eTF}{{e^\mathrm{TF}}}
\newcommand{\eSC}{{e^\mathrm{SC}}}
\newcommand{\eCS}{{e^\mathrm{CS}}}   
\newcommand{\eXY}{{e^\mathrm{XY}}}
\newcommand{\eCX}{{e^\mathrm{CX}}}
\newcommand{\tT}{{t_\mathrm{T}}}
\newcommand{\tF}{{t_\mathrm{F}}} 
\newcommand{\tX}{{t_\mathrm{X}}}
\newcommand{\tY}{{t_\mathrm{Y}}}

\newcommand{\IT}{{I_\mathrm{T}}}
\newcommand{\IC}{{I_\mathrm{C}}} 
\newcommand{\IF}{{I_\mathrm{F}}} 
\newcommand{\ICT}{{I_\mathrm{CT}}} 
\newcommand{\ITC}{{I_\mathrm{TC}}}  
\newcommand{\ICF}{{I_\mathrm{CF}}} 
\newcommand{\ITF}{{I_\mathrm{TF}}}  
\newcommand{\IS}{{I_\mathrm{S}}} 
\newcommand{\IX}{{I_\mathrm{X}}}

\newcommand{\Cld}{\mathrm{Cld}}
\newcommand{\prt}{\mathrm{prt}}
\newcommand{\Dsn}{\mathrm{Dsn}}

\newcommand{\CldT}{\mathrm{Cld}_{\mathrm{T}}}
\newcommand{\CldC}{\mathrm{Cld}_{\mathrm{C}}} 
\newcommand{\CldS}{\mathrm{Cld}_{\mathrm{S}}}  
\newcommand{\CldX}{\mathrm{Cld}_{\mathrm{X}}}
\newcommand{\CldB}{\mathrm{Cld}_{\mathrm{B}}}
\newcommand{\CldF}{\mathrm{Cld}_{\mathrm{F}}}

\newcommand{\Tprc}{\mathcal{T}_\mathrm{prc}}
\newcommand{\Pprc}{P_\mathrm{prc}}
\newcommand{\PprcF}{P_\mathrm{prc,F}}
\newcommand{\PprcT}{P_\mathrm{prc,T}}
\newcommand{\PprcC}{P_\mathrm{prc,C}}  
\newcommand{\PprcS}{P_\mathrm{prc,S}}
\newcommand{\PprcX}{P_\mathrm{prc,X}} 

\newcommand{\LeafT}{\mathrm{Leaf}_{\mathrm{T}}}
\newcommand{\LeafC}{\mathrm{Leaf}_{\mathrm{C}}} 
\newcommand{\LeafS}{\mathrm{Leaf}_{\mathrm{S}}} 
\newcommand{\LeafB}{\mathrm{Leaf}_{\mathrm{B}}} 
\newcommand{\LeafF}{\mathrm{Leaf}_{\mathrm{F}}} 
\newcommand{\LeafX}{\mathrm{Leaf}_{\mathrm{X}}}

\newcommand{\degCint}{{\deg_\mathrm{C}^\mathrm{int}}}
\newcommand{\degTint}{{\deg_\mathrm{T}^\mathrm{int}}}
\newcommand{\degFint}{{\deg_\mathrm{F}^\mathrm{int}}}
\newcommand{\degXint}{{\deg_\mathrm{X}^\mathrm{int}}}

\newcommand{\hyddeg}{\mathrm{hyddeg}}  
\newcommand{\hyddegC}{\mathrm{hyddeg}^\mathrm{C}}  
\newcommand{\hyddegT}{\mathrm{hyddeg}^\mathrm{T}}  
\newcommand{\hyddegF}{\mathrm{hyddeg}^\mathrm{F}}  
\newcommand{\hyddegX}{\mathrm{hyddeg}^\mathrm{X}}

\newcommand{\deghydC}{{\deg_\mathrm{C}^\mathrm{hyd}}}
\newcommand{\deghydT}{{\deg_\mathrm{T}^\mathrm{hyd}}}
\newcommand{\deghydF}{{\deg_\mathrm{F}^\mathrm{hyd}}}
\newcommand{\deghydX}{{\deg_\mathrm{X}^\mathrm{hyd}}}

\newcommand{\degCex}{{\deg_\mathrm{C}^\mathrm{ex}}}
\newcommand{\degTex}{{\deg_\mathrm{T}^\mathrm{ex}}}
\newcommand{\degFex}{{\deg_\mathrm{F}^\mathrm{ex}}}
\newcommand{\degXex}{{\deg_\mathrm{X}^\mathrm{ex}}}
\newcommand{\degF}{{\deg^\mathrm{F}}}
\newcommand{\degT}{{\deg^\mathrm{T}}}
\newcommand{\degC}{{\deg^\mathrm{C}}} 
\newcommand{\degS}{{\deg^\mathrm{S}}} 
\newcommand{\degX}{{\deg^\mathrm{X}}}

\newcommand{\degCT}{\deg_\mathrm{CT}}
\newcommand{\degTC}{\deg_\mathrm{TC}}

\newcommand{\degCTT}{\deg^\mathrm{CT}_\mathrm{T}}
\newcommand{\degTCT}{\deg^\mathrm{TC}_\mathrm{T}}
\newcommand{\degCFF}{\deg^\mathrm{CF}_\mathrm{F}}
\newcommand{\degTFF}{\deg^\mathrm{TF}_\mathrm{F}}

\newcommand{\stc}{\mathrm{sc}}

\newcommand{\scexC}{\mathrm{sc}^\mathrm{ex}_\mathrm{C}}
\newcommand{\scexT}{\mathrm{sc}^\mathrm{ex}_\mathrm{T}}
\newcommand{\scexF}{\mathrm{sc}^\mathrm{ex}_\mathrm{F}}
\newcommand{\scexCT}{\mathrm{sc}^\mathrm{ex}_\mathrm{CT}}
\newcommand{\scexTC}{\mathrm{sc}^\mathrm{ex}_\mathrm{TC}}
\newcommand{\scexCF}{\mathrm{sc}^\mathrm{ex}_\mathrm{CF}}
\newcommand{\scexTF}{\mathrm{sc}^\mathrm{ex}_\mathrm{TF}}
\newcommand{\scexX}{\mathrm{sc}^\mathrm{ex}_\mathrm{X}}
\newcommand{\scexY}{\mathrm{sc}^\mathrm{ex}_\mathrm{Y}}

\newcommand{\dlsc}{\delta_\mathrm{sc}  }
\newcommand{\dlscC}{\delta_\mathrm{sc}^\mathrm{C}}
\newcommand{\dlscT}{\delta_\mathrm{sc}^\mathrm{T}}
\newcommand{\dlscF}{\delta_\mathrm{sc}^\mathrm{F}}
\newcommand{\dlscX}{\delta_\mathrm{sc}^\mathrm{X}}

\newcommand{\scC}{\mathrm{sc}^\mathrm{C}}
\newcommand{\scT}{\mathrm{sc}^\mathrm{T}}
\newcommand{\scF}{\mathrm{sc}^\mathrm{F}}
\newcommand{\scX}{\mathrm{sc}^\mathrm{X}}
\newcommand{\scCTT}{\mathrm{sc}^\mathrm{CT}_\mathrm{T}}
\newcommand{\scCTC}{\mathrm{sc}^\mathrm{CT}_\mathrm{C}}
\newcommand{\scTCT}{\mathrm{sc}^\mathrm{TC}_\mathrm{T}} 
\newcommand{\scTCC}{\mathrm{sc}^\mathrm{TC}_\mathrm{C}} 
\newcommand{\scCFF}{\mathrm{sc}^\mathrm{CF}_\mathrm{F}}  
\newcommand{\scCFC}{\mathrm{sc}^\mathrm{CF}_\mathrm{C}}  
\newcommand{\scTFF}{\mathrm{sc}^\mathrm{TF}_\mathrm{F}}  
\newcommand{\scTFT}{\mathrm{sc}^\mathrm{TF}_\mathrm{T}}  

\newcommand{\tlscC}{{\widetilde{\mathrm{sc}}_\mathrm{C}} }
\newcommand{\tlscT}{{\widetilde{\mathrm{sc}}_\mathrm{T}} }
\newcommand{\tlscF}{{\widetilde{\mathrm{sc}}_\mathrm{F}} }
\newcommand{\tlscCT}{{\widetilde{\mathrm{sc}}_\mathrm{CT}} }
\newcommand{\tlscTC}{{\widetilde{\mathrm{sc}}_\mathrm{TC}} }
\newcommand{\tlscCF}{{\widetilde{\mathrm{sc}}_\mathrm{CF}} }
\newcommand{\tlscTF}{{\widetilde{\mathrm{sc}}_\mathrm{TF}} }
\newcommand{\tlscX}{{\widetilde{\mathrm{sc}}_\mathrm{X}} }
\newcommand{\tlscY}{{\widetilde{\mathrm{sc}}_\mathrm{Y}} }

\newcommand{\tldgC}{{\widetilde{\deg}_\mathrm{C}} }
\newcommand{\tldgT}{{\widetilde{\deg}_\mathrm{T}} }
\newcommand{\tldgF}{{\widetilde{\deg}_\mathrm{F}} }
\newcommand{\tldgCT}{{\widetilde{\deg}_\mathrm{CT}} }
\newcommand{\tldgTC}{{\widetilde{\deg}_\mathrm{TC}} }
\newcommand{\tldgCF}{{\widetilde{\deg}_\mathrm{CF}} }
\newcommand{\tldgTF}{{\widetilde{\deg}_\mathrm{TF}} }
\newcommand{\tldgX}{{\widetilde{\deg}_\mathrm{X}} }
\newcommand{\tldgY}{{\widetilde{\deg}_\mathrm{Y}} }
\newcommand{\tldgXp}{{\widetilde{\deg}_{\mathrm{X}(p)}} }

\newcommand{\lfF}{{\mathrm{\ell}^\mathrm{F}}}
\newcommand{\lfT}{{\mathrm{\ell}^\mathrm{T}}}
\newcommand{\lfC}{{\mathrm{\ell}^\mathrm{C}}} 
\newcommand{\lfS}{{\mathrm{\ell}^\mathrm{S}}} 
\newcommand{\lfX}{{\mathrm{\ell}^\mathrm{X}}}

\newcommand{\cF}{{c_\mathrm{F}}}
\newcommand{\cC}{{c_\mathrm{C}}}
\newcommand{\kC}{{k_\mathrm{C}}}
\newcommand{\cT}{{c_\mathrm{T}}}
\newcommand{\cS}{{c_\mathrm{S}}}
\newcommand{\cX}{{c_\mathrm{X}}}

\newcommand{\chiF}{{\chi^\mathrm{F}}} 
\newcommand{\dclrF}{\delta_{\chi}^\mathrm{F}}
\newcommand{\clrF}{\mathrm{clr}^{\mathrm{F}}} 
 
\newcommand{\chiC}{{\chi^\mathrm{C}}} 
\newcommand{\dclrC}{\delta_{\chi}^\mathrm{C}}
\newcommand{\clrC}{\mathrm{clr}^{\mathrm{C}}} 

\newcommand{\chiT}{{\chi^\mathrm{T}}}
\newcommand{\dclrT}{\delta_{\chi}^\mathrm{T}}
\newcommand{\clrT}{\mathrm{clr}^{\mathrm{T}}} 
\newcommand{\chiX}{{\chi^\mathrm{X}}}
\newcommand{\dclrX}{\delta_{\chi}^\mathrm{X}}
\newcommand{\clrX}{\mathrm{clr}^{\mathrm{X}}}

\newcommand{\ztT}{{\zeta^\mathrm{T}}}
\newcommand{\dztT}{\delta_\mathrm{\zeta}^\mathrm{T}} 
\newcommand{\ztF}{{\zeta^\mathrm{F}}}
\newcommand{\dztF}{\delta_\mathrm{\zeta}^\mathrm{F}} 
\newcommand{\ztX}{{\zeta^\mathrm{X}}}
\newcommand{\dztX}{\delta_\mathrm{\zeta}^\mathrm{X}}

\newcommand{\dclrTF}{\delta_{\chi}^\mathrm{TF}} 
\newcommand{\tail}{\mathrm{tail}} 
\newcommand{\hd}{\mathrm{head}} 

\newcommand{\tailF}{\mathrm{tail}^{\mathrm{F}}} 
\newcommand{\hdC}{\mathrm{head}^{\mathrm{C}}} 
\newcommand{\tailC}{\mathrm{tail}^{\mathrm{C}}} 
\newcommand{\hdT}{\mathrm{head}^{\mathrm{T}}} 
\newcommand{\hdX}{\mathrm{head}^{\mathrm{X}}}
\newcommand{\tailT}{\mathrm{tail}^{\mathrm{T}}} 
\newcommand{\tailX}{\mathrm{tail}^{\mathrm{X}}}

\newcommand{\prtF}{\mathrm{prt}_{\mathrm{F}}}
\newcommand{\prtT}{\mathrm{prt}_{\mathrm{T}}}
\newcommand{\prtC}{\mathrm{prt}_{\mathrm{C}}}
\newcommand{\prtS}{\mathrm{prt}_{\mathrm{S}}}  
\newcommand{\prtX}{\mathrm{prt}_{\mathrm{X}}}

\newcommand{\dlfrF}{\delta_\mathrm{fr}^\mathrm{F}}
\newcommand{\dlfrT}{\delta_\mathrm{fr}^\mathrm{T}}
\newcommand{\dlfrC}{\delta_\mathrm{fr}^\mathrm{C}} 
\newcommand{\dlfrS}{\delta_\mathrm{fr}^\mathrm{S}} 
\newcommand{\dlfrX}{\delta_\mathrm{fr}^\mathrm{X}}
 
\newcommand{\ddgF}{\delta_\mathrm{dg}^\mathrm{F}}
\newcommand{\ddgT}{\delta_\mathrm{dg}^\mathrm{T}}
\newcommand{\ddgC}{\delta_\mathrm{dg}^\mathrm{C}} 
\newcommand{\ddgS}{\delta_\mathrm{dg}^\mathrm{S}} 
\newcommand{\ddgX}{\delta_\mathrm{dg}^\mathrm{X}}
\newcommand{\ddgCT}{\delta_\mathrm{dg}^\mathrm{CT}}
\newcommand{\ddgTC}{\delta_\mathrm{dg}^\mathrm{TC}} 
\newcommand{\ddgTF}{\delta_\mathrm{dg}^\mathrm{TF}}
\newcommand{\ddgCF}{\delta_\mathrm{dg}^\mathrm{CF}} 

\newcommand{\ddgFint}{\delta_\mathrm{dg,F}^\mathrm{int}}
\newcommand{\ddgTint}{\delta_\mathrm{dg,T}^\mathrm{int}}
\newcommand{\ddgCint}{\delta_\mathrm{dg,C}^\mathrm{int}}  
\newcommand{\ddgXint}{\delta_\mathrm{dg,X}^\mathrm{int}} 

\newcommand{\dtdgC}{\delta_\mathrm{\widetilde{dg},C}} 
\newcommand{\dtdgT}{\delta_\mathrm{\widetilde{dg},T}} 
\newcommand{\dtdgF}{\delta_\mathrm{\widetilde{dg},F}} 
\newcommand{\dtdgCT}{\delta_\mathrm{\widetilde{dg},CT}} 
\newcommand{\dtdgTC}{\delta_\mathrm{\widetilde{dg},TC}} 
\newcommand{\dtdgCF}{\delta_\mathrm{\widetilde{dg},CF}} 
\newcommand{\dtdgTF}{\delta_\mathrm{\widetilde{dg},TF}} 
\newcommand{\dtdgX}{\delta_\mathrm{\widetilde{dg},X}} 
\newcommand{\dtdgY}{\delta_\mathrm{\widetilde{dg},Y}} 
\newcommand{\dtdgXp}{\delta_{\mathrm{\widetilde{dg},X}(p)}}

\newcommand{\dgco}{\dg^\co}
\newcommand{\dgnc}{\dg^\nc}

\newcommand{\bF}{\beta^\mathrm{F}}
\newcommand{\bT}{\beta^\mathrm{T}}
\newcommand{\bC}{\beta^\mathrm{C}} 
\newcommand{\bX}{\beta^\mathrm{X}}
\newcommand{\bS}{\beta^\mathrm{S}}  
\newcommand{\bCT}{\beta^\mathrm{CT}}
\newcommand{\bTC}{\beta^\mathrm{TC}} 
\newcommand{\bTF}{\beta^\mathrm{TF}} 
\newcommand{\bCF}{\beta^\mathrm{CF}} 
\newcommand{\bXF}{\beta^\mathrm{XF}} 
\newcommand{\bsF}{\beta^{*\mathrm{F}}} 
\newcommand{\bCTF}{\beta^\mathrm{CTF}}

\newcommand{\bFex}{\beta^\mathrm{F}_\mathrm{ex}} 
\newcommand{\bTex}{\beta^\mathrm{T}_\mathrm{ex}} 
\newcommand{\bCex}{\beta^\mathrm{C}_\mathrm{ex}} 
\newcommand{\bXex}{\beta^\mathrm{X}_\mathrm{ex}}

\newcommand{\delb}{\delta_{\beta}}
\newcommand{\delbF}{\delta_{\beta}^\mathrm{F}}
\newcommand{\delbT}{\delta_{\beta}^\mathrm{T}}
\newcommand{\delbC}{\delta_{\beta}^\mathrm{C}} 
\newcommand{\delbS}{\delta_{\beta}^\mathrm{S}}  
\newcommand{\delbSC}{\delta_{\beta}^\mathrm{SC}}
\newcommand{\delbCS}{\delta_{\beta}^\mathrm{CS}}
\newcommand{\delbCT}{\delta_{\beta}^\mathrm{CT}}
\newcommand{\delbTC}{\delta_{\beta}^\mathrm{TC}}
\newcommand{\delbST}{\delta_{\beta}^\mathrm{ST}} 
\newcommand{\delbTF}{\delta_{\beta}^\mathrm{TF}} 
\newcommand{\delbCF}{\delta_{\beta}^\mathrm{CF}} 
\newcommand{\delbXF}{\delta_{\beta}^\mathrm{XF}} 
\newcommand{\delbsF}{\delta_{\beta}^{*\mathrm{F}}} 
\newcommand{\delbX}{\delta_{\beta}^\mathrm{X}}
\newcommand{\delbY}{\delta_{\beta}^\mathrm{Y}}
\newcommand{\delbXp}{\delta_{\beta}^{\mathrm{X}(p)}}

\newcommand{\aF}{{\alpha}^\mathrm{F}}
\newcommand{\aT}{{\alpha}^\mathrm{T}}
\newcommand{\aC}{{\alpha}^\mathrm{C}}  
\newcommand{\aS}{{\alpha}^\mathrm{S}} 
\newcommand{\aX}{{\alpha}^\mathrm{X}}
\newcommand{\aCT}{{\alpha}^\mathrm{CT}}
\newcommand{\aTC}{{\alpha}^\mathrm{TC}}
\newcommand{\aCTT}{{\alpha}^\mathrm{CT}_\mathrm{T}}
\newcommand{\aCTC}{{\alpha}^\mathrm{CT}_\mathrm{C}}
\newcommand{\aTCT}{{\alpha}^\mathrm{TC}_\mathrm{T}}
\newcommand{\aTCC}{{\alpha}^\mathrm{TC}_\mathrm{C}}
\newcommand{\aCF}{{\alpha}^\mathrm{CF}}  
\newcommand{\aTF}{{\alpha}^\mathrm{TF}}  
\newcommand{\aCFC}{{\alpha}^\mathrm{CF}_\mathrm{C}}  
\newcommand{\aCFF}{{\alpha}^\mathrm{CF}_\mathrm{F}}  
\newcommand{\aTFT}{{\alpha}^\mathrm{TF}_\mathrm{T}}    
\newcommand{\aTFF}{{\alpha}^\mathrm{TF}_\mathrm{F}}

\newcommand{\delaC}{\delta_\mathrm{\alpha}^{\mathrm{C}}}
\newcommand{\delaT}{\delta_\mathrm{\alpha}^{\mathrm{T}}}
\newcommand{\delaF}{\delta_\mathrm{\alpha}^{\mathrm{F}}}
\newcommand{\delaX}{\delta_\mathrm{\alpha}^{\mathrm{X}}}
\newcommand{\delaY}{\delta_\mathrm{\alpha}^{\mathrm{Y}}}
\newcommand{\delaXp}{\delta_{\mathrm{\alpha}^{\mathrm{X}(p)}}}
\newcommand{\delaCT}{\delta_\mathrm{\alpha}^{\mathrm{CT}}}
\newcommand{\delaTC}{\delta_\mathrm{\alpha}^{\mathrm{TC}}}
\newcommand{\delaTF}{\delta_\mathrm{\alpha}^{\mathrm{TF}}}
\newcommand{\delaCF}{\delta_\mathrm{\alpha}^{\mathrm{CF}}}

\newcommand{\dlaC}{\delta_\mathrm{\alpha}^{\mathrm{C}}}
\newcommand{\dlaT}{\delta_\mathrm{\alpha}^{\mathrm{T}}}
\newcommand{\dlaF}{\delta_\mathrm{\alpha}^{\mathrm{F}}}
\newcommand{\dlaX}{\delta_\mathrm{\alpha}^{\mathrm{X}}}
\newcommand{\dlaY}{\delta_\mathrm{\alpha}^{\mathrm{Y}}}
\newcommand{\dlaXp}{\delta_{\mathrm{\alpha}^{\mathrm{X}(p)}}}

\newcommand{\dtaC}{\delta_\mathrm{\alpha,\mathrm{C}}}
\newcommand{\dtaT}{\delta_\mathrm{\alpha,\mathrm{T}}}
\newcommand{\dtaF}{\delta_\mathrm{\alpha,\mathrm{F}}}
\newcommand{\dtaX}{\delta_\mathrm{\alpha,\mathrm{X}}}
\newcommand{\dtaY}{\delta_\mathrm{\alpha,\mathrm{Y}}}
\newcommand{\dtaXp}{\delta_{\mathrm{\alpha,\mathrm{X}(p)}}}

\newcommand{\dltuF}{\delta_{\tau}^\mathrm{F}}
\newcommand{\dltuT}{\delta_{\tau}^\mathrm{T}}
\newcommand{\dltuC}{\delta_{\tau}^\mathrm{C}} 
\newcommand{\dltuS}{\delta_{\tau}^\mathrm{S}}  
\newcommand{\dltuCS}{\delta_{\tau}^\mathrm{CS}}
\newcommand{\dltuSC}{\delta_{\tau}^\mathrm{SC}}
\newcommand{\dltuCT}{\delta_{\tau}^\mathrm{CT}}
\newcommand{\dltuTC}{\delta_{\tau}^\mathrm{TC}}
\newcommand{\dltuST}{\delta_{\tau}^\mathrm{ST}}
\newcommand{\dltuCF}{\delta_{\tau}^\mathrm{CF}} 
\newcommand{\dltuTF}{\delta_{\tau}^\mathrm{TF}} 
\newcommand{\dltuX}{\delta_{\tau}^\mathrm{X}}
\newcommand{\dltuXp}{\delta_{\tau}^{\mathrm{X}(p)}}

\newcommand{\dlnsF}{\delta_{\mathrm{ns}}^\mathrm{F}}
\newcommand{\dlnsT}{\delta_{\mathrm{ns}}^\mathrm{T}}
\newcommand{\dlnsC}{\delta_{\mathrm{ns}}^\mathrm{C}} 
\newcommand{\dlnsX}{\delta_{\mathrm{ns}}^\mathrm{X}}

\newcommand{\dlacTlnk}{\delta_{\mathrm{ac}}^\mathrm{T,lnk}} 

\newcommand{\dlacF}{\delta_{\mathrm{ac}}^\mathrm{F}}
\newcommand{\dlacT}{\delta_{\mathrm{ac}}^\mathrm{T}}
\newcommand{\dlacC}{\delta_{\mathrm{ac}}^\mathrm{C}} 
\newcommand{\dlacS}{\delta_{\mathrm{ac}}^\mathrm{S}}  
\newcommand{\dlacCS}{\delta_{\mathrm{ac}}^\mathrm{CS}}
\newcommand{\dlacSC}{\delta_{\mathrm{ac}}^\mathrm{SC}}
\newcommand{\dlacCT}{\delta_{\mathrm{ac}}^\mathrm{CT}}
\newcommand{\dlacTC}{\delta_{\mathrm{ac}}^\mathrm{TC}}
\newcommand{\dlacST}{\delta_{\mathrm{ac}}^\mathrm{ST}}
\newcommand{\dlacCF}{\delta_{\mathrm{ac}}^\mathrm{CF}} 
\newcommand{\dlacTF}{\delta_{\mathrm{ac}}^\mathrm{TF}} 
\newcommand{\dlacX}{\delta_{\mathrm{ac}}^\mathrm{X}}
\newcommand{\dlacXp}{\delta_{\mathrm{ac}}^{\mathrm{X}(p)}}

\newcommand{\dlecTlnk}{\delta_{\mathrm{ec}}^\mathrm{T,lnk}}
\newcommand{\dlecF}{\delta_{\mathrm{ec}}^\mathrm{F}}
\newcommand{\dlecT}{\delta_{\mathrm{ec}}^\mathrm{T}}
\newcommand{\dlecC}{\delta_{\mathrm{ec}}^\mathrm{C}} 
\newcommand{\dlecS}{\delta_{\mathrm{ec}}^\mathrm{S}}  
\newcommand{\dlecCS}{\delta_{\mathrm{ec}}^\mathrm{CS}}
\newcommand{\dlecSC}{\delta_{\mathrm{ec}}^\mathrm{SC}}
\newcommand{\dlecCT}{\delta_{\mathrm{ec}}^\mathrm{CT}}
\newcommand{\dlecCTT}{\delta_{\mathrm{ec,T}}^\mathrm{CT}}
\newcommand{\dlecCTC}{\delta_{\mathrm{ec,C}}^\mathrm{CT}}
\newcommand{\dlecTC}{\delta_{\mathrm{ec}}^\mathrm{TC}}
\newcommand{\dlecTCC}{\delta_{\mathrm{ec,C}}^\mathrm{TC}}
\newcommand{\dlecTCT}{\delta_{\mathrm{ec,T}}^\mathrm{TC}}
\newcommand{\dlecST}{\delta_{\mathrm{ec}}^\mathrm{ST}}
\newcommand{\dlecCF}{\delta_{\mathrm{ec}}^\mathrm{CF}} 
\newcommand{\dlecCFC}{\delta_{\mathrm{ec,C}}^\mathrm{CF}} 
\newcommand{\dlecCFF}{\delta_{\mathrm{ec,F}}^\mathrm{CF}} 
\newcommand{\dlecTF}{\delta_{\mathrm{ec}}^\mathrm{TF}} 
\newcommand{\dlecTFT}{\delta_{\mathrm{ec,T}}^\mathrm{TF}} 
\newcommand{\dlecTFF}{\delta_{\mathrm{ec,F}}^\mathrm{TF}} 
\newcommand{\dlecX}{\delta_{\mathrm{ec}}^\mathrm{X}}
\newcommand{\dlecXp}{\delta_{\mathrm{ec}}^{\mathrm{X}(p)}}

\newcommand{\DlacFp}{\Delta_{\mathrm{ac}}^\mathrm{F+}}
\newcommand{\DlacTp}{\Delta_{\mathrm{ac}}^\mathrm{T+}}
\newcommand{\DlacCp}{\Delta_{\mathrm{ac}}^\mathrm{C+}} 
\newcommand{\DlacSp}{\Delta_{\mathrm{ac}}^\mathrm{S+}}  
\newcommand{\DlacCSp}{\Delta_{\mathrm{ac}}^\mathrm{CS+}}
\newcommand{\DlacSCp}{\Delta_{\mathrm{ac}}^\mathrm{SC+}}
\newcommand{\DlacCTp}{\Delta_{\mathrm{ac}}^\mathrm{CT+}}
\newcommand{\DlacTCp}{\Delta_{\mathrm{ac}}^\mathrm{TC+}}
\newcommand{\DlacSTp}{\Delta_{\mathrm{ac}}^\mathrm{ST+}}
\newcommand{\DlacCFp}{\Delta_{\mathrm{ac}}^\mathrm{CF+}} 
\newcommand{\DlacTFp}{\Delta_{\mathrm{ac}}^\mathrm{TF+}} 
\newcommand{\DlacXp}{\Delta_{\mathrm{ac}}^\mathrm{X+}}
\newcommand{\DlacXpp}{\Delta_{\mathrm{ac}}^{\mathrm{X}(p)+}}

\newcommand{\DlacFm}{\Delta_{\mathrm{ac}}^\mathrm{F-}}
\newcommand{\DlacTm}{\Delta_{\mathrm{ac}}^\mathrm{T-}}
\newcommand{\DlacCm}{\Delta_{\mathrm{ac}}^\mathrm{C-}} 
\newcommand{\DlacSm}{\Delta_{\mathrm{ac}}^\mathrm{S-}}  
\newcommand{\DlacCSm}{\Delta_{\mathrm{ac}}^\mathrm{CS-}}
\newcommand{\DlacSCm}{\Delta_{\mathrm{ac}}^\mathrm{SC-}}
\newcommand{\DlacCTm}{\Delta_{\mathrm{ac}}^\mathrm{CT-}}
\newcommand{\DlacTCm}{\Delta_{\mathrm{ac}}^\mathrm{TC-}}
\newcommand{\DlacSTm}{\Delta_{\mathrm{ac}}^\mathrm{ST-}}
\newcommand{\DlacCFm}{\Delta_{\mathrm{ac}}^\mathrm{CF-}} 
\newcommand{\DlacTFm}{\Delta_{\mathrm{ac}}^\mathrm{TF-}} 
\newcommand{\DlacXm}{\Delta_{\mathrm{ac}}^\mathrm{X-}}
\newcommand{\DlacXpm}{\Delta_{\mathrm{ac}}^{\mathrm{X}(p)-}}

\newcommand{\dlacqF}{\delta_{\mathrm{ac},q}^\mathrm{F}}
\newcommand{\dlacqT}{\delta_{\mathrm{ac},q}^\mathrm{T}}
\newcommand{\dlacqC}{\delta_{\mathrm{ac},q}^\mathrm{C}} 
\newcommand{\dlacqS}{\delta_{\mathrm{ac},q}^\mathrm{S}}  
\newcommand{\dlacqCS}{\delta_{\mathrm{ac},q}^\mathrm{CS}}
\newcommand{\dlacqSC}{\delta_{\mathrm{ac},q}^\mathrm{SC}}
\newcommand{\dlacqCT}{\delta_{\mathrm{ac},q}^\mathrm{CT}}
\newcommand{\dlacqTC}{\delta_{\mathrm{ac},q}^\mathrm{TC}}
\newcommand{\dlacqST}{\delta_{\mathrm{ac},q}^\mathrm{ST}}
\newcommand{\dlacqCF}{\delta_{\mathrm{ac},q}^\mathrm{CF}} 
\newcommand{\dlacqTF}{\delta_{\mathrm{ac},q}^\mathrm{TF}} 
\newcommand{\dlacqX}{\delta_{\mathrm{ac},q}^\mathrm{X}}
\newcommand{\dlacqXp}{\delta_{\mathrm{ac},q}^{\mathrm{X}(p)}}

\newcommand{\DlecFp}{\Delta_{\mathrm{ec}}^\mathrm{F+}}
\newcommand{\DlecTp}{\Delta_{\mathrm{ec}}^\mathrm{T+}}
\newcommand{\DlecCp}{\Delta_{\mathrm{ec}}^\mathrm{C+}} 
\newcommand{\DlecSp}{\Delta_{\mathrm{ec}}^\mathrm{S+}}  
\newcommand{\DlecCSp}{\Delta_{\mathrm{ec}}^\mathrm{CS+}}
\newcommand{\DlecSCp}{\Delta_{\mathrm{ec}}^\mathrm{SC+}}
\newcommand{\DlecCTp}{\Delta_{\mathrm{ec}}^\mathrm{CT+}}
\newcommand{\DlecTCp}{\Delta_{\mathrm{ec}}^\mathrm{TC+}}
\newcommand{\DlecSTp}{\Delta_{\mathrm{ec}}^\mathrm{ST+}}
\newcommand{\DlecCFp}{\Delta_{\mathrm{ec}}^\mathrm{CF+}} 
\newcommand{\DlecTFp}{\Delta_{\mathrm{ec}}^\mathrm{TF+}} 
\newcommand{\DlecXp}{\Delta_{\mathrm{ec}}^\mathrm{X+}}
\newcommand{\DlecXpp}{\Delta_{\mathrm{ec}}^{\mathrm{X}(p)+}}

\newcommand{\DlecFm}{\Delta_{\mathrm{ec}}^\mathrm{F-}}
\newcommand{\DlecTm}{\Delta_{\mathrm{ec}}^\mathrm{T-}}
\newcommand{\DlecCm}{\Delta_{\mathrm{ec}}^\mathrm{C-}} 
\newcommand{\DlecSm}{\Delta_{\mathrm{ec}}^\mathrm{S-}}  
\newcommand{\DlecCSm}{\Delta_{\mathrm{ec}}^\mathrm{CS-}}
\newcommand{\DlecSCm}{\Delta_{\mathrm{ec}}^\mathrm{SC-}}
\newcommand{\DlecCTm}{\Delta_{\mathrm{ec}}^\mathrm{CT-}}
\newcommand{\DlecTCm}{\Delta_{\mathrm{ec}}^\mathrm{TC-}}
\newcommand{\DlecSTm}{\Delta_{\mathrm{ec}}^\mathrm{ST-}}
\newcommand{\DlecCFm}{\Delta_{\mathrm{ec}}^\mathrm{CF-}} 
\newcommand{\DlecTFm}{\Delta_{\mathrm{ec}}^\mathrm{TF-}} 
\newcommand{\DlecXm}{\Delta_{\mathrm{ec}}^\mathrm{X-}}
\newcommand{\DlecXpm}{\Delta_{\mathrm{ec}}^{\mathrm{X}(p)-}}

\newcommand{\DsnF}{\mathrm{Dsn}_\mathrm{F}}  
\newcommand{\DsnT}{\mathrm{Dsn}_\mathrm{T}}
\newcommand{\DsnC}{\mathrm{Dsn}_\mathrm{C}}  
\newcommand{\DsnS}{\mathrm{Dsn}_\mathrm{S}} 
\newcommand{\DsnX}{\mathrm{Dsn}_\mathrm{X}} 
\newcommand{\DsnY}{\mathrm{Dsn}_\mathrm{Y}} 
\newcommand{\DsnXp}{\mathrm{Dsn}_{\mathrm{X}(p)}} 

\newcommand{\ColB}{\mathrm{Col}_\mathrm{B}}
\newcommand{\AncB}{\mathrm{Anc}_\mathrm{B}}

\section{Preliminary}\label{sec:preliminary}
We give some notions and terminologies on graphs in Section~\ref{sec:graphs}
and review the framework \molinfer\ in Section~\ref{sec:frame_all},
a modeling of chemical compounds in Section~\ref{sec:chemical_model},
and the two-layered model, the standard model in \molinfer, in Section~\ref{sec:2LM}.
Some necessary modifications introduced by Ido~et~al.~\cite{Ido:2024aa} when the molecule is a polymer 
will be covered in Section~\ref{sec:polymer}.


Let $\bbR$, $\bbR_+$, $\bbZ$  and $\bbZ_+$ 
represent the sets of reals,  non-negative reals, 
integers, and non-negative integers, respectively.
For two integers $a$ and $b$ such that $a \leq b$, $[a,b]$ is defined as the set of 
integers $i$ such that $a\leq i\leq b$.

\subsection{Graphs}\label{sec:graphs}

When referring to a {\em graph} $G$, it is assumed that $G$ is a connected and simple graph.
The sets of vertices and edges of a given graph $G$ are denoted
by $V(G)$ and $E(G)$, respectively.
For any vertex $v\in V(G)$, we denote the set of its neighbors in $G$ by $N_G(v)$,
and the {\em degree} $\mathrm{deg}_G(v)$ of $v$ is
$\mathrm{deg}_G(v)=|N_G(v)|$.

A vertex designated in a graph $G$ is called a {\em root},
and a graph with such a vertex is referred to as a {\em rooted graph}. 
 For a graph $G$ (possibly rooted),
 a {\em leaf-vertex} is a non-root vertex  $v$ with degree 1.
 For any subset $V'\subseteq V(G)$, 
the graph $G-V'$ is obtained by removing all vertices in $V'$ along with any edges incident to them. 
An edge $uv$ incident to a leaf-vertex $v$ is called a {\em leaf-edge}.
 We denote the sets of leaf-vertices and leaf-edges in $G$ by $\Vleaf(G)$ and $\Eleaf(G)$, respectively.
 For a graph $G$ (possibly rooted),
 a sequence of graphs $G_i, i\in \mathbb{Z}_+$ is defined by iteratively removing all leaf-vertices
 $i$ times as follows:
\[ G_0:=G; ~~ G_{i+1}:=G_i - \Vleaf(G_i). \]
A vertex $v$ is called a {\em tree vertex} if $v\in \Vleaf(G_i)$
for some $i\geq 0$. 
We define the {\em height} $\h(v)$ of a tree vertex $v\in \Vleaf(G_i)$ to be $i$; 
and for a non-tree vertex $v$ adjacent to a tree vertex, we define the height
$\h(v)$ to be $\h(u)+1$, 
where $u$ is the tree vertex with the maximum height $\h(u)$ among those adjacent to $v$.
The heights of other vertices are left undefined. 
Finally, the {\em height} $\h(T)$ of a rooted tree $T$ is defined
to be the maximum of $\h(v)$ among all vertices $v\in V(T)$.

\subsection{\molinfer : An Inverse QSAR/QSPR Framework Based on Machine Learning and MILP}\label{sec:frame_all}


The computation process of an artificial neural network (ANN)
with ReLU activation functions can be 
represented through a mixed integer linear programming (MILP) formulation,
as demonstrated by Akutsu and Nagamochi~\cite{Akutsu:2019aa}.
%
Based on this concept, 
a two-phase inverse QSAR/QSPR framework,
 called \molinfer,
has been proposed and subsequently refined~\cite{Azam:2020aa,Zhang:2022aa, Shi:2021aa, Zhu:2022ad, Ido:2024aa, Zhu:2023aa},
as depicted in Figure~\ref{fig:framework}.
This framework is mainly based on using the \emph{mixed integer linear programming} (MILP) formulation
to simulate the computational process of machine learning methods and describe 
the necessary and sufficient conditions to ensure such a chemical graph exists,
utilizing only 2D structural information.
The advantage of \molinfer\ compared to other methods is that it
guarantees both optimality and exactness.
Here, optimality refers to the quality of the solution in addressing 
the inverse problem of learning methods,
while exactness ensures that the solution corresponds to a valid chemical graph.
This framework was first introduced for general molecules~\cite{Shi:2021aa, Zhu:2022ad} and then extended to
polymers recently~\cite{Ido:2024aa}.
This subsection provides an overview of the core ideas for \molinfer\ for completeness.

\subsubsection{Phase~1}
Phase~1 is the QSAR/QSPR phase, aiming to construct a prediction function between chemical compounds and their observed property values, 
and consists of three stages. 
Here we denote by $\calG$ the set of all possible chemical graphs.
\begin{itemize}
\item[-] Stage~1: Given a chemical property $\pi$, we collect a data set $D_\pi \subseteq \calG$ of chemical graphs such that
for every chemical graph $\bbC \in D_\pi$, the observed value $a(\bbC)$ of property $\pi$ 
is available.
\item[-] Stage~2: A feature function $f: \calG \to \bbR^K$ ($K$ is a positive integer) is defined. 
This feature function consists of only graph-theoretic descriptors, 
mainly based on the local graph-theoretic structures of the chemical graph
 so that $f$ is tractable by MILP formulations in Phase~2. 
\item[-] Stage~3: A prediction function $\eta$ is constructed by some machine learning methods
in order to produce an output $y=\eta(x)\in\bbR$ based
on the feature vector $x=f(\bbC)\in\bbR^K$ for each $\bbC \in \mathcal{G}$.
\end{itemize}

\subsubsection{Phase~2}
Phase~2 is devoted to the inverse QSAR/QSPR phase, 
designed to infer chemical graphs with a specified property value
based on the prediction function $\eta$ constructed in Phase~1. 
It consists of two stages.

\begin{itemize}
\item[-] Stage~4: Given a set of rules called topological specification $\sigma$ (see Section~\ref{sec:2LM} for more details) that specifies
the desired structure of the inferred chemical graphs, and a desired range $[\ylb, \yub]$ of the target value, 
Stage 4 is to infer a chemical graph $\bbC^\dagger$ that satisfies the rules $\sigma$ and
$\eta(f(\bbC^\dagger)) \in [\ylb, \yub]$.
To achieve this,
an MILP formulation $\mathcal{M}(g,x,y;\mathcal{C}_1,\mathcal{C}_2)$ is formulated, 
which consists of two parts:
\begin{itemize}
\item[(i)] $\mathcal{M}(x,y;\mathcal{C}_1)$: the computation process of $y := \eta(x)$ from a vector $x \in \bbR^K$; and
\item[(ii)] $\mathcal{M}(g,x;\mathcal{C}_2)$: that of $x := f(\bbC)$ and the constraints for $\bbC \in \calG_\sigma$,
\end{itemize}
where $\calG_\sigma$ denotes the set of all chemical graphs satisfying $\sigma$.
We solve the MILP $\mathcal{M}(g,x,y;\mathcal{C}_1,\mathcal{C}_2)$ 
for a given $\sigma$ and $[\ylb, \yub]$
to find
a feature vector $x^* \in \bbR^K$ and a chemical graph $\bbC^\dagger \in \mathcal{G}_\sigma$
such that $f(\bbC^\dagger) = x^*$ and $\eta(x^*) \in [\ylb, \yub]$.
If the MILP is infeasible, no chemical graph in $\calG_\sigma$ satisfies the specified demand.
\item[-] Stage~5:
The final stage is to generate the isomers of the inferred chemical graphs $\bbC^\dagger$
by using a dynamic programming-based graph enumeration algorithm developed by Ido~et~al.~\cite{Ido:2025aa}. 
A {\em chemical isomer} of $\bbC^\dagger$ under
a topological specification $\sigma$ is defined as 
a chemical graph $\bbC^*$  such that
$f(\bbC^*)=f(\bbC^\dagger)$ and $\bbC^*\in \calG_\sigma$.
This graph enumeration algorithm operates by decomposing $\bbC^\dagger$
into trees and generating their isomers respectively. These isomers
are then combined to produce a set of chemical isomers $\bbC^*$ that belong to the desired
chemical graph space $\calG_\sigma$ and have exactly the same feature vector as $\bbC^\dagger$.
\end{itemize}

\subsection{Modeling of Chemical Compounds}\label{sec:chemical_model}

This subsection reviews a modeling of chemical compounds 
introduced by Zhu~et~al.\cite{Zhu:2022ad}.
Let $\Lambda$ represent the set of chemical elements;
for example,  $\Lambda=\{\tH,  \tC, \tO, \tN, \tP, \tS_{(2)}, \tS_{(4)}, \tS_{(6)}\}$. 
Elements $\ta$ with multiple valence states are distinguished with a suffix, i.e., 
we denote an element $\ta$ with valence $i$ as $\ta_{(i)}$.


A chemical compound $\bbC$ is represented as a {\em chemical graph}, which is defined as
a triplet $\bbC=(H,\alpha,\beta)$, where
$H$ is a graph $H$,
$\alpha:V(H)\to \Lambda$ assigns chemical elements to vertices, and
$\beta:E(H)\to [1,3]$ assigns bond multiplicities to edges.
Two chemical graphs $(H_1, \alpha_1, \beta_1)$ and $(H_2,\alpha_2,\beta_2)$ are
{\em isomorphic} if there exists
an isomorphism $\phi$,
i.e., a bijection $\phi: V(H_1)\to V(H_2)$
such that
 $uv\in E(H_1), \alpha_1(u)=\ta, \alpha_1(v)=\tb, \beta_1(uv)=m$
if and only if
 $\phi(u)\phi(v) \in E(H_2), \alpha_2(\phi(u))=\ta, 
 \alpha_2(\phi(v))=\tb, \beta_2(\phi(u)\phi(v))=m$. 
 If $H_1$ and $H_2$ are rooted graphs with roots $r_1$ and $r_2$, respectively,
 the chemical graphs are considered
{\em rooted-isomorphic} 
if there exists an isomorphism $\phi$ such that $\phi(r_1)=r_2$ also holds. 
  
For a chemical graph  $\bbC=(H,\alpha,\beta)$, 
  let  $V_{\ta}(\bbC)$ ($\ta\in \Lambda$)
represent the set of vertices $v\in V(H)$ such that $\alpha(v)=\ta$. 
The {\em hydrogen-suppressed chemical graph} of $\bbC$, denoted as $\anC$,
is obtained by removing all vertices $v\in \VH(\bbC)$ from $H$.

\subsubsection{Two-layered Model}\label{sec:2LM}
Shi~et~al.~\cite{Shi:2021aa} introduced the
two-layered model for chemical graphs
to efficiently capture the graph-theoretic information and develop descriptors based on it.
We summarize the key concepts of this model for completeness.
 
\begin{figure}[t!] \begin{center}
\includegraphics[width=.90\columnwidth]{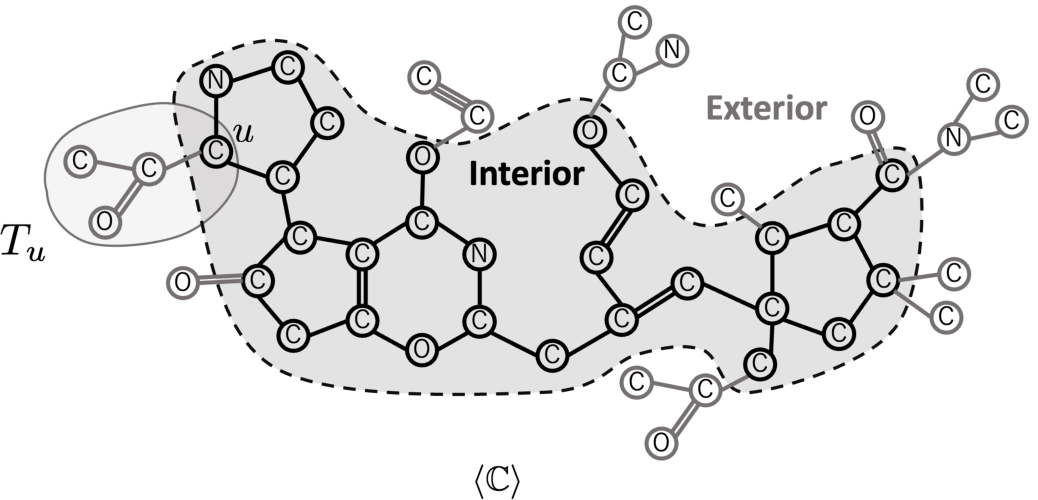}
\end{center}
\caption{
An illustration of the two-layered model.
The interior region is represented by the shaded area enclosed by black dashed lines,
while the remaining parts form the exterior.
$T_u$ is the chemical tree rooted at $u$ and is outlined by a thin gray line.
 }
\label{fig:two_layer} \end{figure}

Consider a chemical graph $\bbC=(H,\alpha,\beta)$ and an integer $\rho \geq 1$,
referred to as the {\em branch-parameter}.
For this study, the standard value of $\rho = 2$ is used.
The {\em two-layered model} of $\bbC$ is a partition of
 the hydrogen-suppressed chemical graph $\anC$ into
two regions: the ``interior'' and the ``exterior'' based on the branch-parameter $\rho$.
A vertex $v \in V(\anC)$ (resp., an edge $e \in E(\anC)$) of $\bbC$
is classified as an {\em exterior-vertex} (resp., {\em exterior-edge})
if $\h(v)< {\rho}$ (resp., $e$ is incident to an exterior-vertex).
We denote 
the sets of exterior-vertices and exterior-edges of $\bbC$
by $V^\ex(\bbC)$ and $E^\ex(\bbC)$, respectively. 
The remaining vertices and edges, defined as
$V^\inte(\bbC)=V(\anC)\setminus  V^\ex(\bbC)$ and 
$E^\inte(\bbC)=E(\anC)\setminus E^\ex(\bbC)$,
are called  {\em interior-vertices} and {\em interior-edges}, respectively.
Notice that the set  $E^\ex(\bbC)$  forms 
a collection of connected graphs, each can be treated as a rooted tree $T$ with the root being 
the vertex $v\in V(T)$ with the maximum height $\h(v)$. 
Let $\mathcal{T}^\ex(\anC)$ denote 
the set of these rooted trees in $\anC$. 
The {\em interior} of $\bbC$ is defined to be the subgraph
 $(V^\inte(\bbC),E^\inte(\bbC))$ of $\anC$. 
See Figure~\ref{fig:two_layer} for an example.
 


For each interior-vertex $u\in V^\inte(\bbC)$,
let $T_u\in \mathcal{T}^\ex(\anC)$ represent the chemical tree rooted at $u$
(where $T_u$ may consist solely of the vertex $u$).
The {\em $\rho$-fringe-tree} $\bbC[u]$  is defined
as the chemical rooted tree obtained by restoring the hydrogens which are originally attached
to $T_u$ in $\bbC$.

For a given integer $K$, a feature vector $f(\bbC)$ for a chemical graph $\bbC$
is defined by a {\em feature function} $f$ which comprises $K$ descriptors 
based on the two-layered model.
A comprehensive list and detailed explanation of the feature function $f$ used in this study
 can be found in Section~\ref{sec:descriptor}.

Furthermore, in order to allow the usage of domain knowledge for inference of chemical graphs,
such as some abstract structures or limits on the available 2-fringe-trees,
a set of rules called {\em topological specification} is used.
A topological specification includes the following components:
\begin{itemize}
\item[-] A {\em seed graph} $\GC$, which serves as an abstract form of the target chemical graph $\bbC$.
\item[-] A set $\mathcal{F}$ of chemical rooted trees, which serve as candidates 
for the tree  $\bbC[u]$ rooted at each interior-vertex $u$ in $\bbC$.
\item[-] Lower and upper bounds that constrain the number of various components 
in the target chemical graph, such as the interior-vertices, 
double/triple bonds, and chemical elements in $\bbC$. 
\end{itemize}


Refer~\cite{Zhu:2022ad} or Section~\ref{sec:specification}
 for a more detailed description of the topological specification.

\subsubsection{Modeling of Polymers}\label{sec:polymer}

In this subsection, we review the way of representing a polymer as a form of monomer that is proposed by Ido~et~al.~\cite{Ido:2024aa},
and the necessary modification for polymers in the two-layered model.

\begin{figure}[t!] \begin{center}
\includegraphics[width=.96\columnwidth]{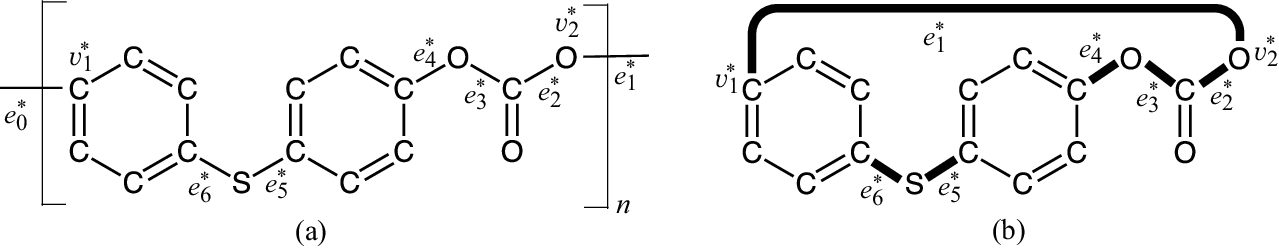}
\end{center}
\caption{(a) The repeating unit of the polymer thioBis(4-phenyl)carbonate,
where $v^*_1$ and $v^*_2$ are the connecting-vertices
and $e^*_0$ and $e^*_1$ are the connecting-edges;
(b) The monomer representation of the polymer in (a), where 
$v^*_1$ and $v^*_2$ are the connecting-vertices
and the link-edges are depicted with thick lines.
}
\label{fig:polymer_example}  \end{figure}

For polymers,
we mainly focus on the case of homopolymer, i.e., a linear sequence
of identical repeating units connected by two specific edges, $e^*_0$ and $e^*_1$,
such that two adjacent units in the sequence are joined with them.
The two edges are referred to as the {\em connecting-edges},
and the two vertices incident 
to the two connecting-edges are called the {\em connecting-vertices}.  
An example of these concepts can be found in Figure~\ref{fig:polymer_example}(a).
 
We call an edge $e$ a  {\em link-edge} 
 in a repeating unit of a polymer if it is traversed by every path 
connecting $e^*_0$ and $e^*_1$,
and denote the set of link-edges in $\bbC$ by $\Elnk(\bbC)$.
For instance, in  Figure~\ref{fig:polymer_example}(a), the link-edges 
 are $e^*_2,e^*_3,\ldots,e^*_6$. 
Following Ido et al.~\cite{Ido:2024aa},
we treat the two connecting-edges as a single edge $e^*_1$ 
to simplify the representation of the polymer,
as illustrated in Figure~\ref{fig:polymer_example}(b). 
The resulting graph is called
the {\em monomer representation} of the polymer,
and edge $e^*_1$ is also called a {\em link-edge}.
In what follows, we represent polymers by their monomer representations $\bbC$.


The link-edges and connecting-vertices are both used in the feature function defined for polymers.
See Section~\ref{sec:descriptor}
 for more details about this.
Also, we specify the set of link-edges in the seed graph $\GC$ and the lower and upper bounds on the number
of components such as the  link-edges and connecting-vertices
in the topological specification used for a polymer.
See Section~\ref{sec:specification} 
for a more detailed explanation of the topological specification for polymers.

\section{A Full Description of Descriptors}\label{sec:descriptor}

Our definition of feature function is analogous to the one  by 
 Zhu~et~al.~\cite{Zhu:2022ad}. When the molecule is a polymer, we follow Ido~et~al.~\cite{Ido:2024aa}
 to add some additional descriptors.

Associated with the two functions 
$\alpha$ and $\beta$ in a chemical graph $\Co=(H,\alpha,\beta)$,
we introduce   functions  
 $\ac: E(H)\to (\Lambda\setminus\{\ttH\})\times (\Lambda\setminus\{\ttH\})\times [1,3]$, 
 $\cs: V(H)\to (\Lambda\setminus\{\ttH\})\times [1,6]$ and
$\ec: E(H)\to ((\Lambda\setminus\{\ttH\})\times [1,6])
\times ((\Lambda\setminus\{\ttH\})\times [1,6])\times [1,3]$
in the following.

We define a method for featuring interior-edges  as follows.
Let $e=uv\in E^\inte(\Co)$  be 
 an interior-edge 
 such that $\alpha(u)=\ta$, $\alpha(v)=\tb$ and $\beta(e)=m$ 
   in a chemical graph  $\Co=(H,\alpha,\beta)$.
To feature this edge $e$, 
 we use a tuple $(\ta,\tb,m)\in (\Lambda\setminus\{{\tt H}\})
    \times (\Lambda\setminus\{{\tt H}\})\times [1,3]$,
 which we call the {\em adjacency-configuration} $\ac(e)$ of the edge $e$. 
 We introduce a total order $<$ over the elements in $\Lambda$
 to distinguish  between $(\ta,\tb, m)$ and $(\tb,\ta, m)$ 
 $(\ta\neq \tb)$ notationally.
 For a tuple  $\nu=(\ta,\tb, m)$,
 let $\overline{\nu}$ denote the tuple $(\tb,\ta, m)$.  

To  represent a feature of an interior-vertex $v\in V^\inte(\Co)$ such that
$\alpha(v)=\ta$  and  $\deg_{\anC}(v)=d$
(i.e., the number of non-hydrogen atoms adjacent to $v$ is $d$) 
   in a chemical   graph  $\Co=(H,\alpha,\beta)$,
 we use  a pair $(\ta, d)\in (\Lambda\setminus\{{\tt H}\})\times [1,4]$,
 which we call the {\em chemical symbol} $\cs(v)$ of the vertex $v$.
 We treat $(\ta, d)$ as a single symbol $\ta d$,  and  
define $\Ldg$   to be  the set of all chemical symbols
$\mu=\ta d\in  (\Lambda\setminus\{{\tt H}\})\times [1,4]$.  

Let $e=uv\in E^\inte(\Co)$  be 
an  interior-edge 
 such that $\cs(u)=\mu$, $\cs(v)=\mu'$ and $\beta(e)=m$ 
   in a chemical  graph  $\Co=(H,\alpha,\beta)$.
To feature this edge $e$, 
 we use a tuple $(\mu,\mu',m)\in \Ldg\times \Ldg\times [1,3]$, 
 which we call  the {\em edge-configuration} $\ec(e)$ of the edge $e$. 
 We introduce a total order $<$ over the elements in $\Ldg$
 to distinguish between $(\mu,\mu', m)$ and $(\mu', \mu, m)$ 
 $(\mu \neq \mu')$ notationally. 
 For a tuple  $\gamma=(\mu,\mu', m)$,
 let $\overline{\gamma}$ denote the tuple $(\mu', \mu, m)$.

 To represent  a feature of the exterior  of  $\Co$, 
  a  chemical rooted tree in $\mathcal{T}(\Co)$ is
  called a {\em fringe-configuration} of $\Co$. 
We also represent leaf-edges in the exterior of $\Co$.
For a leaf-edge $uv\in E(\anC)$ with $\deg_{\anC}(u)=1$, we define
the {\em adjacency-configuration} of $e$ to be an ordered tuple
$(\alpha(u),\alpha(v),\beta(uv))$. 
Define 
\[ \Gac^\lf\triangleq \{(\ta,\tb,m)\mid \ta,\tb\in\Lambda, 
m\in[1,\min\{\val(\ta),\val(\tb)\}]\} \]
as a set of possible adjacency-configurations for leaf-edges. 
   
Let $\pi$ be a chemical property for which we will construct
a prediction function $\eta$ from a feature
vector $f(\C)$ of a chemical graph $\C$ 
to a predicted value $y\in \mathbb{R}$
for the  chemical property of $\C$.

We first choose a set $\Lambda$ of chemical elements
 and then collect a data set  $D_{\pi}$ of
  chemical compounds  $C$ whose 
  chemical elements belong to $\Lambda$,
  where we regard  $D_{\pi}$ as a set of chemical graphs $\C$
  that represent the chemical compounds $C$  in  $D_{\pi}$.
To define the interior/exterior of 
chemical graphs  $\C\in D_{\pi}$,
we  next choose a branch-parameter ${\rho}$, where
 we recommend ${\rho}=2$.  
 
Let $\Lambda^\inte(D_\pi)\subseteq \Lambda$  
(resp., 
$\Lambda^\ex(D_\pi)\subseteq \Lambda$)
denote the set  of chemical elements  used in
the set $V^\inte(\C)$ of interior-vertices
(resp., the set $V^\ex(\C)$ of  exterior-vertices) of $\C$
 over all chemical graphs $\C\in D_\pi$, 
and $\Gamma^\inte(D_\pi)$
(resp., $\Gamma^\lnk(D_\pi)$) 
denote the set of edge-configurations used in
the set $E^\inte(\C)$  of interior-edges
(resp., the set $\Elnk(\C)$ of linked-edges) in $\C$
 over all chemical graphs $\C\in D_\pi$. 
Let $\mathcal{F}(D_\pi)$ denote the set of
chemical rooted trees $\psi$  
r-isomorphic to a chemical rooted tree in $\mathcal{T}(\C)$
  over all chemical graphs $\C\in D_\pi$,
  where possibly a chemical rooted tree $\psi\in \mathcal{F}(D_\pi)$
  consists of a single chemical element $\ta\in \Lambda\setminus \{{\tt H}\}$.
  
We define an integer encoding of a finite set $A$ of elements
to be a bijection $\sigma: A \to [1, |A|]$, 
where we denote by $[A]$   the set $[1, |A|]$ of integers.
Introduce  an integer coding of each of the   sets 
$\Lambda^\inte(D_\pi)$, $\Lambda^\ex(D_\pi)$, 
$\Gamma^\inte(D_\pi)$ and $\mathcal{F}(D_\pi)$. 
Let $[\ta]^\inte$  
(resp., $[\ta]^\ex$)  denote   
the coded integer of  an element $\ta\in \Lambda^\inte(D_\pi)$
(resp., $\ta\in \Lambda^\ex(D_\pi)$),  
$[\gamma]$   denote  
the coded integer of  an element $\gamma$ in $\Gamma^\inte(D_\pi)$
and 
$[\psi]$   denote  an element $\psi$ in $\mathcal{F}(D_\pi)$. 
 
We assume that a chemical graph $\C$
 treated in this paper satisfies  $\deg_{\anC}(v)\leq 4$
in the hydrogen-suppressed graph $\anC$.
 
In our model, we  use an integer 
  $\mathrm{mass}^*(\ta)=\lfloor 10\cdot \mathrm{mass}(\ta)\rfloor$, 
 for each $\ta\in \Lambda$.
 
  We define the {\em feature vector} $f(\C)$ 
  of a molecule $\C=(H,\alpha,\beta)\in D_{\pi}$ 
  to be a vector that consists of the following  
non-negative integer descriptors $\dcp_i(\C)$, $i\in [1,K]$, where 
$K=14+ |\Lambda^\inte(D_\pi)|+|\Lambda^\ex(D_\pi)|
+|\Gamma^\inte(D_\pi)|+|\Gamma^\lnk(D_\pi)|+|\Ldg^\inte|
+|\mathcal{F}(D_\pi)|+|\Gac^\lf|$. 
Notice that some descriptors are used for the case of polymers only.


\begin{enumerate}  
\item   
$\dcp_1(\C)$: the number  $|V(H)|-|\VH|$ of non-hydrogen atoms  in  $\C$.  
 
\item 
$\dcp_2(\C)$:  the number $|V^\inte(\C)|$ of interior-vertices in  $\C$.
  
\item 
$\dcp_3(\C)$:  the number $|\Elnk(\C)|$ of link-edges in  $\C$.
This descriptor is only for the case of polymers.

\item 
$\dcp_4 (\C)$: 
the average $\overline{\mathrm{ms}}(\C)$ of mass$^*$ 
over all atoms in $\C$; \\
 i.e., $\overline{\mathrm{ms}}(\C)\triangleq 
 \frac{1}{|V(H)|}\sum_{v\in V(H)}\mathrm{mass}^*(\alpha(v))$. 

\item 
$\dcp_i(\C)$,  $i=4+d,   d\in [1,4]$: 
the number $\dg_d^{\oH} (\C)$  of non-hydrogen vertices $v\in V(H)\setminus \VH$
 of degree $\deg_{\anC}(v)=d$
 in the hydrogen-suppressed chemical graph $\anC$.  
 
\item 
$\dcp_i(\C)$,  $i=8+d,   d\in [1,4]$: 
the number $\dg_d^\inte(\C)$
 of interior-vertices of interior-degree  $\deg_{\C^\inte}(v)=d$
  in the interior $\C^\inte=(V^\inte(\C),E^\inte(\C))$ of  $\C$. 
  
   
\item $\dcp_i(\C)$, $i=12+m$,  $m\in[2,3]$: 
the number $\bd_m^\inte(\C)$
 of  interior-edges with bond multiplicity $m$ in  $\C$; 
 i.e., $\bd_m^\inte(\C)\triangleq | \{e\in E^\inte(\C)\mid \beta(e)=m\} |$.

\item $\dcp_i(\C)$, $i=14+[\ta]^\inte$, 
 $\ta\in \Lambda^\inte(D_\pi)$: 
 the frequency $\na_\ta^\inte(\C)=|V_\ta(\C)\cap V^\inte(\C) |$ 
 of chemical element $\ta$ in
 the set $V^\inte(\C)$ of  interior-vertices in  $\C$. 
 
\item $\dcp_i(\C)$, 
$i=14+|\Lambda^\inte(D_\pi)|+[\ta]^\ex$, 
 $\ta\in \Lambda^\ex(D_\pi)$: 
 the frequency $\na_\ta^\ex(\C)=|V_\ta(\C)\cap V^\ex(\C) |$
  of chemical element $\ta$ in
 the set $V^\ex(\C)$ of  exterior-vertices in  $\C$. 
 
\item $\dcp_i(\C)$, 
$i=14+|\Lambda^\inte(D_\pi)|+|\Lambda^\ex(D_\pi)|+ [\gamma]$, 
$\gamma \in \Gamma^\inte(D_\pi)$: 
the frequency $\ec_{\gamma} (\C)$ of edge-configuration $\gamma$
in the set $E^\inte(\C)$ of interior-edges   in  $\C$. 

\item $\dcp_i(\C)$, 
$i=14+|\Lambda^\inte(D_\pi)|+|\Lambda^\ex(D_\pi)|+ |\Gamma^\inte(D_\pi)|
+ [\gamma]$, 
$\gamma \in \Gamma^\lnk(D_\pi)$: 
the frequency $\ec_{\gamma} (\C)$ of edge-configuration $\gamma$
in the set $\Elnk(\C)$ of link-edges   in  $\C$. 
This descriptor is only for the case of polymers.

\item $\dcp_i(\C)$, 
$i=14+|\Lambda^\inte(D_\pi)|+|\Lambda^\ex(D_\pi)|+ |\Gamma^\inte(D_\pi)| + |\Gamma^\lnk(D_\pi)|
+ [\mu]$, 
$\mu\in \Ldg^\inte$: 
the frequency of chemical symbols $\mu=\alpha(u)\deg_{\anC}(u)$ 
 of connecting-vertices $u$   in  $\C$.
 This descriptor is only for the case of polymers.

\item $\dcp_i(\C)$, 
$i= 14+|\Lambda^\inte(D_\pi)|+|\Lambda^\ex(D_\pi)|
+|\Gamma^\inte(D_\pi)|+|\Gamma^\lnk(D_\pi)|+|\Ldg^\inte|+ [\psi]$,  
 $\psi \in \mathcal{F}(D_\pi)$: 
the frequency $\fc_{\psi}(\C)$ of fringe-configuration $\psi $
in the set of ${\rho}$-fringe-trees in  $\C$. 

\item $\dcp_i(\C)$, 
$i= 14+|\Lambda^\inte(D_\pi)|+|\Lambda^\ex(D_\pi)|
+ |\Gamma^\inte(D_\pi)|+|\Gamma^\lnk(D_\pi)|+|\Ldg^\inte|
+|\mathcal{F}(D_\pi)|+ [\nu]$,  
 $\nu \in \Gac^\lf$: 
the frequency $\ac_{\nu}^\lf(\C)$ of adjacency-configuration $\nu$
in the set of leaf-edges in  $\anC$. 
\end{enumerate}

\section{Specifying Target Chemical Graphs}\label{sec:specification}

Our definition of the topological specification is analogous to the one  by 
 Zhu~et~al.~\cite{Zhu:2022ad}.
 Here we review the one particularly modified for polymers proposed by~Ido~et~al.~\cite{Ido:2024aa}, 
 which is the instance $I_a$ used in Section~\ref{sec:experiment_phase2}.

\subsection*{Seed Graph}

A  {\em seed graph} for a polymer  is defined
to be a graph $\GC=(\VC,\EC)$  with a specified edge subset $\EC^\lnk$
 such that 
the edge set $\EC$ consists of four sets 
$\Et$, $\Ew$, $\Ez$ and $\Eew$, 
where each of them can be empty, and
 $\EC^\lnk$ is a circular  set in $\GC$ such that 
  $\emptyset\neq \EC^\lnk\subseteq \Et\cup \Ew\cup \Eew$ (only for polymer). 
Figure~\ref{fig:seed-graph-abc}(a) 
 illustrates an example of a seed graph,
where $\VC=\{u_1,u_2,\ldots,u_{14}\}$, 
$\Et=\{a_1,a_2,a_3,a_4\}$, 
$\Ew=\{a_5,a_6,\ldots,a_9\}$,
$\Ez=\{a_{10}\}$,
$\Eew=\{a_{11},a_{12},\ldots,a_{18}\}$ and 
$\EC^\lnk=\{a_1,a_2\}$.

 A {\em subdivision} $S$ of $\GC$  
is a graph constructed from a seed graph $\GC$ 
according to the following rules:
\begin{enumerate}[leftmargin=*]
\item[-]
Each edge $e=uv\in \Et$ is replaced
with a $u,v$-path $P_e$ of length at least 2;

\item[-] 
Each edge $e=uv\in \Ew$ is replaced
with a $u,v$-path $P_e$ of length at least 1
(equivalently $e$ is directly used or replaced with
a $u,v$-path $P_e$ of length at least 2);

\item[-] 
Each edge $e\in \Ez$ is either used or discarded;   and

\item[-]
Each edge $e\in \Eew$ is always used directly.
\end{enumerate}

The set of link-edges in the monomer representation  $\C$ of 
an inferred polymer 
consists of edges in $\EC^\lnk\cap( \Eew\cup \Ew)$
or edges  in   paths $P_e$ for all edges $e=uv\in \EC^\lnk\cap (\Ew\cup \Et)$
in a  subdivision  $S$ of $\GC$. 
 
A target chemical graph $\C=(H,\alpha,\beta)$ will contain  $S$  as a subgraph
of the interior $H^\inte$ of $\C$.


\subsection*{Interior-specification}

A graph $H^*$ that serves as the interior $H^\inte$ of
a target chemical graph $\C$ will be constructed as follows.
First construct a subdivision  $S$ of a seed graph $\GC$ 
by replacing each edge $e=u u'\in \Et\cup\Ew$
with a pure $u,u'$-path $P_e$.
Next construct a supergraph $H^*$ of $S$ by 
attaching a leaf path $Q_v$ at each vertex $v\in \VC$ or
at an internal vertex $v\in V(P_e)\setminus\{u,u'\}$ 
of each pure $u,u'$-path $P_e$ for some edge $e=uu'\in \Et\cup\Ew$,
where possibly $Q_v=(v), E(Q_v)=\emptyset$ 
(i.e., we do not attach any new edges to $v$).
We introduce the following rules for specifying
 the size of $H^*$, the length $|E(P_e)|$  of
a pure path  $P_e$,  the length $|E(Q_v)|$ of
a   leaf path $Q_v$, the number of  leaf paths $Q_v$
and a bond-multiplicity of each interior-edge,
where we call the set of prescribed constants  
 an  {\em interior-specification}   $\sint$: 
\begin{enumerate}[leftmargin=*]
 \item[-]
  Lower and upper bounds $\nint_\LB, \nint_\UB\in \Z_+$ 
  on   the number of interior-vertices 
of a target chemical graph~$\C$. 

 \item[-]
  Lower and upper bounds $\nlnk_\LB, \nlnk_\UB\in \Z_+$ 
  on   the number of link-edges 
of a target chemical graph~$\C$ (only for polymer). 
  
\item[-] 
For each edge $e=u u'\in \Et\cup\Ew$, 
\begin{description} 
\item[]
 a lower bound $\ell_{\LB}(e)$ and 
 an upper bound $\ell_{\UB}(e)$  on the length $|E(P_e)|$ of
 a pure $u,u'$-path $P_e$. 
(For a notational convenience, set 
$\ell_\LB(e):=0$, $\ell_\UB(e):=1$, $e\in \Ez$ and
$\ell_\LB(e):=1$, $\ell_\UB(e):=1$, $e\in \Eew$.)
   
\item[]  
 a lower bound $\bl_{\LB}(e)$ and 
 an upper bound $\bl_{\UB}(e)$ on the number of leaf paths $Q_v$ attached 
 at  internal vertices $v$ of a pure $u,u'$-path $P_e$.   

\item[] 
 a lower bound $\chch_{\LB}(e)$ and 
 an upper bound $\chch_{\UB}(e)$  on the maximum 
 length  $|E(Q_v)|$ of a leaf path $Q_v$ attached  
 at an internal vertex $v\in V(P_e)\setminus\{u,u'\}$ 
 of a pure $u,u'$-path $P_e$.   
\end{description} 

\item[-]
For each vertex $v\in \VC$, 
\begin{description} 
\item[]
 a lower bound $\chch_{\LB}(v)$ and 
 an upper bound $\chch_{\UB}(v)$  on  
 the number of leaf paths $Q_v$ attached to $v$,
 where $0\leq \chch_{\LB}(v)\leq \chch_{\UB}(v)\leq 1$.
 
\item[]
 a lower bound $\chch_{\LB}(v)$ and 
 an upper bound $\chch_{\UB}(v)$  on the
 length $|E(Q_v)|$ of a leaf path $Q_v$ attached to $v$. 
\end{description}  

\item[-] 
For each edge $e=u u'\in \EC$, 
a lower bound $\bd_{m, \LB}(e)$ 
and an  upper bound $\bd_{m, \UB}(e)$  on
the number of edges with bond-multiplicity $m\in [2,3]$ in
$u,u'$-path $P_e$, where we regard $P_e$, $e  \in \Ez\cup \Eew$ 
as single edge $e$.
\end{enumerate}

We call a graph $H^*$ that satisfies an interior-specification $\sint$
a {\em $\sint$-extension of $\GC$}, 
where the bond-multiplicity of each edge has been determined.

Table~\ref{table:interior-spec}  shows an example of 
an interior-specification  $\sint$ to the seed graph  $\GC$ in 
Figure~\ref{fig:seed-graph-abc}(a). 

\begin{table}[h!]\caption{Example~1 of an interior-specification  $\sint$. }
\begin{tabular}{ |  c | c | c | c |  } \hline 
$\nint_\LB=20$ & $\nint_\UB = 30$ & 
$\nlnk_\LB=2$ & $\nlnk_\UB = 24$ \\\hline 
\end{tabular}

 \begin{tabular}{ |  c | c c c c c c c c c |  } \hline
              & $a_1$ &  $a_2$ &   $a_3$ &   $a_4$ &   $a_5$ &   $a_6$ &   $a_7$ &   $a_8$  &   $a_9$   \\\hline
 $\ell_\LB(a_i)$ &  2 &  4 &  3 & 2 &  2 &  1  &  1 &  1 &   1\\ \hline
 $\ell_\UB(a_i)$ &  3 & 6 &  6 & 5 &  3 &  3  &  6 &  2 &   6 \\\hline
 $\bl_\LB(a_i)$ &   0 &  1 & 1 & 0 &  0 &   0 &  0 &   0 &  0 \\ \hline
 $\bl_\UB(a_i)$ &  1 &  4 &  4 & 3 &  2 &   1 &  1 &  1  &  1 \\\hline
 $\chch_\LB(a_i)$ &   0 &  2 &  1 & 0 &  0 &  0 &  0 &   0 &   0 \\ \hline
 $\chch_\UB(a_i)$ &  3 &  6 &  6 & 3 &  3 &   3 &  3 &   0 &   0 \\\hline
\end{tabular} 

\begin{tabular}{ |  c | c c c c c c   c c c c  c c c  c|  } \hline
                        & $u_1$ &  $u_2$ &   $u_3$ &   $u_4$ &   $u_5$ &   $u_6$ 
                       & $u_7$ &   $u_8$ &   $u_9$ &   $u_{10}$ &   $u_{11}$ 
                       &   $u_{12}$ &   $u_{13}$ &   $u_{14}$ \\\hline 
 $\bl_\LB(u_i)$ &  0 & 0 & 0 & 0 & 0 &  0 & 0 & 0 & 1 & 0 & 0 & 0 & 0 & 0 \\ \hline
 $\bl_\UB(u_i)$&  1 & 1 & 1 & 1 & 1 &  1 & 1 & 1 & 1 & 1 & 1 & 1 & 1 & 1 \\\hline
 $\chch_\LB(u_i)$&  0 & 0 & 0 & 0 & 0 &  0 & 0 & 0 & 1 & 0 & 0 & 0 & 0 & 0 \\\hline
 $\chch_\UB(u_i)$& 4 & 4 & 4 & 4 & 4 &  4 & 4 & 4 & 6 & 4 & 4 & 4 & 4 & 4 \\\hline
\end{tabular} 

\begin{tabular}{ |  c | c c c c c c   c c c c c c  c c c c c c |  } \hline
                               & $a_1$ &  $a_2$ &   $a_3$ &   $a_4$ &   $a_5$ &   $a_6$ 
                               & $a_7$ &  $a_8$ &   $a_9$ &   $a_{10}$ &   $a_{11}$ &   $a_{12}$ 
                               & $a_{13}$ &   $a_{14}$ &   $a_{15}$ &   $a_{16}$   &   $a_{17}$ &   $a_{18}$
                                    \\\hline
 $\bd_{2, \LB}(a_i)$ &  0 & 0 & 0 & 0 & 0 &  0 & 0 & 0 & 0 & 0 & 0 & 0 & 0 & 1  & 0 & 0 & 0 & 0\\ \hline 
 $ \bd_{2, \UB}(a_i)$& 1 & 2 & 1 & 1 & 1 &  1 & 1 & 1 & 1 & 1 & 1 &  1 & 1 & 1 & 1 & 1 &  1 & 1\\ \hline
 $\bd_{3, \LB}(a_i)$ &  0 & 0 & 0 & 0 & 0 &  0 & 0 & 0 & 0 & 0 & 0 & 0 & 0 & 0  & 0 & 0 & 0 & 0\\ \hline
 $ \bd_{3, \UB}(a_i)$& 1 & 1 & 1 & 1 & 1 &  1 & 1 & 1 & 1 & 1 & 1 &  1 & 1 & 1 & 1 &  1 & 1 & 1\\ \hline
\end{tabular} 
\label{table:interior-spec}  
\end{table}

\begin{figure}[t!]
\begin{center}
\includegraphics[width=.89\columnwidth]{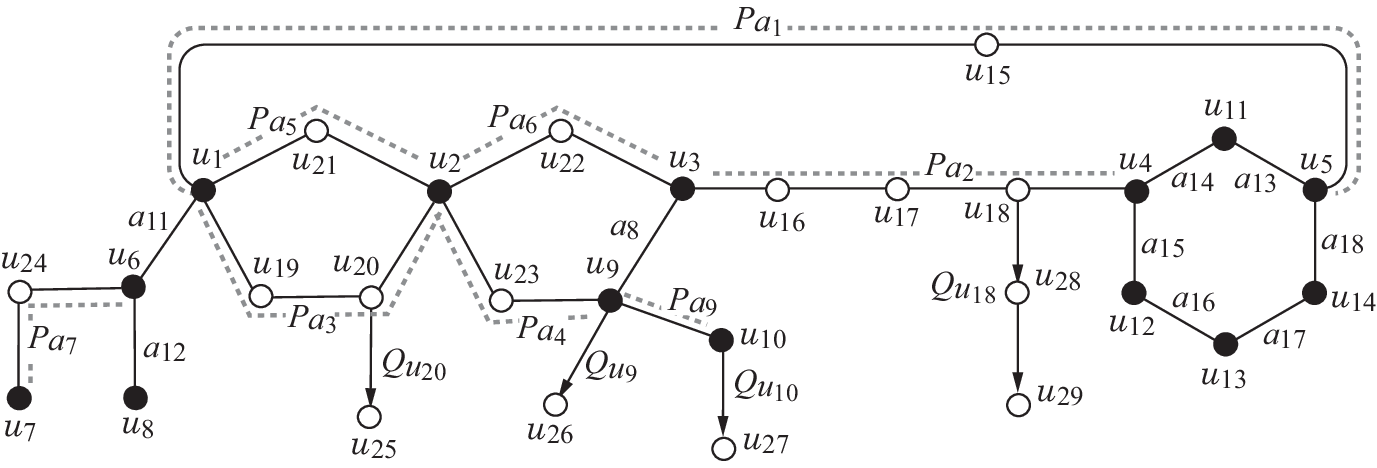}
\end{center}
\caption{A graph  obtained from the seed graph $\GC$  
in Figure~\ref{fig:seed-graph-abc}(a), 
where each path $Q_u$ rooted at a vertex $u$ is
depicted with arrows and the vertices newly introduced from $\GC$ are depicted with white circles. 
}
\label{fig:test_subgraph_polymer}
\end{figure}

Figure~\ref{fig:test_subgraph_polymer} 
illustrates an example of 
an $\sint$-extension $H^*$ of seed graph  $\GC$ in 
Figure~\ref{fig:seed-graph-abc}(a) 
under the interior-specification  $\sint$ in 
Table~\ref{table:interior-spec}.


\subsection*{Chemical-specification}
 
 \begin{figure}[t!]
\begin{center} 
 \includegraphics[width=.45\columnwidth]{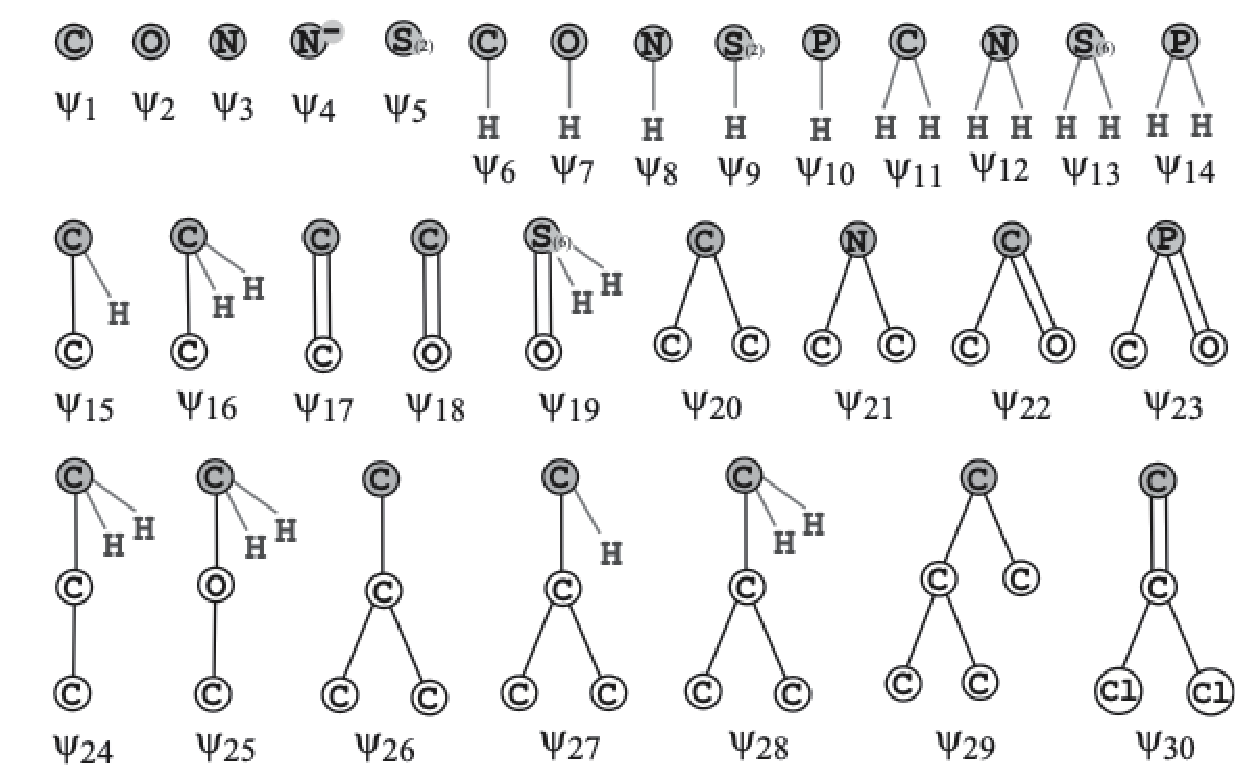}
\end{center}
\caption{Illustrations of the set of chemical rooted trees for the instance $I_a$.}
\label{fig:fringe-tree-set-a}  
\end{figure} 
 
 Let $H^*$ be a graph that serves as 
 the interior $H^\inte$ of a target chemical graph $\C$,
 where the bond-multiplicity of each edge in $H^*$ has be determined.
 Finally we introduce a set of rules for constructing 
   a target chemical graph $\C$ from $H^*$ 
   by choosing  a chemical element $\ta\in \Lambda$ 
and assigning a ${\rho}$-fringe-tree $\psi$
 to each interior-vertex $v\in V^\inte$. 
We introduce the following rules for specifying
the size of $\C$, a set of chemical rooted trees  
that are allowed to use as  ${\rho}$-fringe-trees 
and lower and upper bounds on the frequency of
a chemical element, a chemical symbol, 
 an edge-configuration, and a fringe-configuration
where we call the set of prescribed constants   
 a  {\em chemical specification} $\sce$.
 Notice that those involving link-edges and connecting-vertices are only used for the inference of polymers.
 
\begin{enumerate}[leftmargin=*]
\item[-] 
Lower and upper bounds $n_\LB,  n^*\in \Z_+$
on the number of vertices, where $\nint_\LB \leq n_\LB\leq n^*$.
 
\item[-] 
A subset $\mathcal{F}^* \subseteq \mathcal{F}(D_\pi)$  
 of chemical rooted trees $\psi$ with $\h(\anpsi)\leq {\rho}$, where 
 we require that 
 every ${\rho}$-fringe-tree $\C[v]$ rooted at an interior-vertex $v$ 
    in  $\C$  belongs to $\mathcal{F}^*$.  
    Figure~\ref{fig:fringe-tree-set-a} illustrates the corresponding set of chemical rooted trees for the instance $I_a$
     whose seed graph is illustrated in Figure~\ref{fig:seed-graph-abc}(a).
Let   
$\Lambda^\ex$ denote the set of  chemical elements assigned to non-root
vertices over all chemical rooted trees in $\mathcal{F}^*$.  
 
\item[-] 
A subset  $\Lambda^\inte\subseteq \Lambda^\inte(D_\pi)$, where 
 we require that every chemical element $\alpha(v)$ 
 assigned to an interior-vertex  $v$ in $\C$ belongs to $\Lambda^\inte$.
Let $\Lambda:= \Lambda^\inte\cup \Lambda^\ex$ and
 $\na_\ta(\C)$ (resp., $\na_\ta^\inte(\C)$ and $\na_\ta^\ex(\C)$) 
 denote the number of vertices   (resp.,   interior-vertices and  exterior-vertices)
  $v$ such that $\alpha(v)=\ta$   in  $\C$.
 
\item[-] 
A set $\Ldg^\inte\subseteq \Lambda\times [1,4]$  of chemical  symbols.

\item[-] 
Subsets $\Gamma^\lnk\subseteq \Gamma^\inte$ of $\Gamma^\inte(D_\pi)$  
of  edge-configurations  $(\mu,\mu' ,m)$ with $\mu \leq \mu'$, where 
 we require that the edge-configuration $\ec(e)$ of an interior-edge (resp., a link-edge) $e$ in $\C$ 
 belongs to $\Gamma^\inte$ (resp.,    $\Gamma^\lnk$).
We do not distinguish  $(\mu,\mu' ,m)$ and $(\mu' , \mu,m)$.  
  
\item[-] 
Define  $\Gac^\inte$  (resp.,    $\Gac^\lnk$)  to be the set of   adjacency-configurations such that  
$\Gac^\typ:=\{(\ta, \tb, m) \mid (\ta d, \tb d',m)\in \Gamma^\typ\}, \typ\in\{\inte,\lnk\}$.   
Let  $\ac_\nu^\inte(\C), \nu\in \Gac^\inte$  
(resp.,  $\ac_\nu^\lnk(\C), \nu\in \Gac^\lnk$)   
denote  the number of  interior-edges (resp.,  link-edges) $e$
 such that $\ac(e)=\nu$  in $\C$.
  
\item[-] 
 Subsets $\Lambda^*(v)\subseteq \{\ta\in \Lambda^\inte\mid \val(\ta)\geq 2\}$, 
 $v\in \VC$,  
 we require that every chemical element $\alpha(v)$ 
 assigned to   a vertex $v\in  \VC$
 in the seed graph  belongs to $\Lambda^*(v)$.  

\item[-] Lower and upper bound functions 
$\na_\LB,\na_\UB: \Lambda\to  [0,n^*]$  and 
$\na_\LB^\inte,\na_\UB^\inte: \Lambda^\inte\to  [0,n^*]$ 
on the number of   interior-vertices  $v$ such that  $\alpha(v)=\ta$  in $\C$. 

\item[-] Lower and upper bound functions  
$\ns_\LB^\inte,\ns_\UB^\inte: \Ldg^\inte\to  [0,n^*]$ 
  on the number of   interior-vertices $v$ such that $\cs(v)=\mu$  in $\C$.   

\item[-] Lower and upper bound functions  
$\ns_\LB^\cnt,\ns_\UB^\cnt: \Ldg^\inte\to  [0,2]$ 
  on the number of connecting-vertices $v$ such that $\cs(v)=\mu$  in $\C$.   
  
\item[-] Lower and upper bound functions  
$\ac_\LB^\inte,\ac_\UB^\inte: \Gac^\inte \to  \Z_+$ 
($\ac_\LB^\lnk,\ac_\UB^\lnk: \Gac^\lnk \to  \Z_+$)
 on the number of  interior-edges (resp., link-edges) $e$ such that $\ac(e)=\nu$  in $\C$. 

\item[-] Lower and upper bound functions  
$\ec_\LB^\inte,\ec_\UB^\inte: \Gamma^\inte \to  \Z_+$ 
(resp., $\ec_\LB^\lnk,\ec_\UB^\lnk: \Gamma^\lnk \to  \Z_+$)  
 on the number of  interior-edges  (resp., link-edges)  $e$ such that $\ec(e)=\gamma$  in $\C$.  
 
\item[-] Lower and upper bound functions  
$\fc_\LB,\fc_\UB: \mathcal{F}^*\to  [0,n^*]$ 
  on the number of   interior-vertices $v$ 
  such that $\C[v]^\fr$ is r-isomorphic to $\psi\in \mathcal{F}^*$  in $\C$.  
  
 \item[-] Lower and upper bound functions  
$\ac^\lf_\LB,\ac^\lf_\UB: \Gac^\lf \to  [0,n^*]$ 
  on the number of  leaf-edges $uv$ in $\acC$
  with adjacency-configuration $\nu$.  
\end{enumerate}
 
We call a chemical graph $\C$ that satisfies a chemical specification $\sce$
a {\em $(\sint,\sce)$-extension of $\GC$},
and denote by $\mathcal{G}(\GC, \sint,\sce)$ the set of
all $(\sint,\sce)$-extensions of $\GC$. 

Table~\ref{table:chemical_spec}  shows an example of 
a chemical-specification  $\sce$ to the seed graph  $\GC$
 in Figure~\ref{fig:seed-graph-abc}(a). 

\begin{table}[H]\caption{Example~2 of a chemical-specification  $\sce$.  
}
\begin{tabular}{ |  l |  } \hline
 $n_\LB=30$,  $n^* =50$. \\\hline
  branch-parameter:   ${\rho}=2$  \\\hline
\end{tabular}

\begin{tabular}{ |  l |  } \hline
 Each of sets $\mathcal{F}(v), v\in \VC$ and
 $\mathcal{F}_E$ is set to be \\
 the set $\mathcal{F}$  of chemical rooted trees $\psi$ with $\h(\anpsi)\leq {\rho}=2$
in Figure~\ref{fig:fringe-tree-set-a}. \\\hline  
\end{tabular}

\begin{tabular}{ |  c | c |   } \hline
  $\Lambda=\{ \ttH,\ttC,\ttN,\ttO, \ttS_{(2)},\ttS_{(6)}, \ttP=\ttP_{(6)},\ttCl\}$ & 
  $\Ldg^\inte =\{ \ttC2 , \ttC3,  \ttC4, \ttN2, \ttN3, \ttO2,
    \ttS_{(2)}2,  \ttS_{(6)}3, \ttP4   \}$  
\\\hline
\end{tabular}

\begin{tabular}{ |  c | l |  } \hline
  $\Gac^{\inte}$ &
  $ \nu_1 \!=\!(\ttC   , \ttC  , 1) ,   \nu_2 \!=\!(\ttC   , \ttC  , 2) ,   
   \nu_3 \!=\!(\ttC   , \ttN  , 1) ,  \nu_4 \!=\!(\ttC  , \ttO  , 1), 
    \nu_5 \!=\! (\ttC, \ttS_{(2)}, 1),\nu_6 \!=\!(\ttC  , \ttS_{(6)}, 1), 
    \nu_7 \!=\! (\ttC  , \ttP  , 1) $  \\ \hline
\end{tabular}

\begin{tabular}{ |  c | l |  } \hline
  $\Gamma^{\inte}$ &
  $ \gamma_1 \!=\! (\ttC 2 , \ttC 2, 1) ,
    \gamma_{2} \!=\!(\ttC 2 , \ttC 2, 2),  
   \gamma_3 \!=\!(\ttC 2 , \ttC 3, 1) ,  
   \gamma_4 \!=\!(\ttC 2 , \ttC 3, 2) ,  
   \gamma_5 \!=\!(\ttC 2 , \ttC 4, 1) , 
   \gamma_6 \!=\!(\ttC 3 , \ttC 3, 1) , 
 $ \\
   &
  $  \gamma_7 \!=\!(\ttC 3 , \ttC 3, 2) ,
    \gamma_8 \!=\!(\ttC 3 , \ttC 4, 1), 
   \gamma_9 \!=\!(\ttC 2 , \ttN 3, 1) ,  
   \gamma_{10} \!=\!(\ttC 3 , \ttN 2, 1) ,   
    \gamma_{11} \!=\!(\ttC 4 , \ttN2, 1), 
   \gamma_{12} \!=\!(\ttC 2 , \ttO 2, 1), $ \\ 
   &
  $  
    \gamma_{13} \!=\!(\ttC 3 , \ttO 2, 1) ,    
    \gamma_{14} \!=\!(\ttC 2, \ttS_{(2)} 2, 1),  
    \gamma_{15} \!=\!(\ttC 3, \ttS_{(2)} 2, 1),  
    \gamma_{16} \!=\!(\ttC 4, \ttS_{(2)} 2, 1),  
    \gamma_{17} \!=\!(\ttC 3 , \ttS_{(6)}3, 1),   $ \\ 
   &
  $  
   \gamma_{18} \!=\!(\ttC 4, \ttS_{(6)}3, 1), 
    \gamma_{19} \!=\!(\ttC 2, \ttP4, 1), 
    \gamma_{20} \!=\!(\ttC 3, \ttP4, 1)  
     $ \\ \hline
\end{tabular}

\begin{tabular}{ |  c | l |  } \hline
  $\Gac^{\lnk}$ &
  $ \nu'_1 \!=\!(\ttC   , \ttC  , 1) ,   \nu'_2 \!=\!(\ttC   , \ttC  , 2) ,   
   \nu'_3 \!=\!(\ttC   , \ttN  , 1),  \nu'_4 \!=\!(\ttC   , \ttS_{(2)}  , 1)  $  \\ \hline
\end{tabular}

\begin{tabular}{ |  c | l |  } \hline
  $\Gamma^{\lnk}$ &
  $ \gamma'_1 \!=\! (\ttC 2 , \ttC 2, 1) ,
   \gamma'_2 \!=\!(\ttC 2 , \ttC 3, 1) ,  
   \gamma'_3 \!=\!(\ttC 2 , \ttC 4, 1) ,  
   \gamma'_4 \!=\!(\ttC 3 , \ttC 3, 1) , 
   \gamma'_5 \!=\!(\ttC 3 , \ttC 3, 2) ,
   \gamma'_6 \!=\!(\ttC 2 , \ttN 3, 1) ,   $ \\
   &
  $   
   \gamma'_7 \!=\!(\ttC 3 , \ttN 2, 1), 
    \gamma'_8 \!=\!(\ttC 2, \ttS_{(2)} 2, 1),  
    \gamma'_9 \!=\!(\ttC 3, \ttS_{(2)} 2, 1),  
    \gamma'_{10} \!=\!(\ttC 4, \ttS_{(2)} 2, 1) $  \\\hline
\end{tabular}

\begin{tabular}{ |  l|  } \hline
$\Lambda^*(u_i)=\{\ttC\}, i\in\{1,2,3,4,5,6,9\}$, 
$\Lambda^*(u_8)=\{{\ttO}\}$, 
$\Lambda^*(u_{12})=\{{\tt C, P}\}$, \\
$\Lambda^*(u_i)=\{\ttC,\ttO,\ttN\}$, $i\in [1,14]\setminus\{1,2,3,4,5,6,8,9,12\}$
   \\\hline
\end{tabular}
  
\begin{tabular}{ |  c | c c c c  c c c c |  } \hline
                         & ${\tt H}$  & ${\tt C}$ &   ${\tt N}$ &     ${\tt O}$ 
                         & $\ttS_{(2)}$ & $\ttS_{(6)}$ & $\ttP$ & $\ttCl$ \\\hline
 $\na_\LB(\ta)$ & 40 &  25 & 1 &  1 & 0 & 0 & 0  & 0 \\ \hline 
 $\na_\UB(\ta)$ & 80 & 50 & 8 &  8  & 4 & 4 & 4 & 4 \\\hline
\end{tabular} 
\begin{tabular}{ |  c | c c c  c c c   |  } \hline
   & $\ttC$ &   $\ttN$ &     $\ttO$  & $\ttS_{(2)}$ & $\ttS_{(6)}$ & $\ttP$  \\\hline
 $\na_\LB^{\inte}(\ta)$ &  10 &  1 & 0  & 0 & 0 & 0      \\ \hline
 $\na_\UB^{\inte}(\ta) $&  25 & 4 & 5 & 2 & 2 & 2  \\\hline
\end{tabular} 

\begin{tabular}{ |  c | c c c c c c  c c c   |  } \hline
    & $\ttC2$ &  $\ttC3$ &   $\ttC4$ & $\ttN2$ &   $\ttN3$ &   $\ttO2$
   & $\ttS_{(2)}2$ & $\ttS_{(6)}3$ & $\ttP4$  \\\hline
 $\ns_\LB^{\inte}(\mu)$ &  3 &   5 & 0  & 0 &  0 &  0 & 0 &  0 &  0    \\ \hline
 $\ns_\UB^{\inte}(\mu) $& 12 & 15 & 5 & 5 &  3 &  5  & 1 & 1 &  1   \\\hline
\end{tabular} 

\begin{tabular}{ |  c | c c c c c c  c c c   |  } \hline
    & $\ttC2$ &  $\ttC3$ &   $\ttC4$ & $\ttN2$ &   $\ttN3$ &   $\ttO2$
   & $\ttS_{(2)}2$ & $\ttS_{(6)}3$ & $\ttP4$  \\\hline
 $\ns_\LB^{\cnt}(\mu)$ &  0 &   0 & 0  & 0 &  0 &  0 & 0 &  0 &  0    \\ \hline
 $\ns_\UB^{\cnt}(\mu) $& 2 & 2 & 2 & 2 & 2 &  2  & 1 & 1 &  0   \\\hline
\end{tabular}   

\begin{tabular}{ |  c | c c c c c c c |  } \hline
         & $\nu_1 $ &   $\nu_2 $ & $\nu_3 $   & $\nu_4 $
         &   $\nu_5 $ & $\nu_6 $   & $\nu_7 $ \\\hline
 $\ac_\LB^{\inte}(\nu)$  &  0  &  0  & 0  &  0  & 0  &  0 & 0     \\ \hline
 $\ac_\UB^{\inte}(\nu)$ & 30 & 10 & 10 &  10 & 2 &  3 &  3 \\\hline
\end{tabular} 

\begin{tabular}{ |  c | c c c c c  c c  |  } \hline
    & $\gamma_1 $ &   $\gamma_2 $ & $\gamma_3 $   & $\gamma_4 $  & $\gamma_5 $
    & $\gamma_i, i\in[6,13]$ &   $\gamma_i, i\in[14,20]$    \\\hline
 $\ec_\LB^{\inte}(\gamma)$ &
    0 &  0 & 0 &  0  & 0 &  0 &  0    \\ \hline
 $\ec_\UB^{\inte}(\gamma) $ &
   4 & 15 & 5 &  5  & 10 & 5  &   2  \\\hline
\end{tabular} 

\begin{tabular}{ |  c | c c c  c  |  } \hline
         & $\nu'_1 $ &   $\nu'_2 $ & $\nu'_3 $  & $\nu'_4 $  \\\hline
 $\ac_\LB^{\lnk}(\nu')$  &  0  &  0  & 0  & 0      \\ \hline
 $\ac_\UB^{\lnk}(\nu')$ &  10 &  5 &  5  &  5   \\\hline
\end{tabular} 
\begin{tabular}{ |  c | c    |  } \hline
    & $\gamma'_i, i\in[1,10]$   \\\hline
 $\ec_\LB^{\lnk}(\gamma')$ &  0    \\ \hline
 $\ec_\UB^{\lnk}(\gamma') $& 4   \\\hline
\end{tabular}

\begin{tabular}{ |  c | c   c   |  } \hline 
& $\psi\in\{\psi_i\mid i=1,6,11\}$ 
& $\psi\in \mathcal{F}^*\setminus \{\psi_i\mid i=1,6,11\}$ \\\hline
 $\fc_\LB(\psi)$  &  1 &    0   \\ \hline 
 $\fc_\UB(\psi)$ &  10 &  3\\\hline
\end{tabular} 

\begin{tabular}{ |  c | c   c   |  } \hline 
& $\nu\in\{(\ttC,\ttC,1),(\ttC,\ttC,2)  \}$ 
& $\nu\in \Gac^\lf \setminus \{(\ttC,\ttC,1),(\ttC,\ttC,2)  \}$   \\\hline
 $\ac^\lf_\LB(\nu)$  &  0 &    0   \\ \hline 
 $\ac^\lf_\UB(\nu)$ &  10 &  8 \\\hline
\end{tabular} 

\label{table:chemical_spec}
\end{table}

\begin{figure}[t!]
\begin{center}
\includegraphics[width=.89\columnwidth]{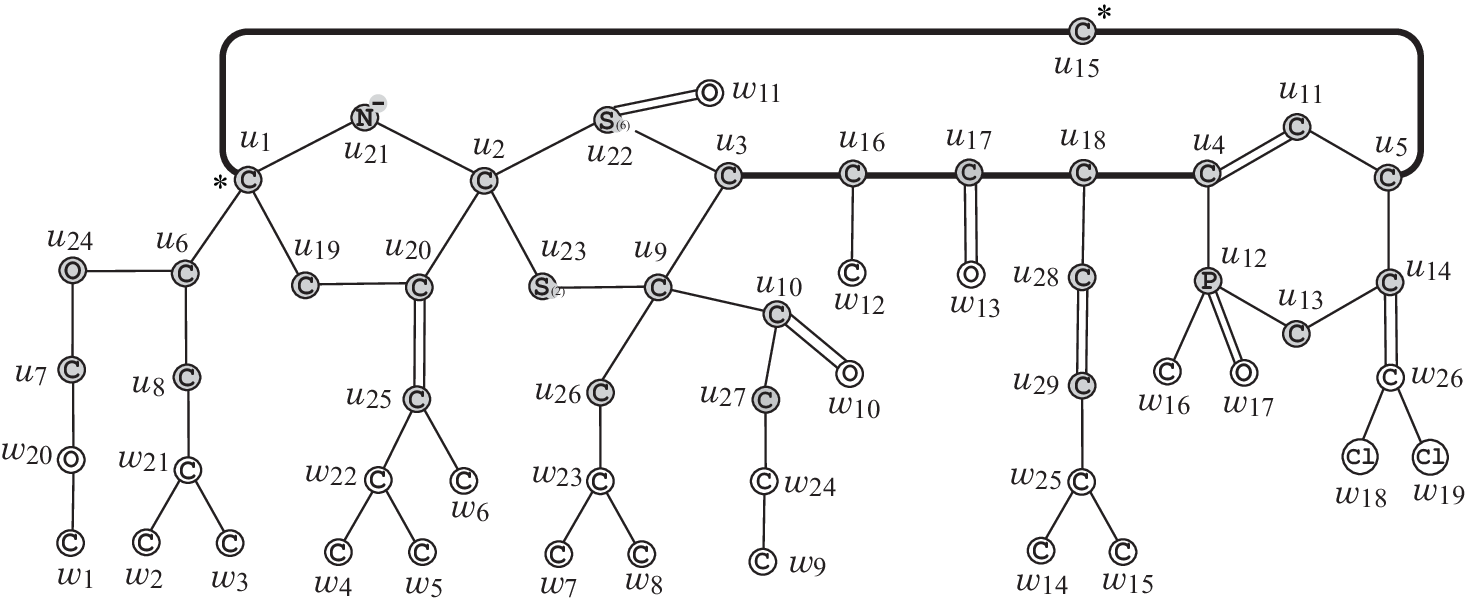}
\end{center}
\caption{
An illustration 
of  the hydrogen-suppressed  monomer representation $\anC$  
obtained from a polymer $\C$ by removing all the 
 hydrogens, 
where the link-edges are depicted with thick lines and 
$V^\ex(\C)=\{w_i \mid i\in [1,26]\}$ and
$V^\inte(\C)=\{u_i \mid i\in [1,29]\}$    for  ${\rho}=2$
 and the connecting-vertices are marked with asterisks. 
}
\label{fig:example_polymer}
\end{figure}
 
Figure~\ref{fig:example_polymer} 
 illustrates an example of 
a   $(\sint,\sce)$-extension of $\GC$   obtained 
from the  $\sint$-extension $H^*$  
 in Figure~\ref{fig:test_subgraph_polymer} 
under the chemical-specification $\sce$ in Table~\ref{table:chemical_spec}.  
  


\section{Test Instances for Phase~2}\label{sec:test_instances} 

We prepared the following instances $I_{\mathrm{a}}$, $I_{\mathrm{b}}$, $I_{\mathrm{c}}$, $I_{\mathrm{d1}}$, $I_{\mathrm{d2}}$, and $I_{\mathrm{d3}}$
 for conducting experiments
in Sections~\ref{sec:experiment_phase2} and~\ref{sec:experiment_phase2_compare}.
 

\begin{figure}[t!]
\begin{center} 
 \includegraphics[width=.89\columnwidth]{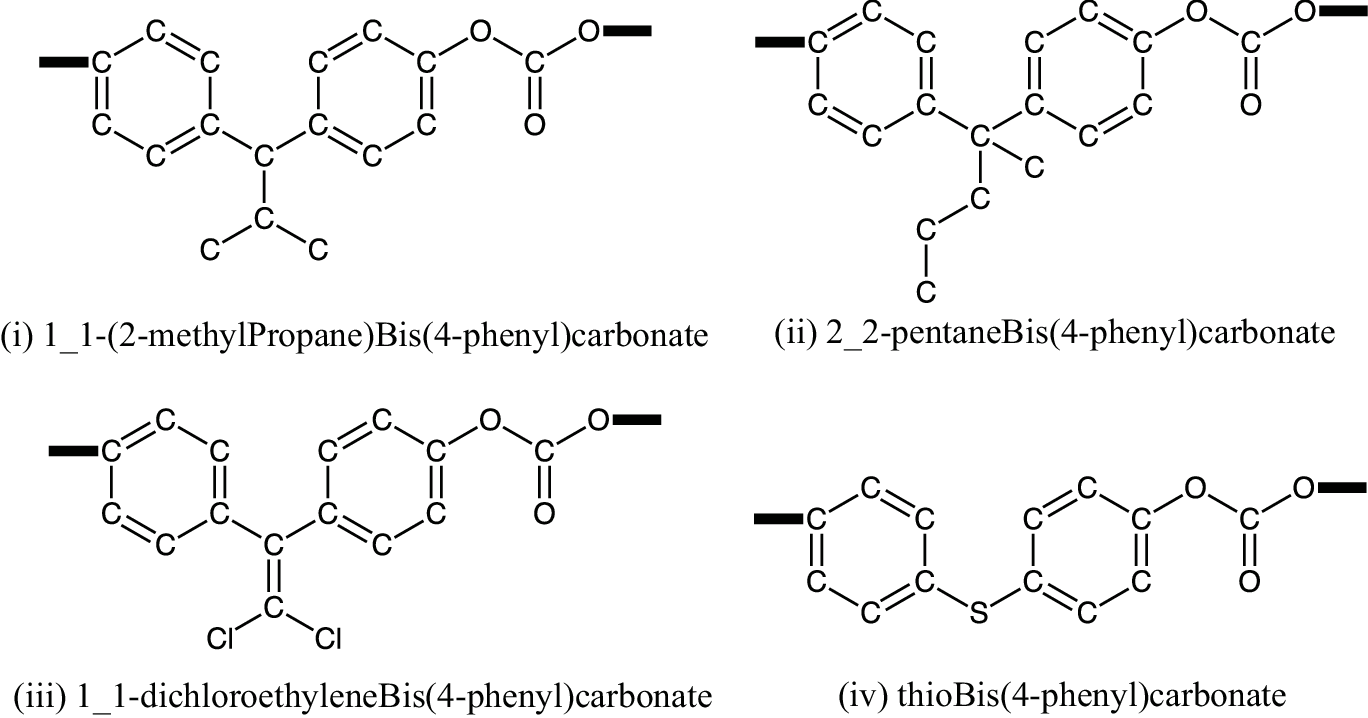}
\end{center}
\caption{Illustrations of  four polymers: 
(i)  1$\underline{~}$1-(2-methylPropane)Bis(4-phenyl)carbonate;
(ii) 2$\underline{~}$2-pentaneBis(4-phenyl)carbonate;
(iii)  1$\underline{~}$1-dichloroethyleneBis(4-phenyl)carbonate;
(iv) thioBis(4-phenyl)carbonate.
Hydrogens are omitted and connecting-edges are depicted with thick lines. 
}
\label{fig:four_polymers}  
\end{figure} 

\begin{figure}[t!]
\begin{center} 
 \includegraphics[width=.45\columnwidth]{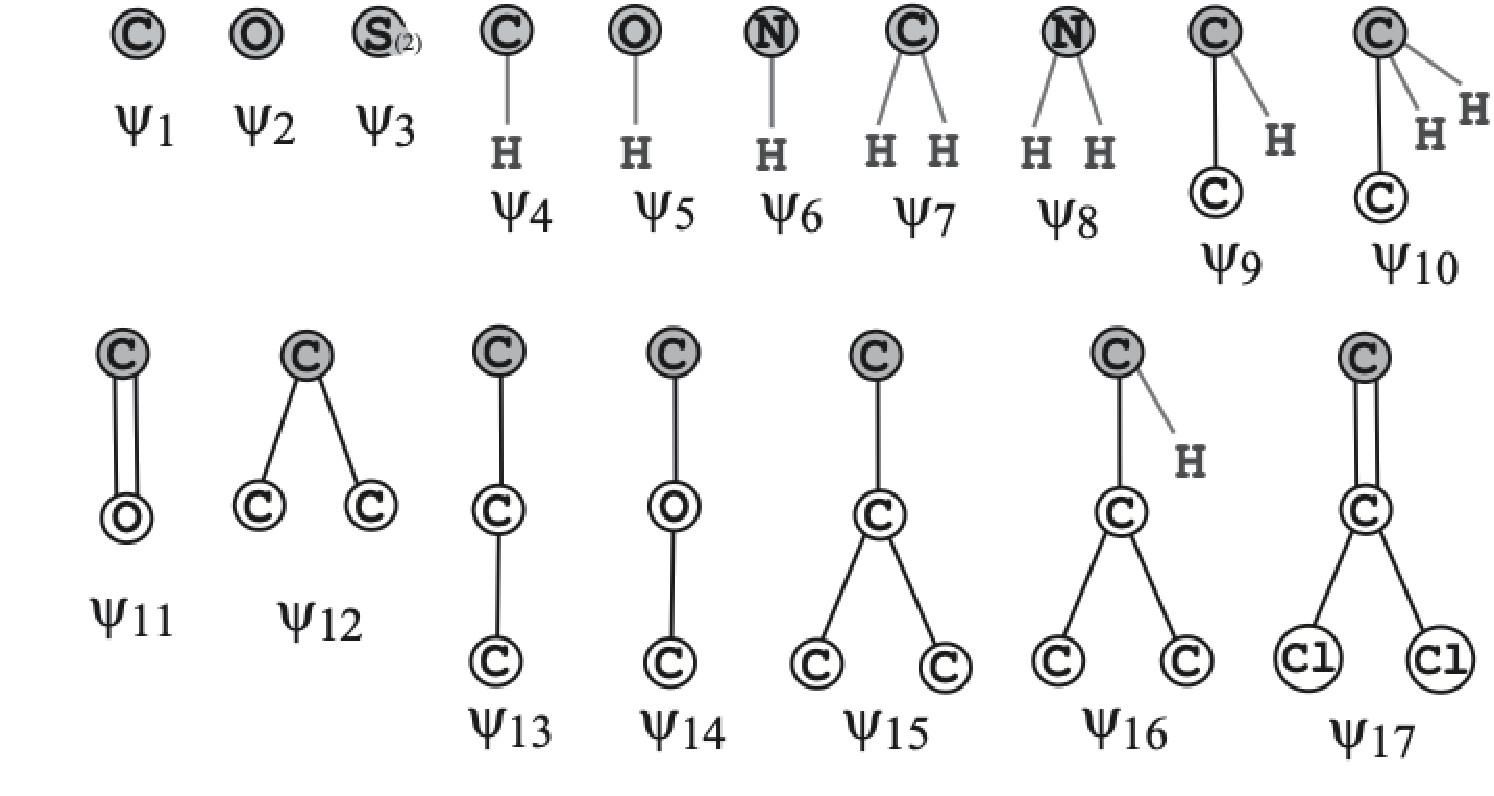}
\end{center}
\caption{Illustrations of the set of chemical rooted trees for the instance $I_b$.}
\label{fig:fringe-tree-set-b}  
\end{figure} 
 
\begin{itemize} 
  \item[(a)]  $I_{\mathrm{a}} =(\GC,\sint,\sce)$: The instance
  used in Section~\ref{sec:specification} to explain the topological specification. The seed graph is illustrated in Figure~\ref{fig:seed-graph-abc}(a).
 
 \end{itemize}
 
\begin{itemize} 
  \item[(b)]  $I_{\mathrm{b}} =(\GC,\sint,\sce)$: An instance that
  represents  a set of polymers that includes the four examples of polymers in 
Figure~\ref{fig:four_polymers}.
We set a seed graph  $\GC=(\VC,\EC)$ to be the graph  
 with two cycles $C_1$ and $C_2$ in Figure~\ref{fig:seed-graph-abc}(b), 
 where we set  
$\Et=\EC^\lnk=\{a_1,a_2\}$ and 
$\Eew=\{a_i \mid i\in[3,14]\}$. \\
Set  $\Lambda:= \{ \ttH, \ttC, \ttN, \ttO, \ttS_{(2)}, \ttCl \}$ 
and  set $\Ldg^\inte$ to be
the set of all possible chemical symbols in $\Lambda\times[1,4]$.\\
Set 
$\Gamma^\inte$ (resp.,  $\Gamma^\lnk$)
to be the set of the edge-configurations of the interior-edges
(resp.,  the link-edges)
used in the four examples of polymers in Figure~\ref{fig:four_polymers}.  
Set 
$\Gamma^\inte_\ac$ (resp.,  $\Gamma^\lnk_\ac$) to be
 the set of the adjacency-configurations of the edge-configurations in 
$\Gamma^\inte$ (resp.,  $\Gamma^\lnk$). \\
We specify $n_\LB:=25$ 
and 
set 
$\nint_\LB:=14$, $\nint_\UB:=n^*:=n_\LB+10$,  
$\nlnk_\LB:=2$,  $\nlnk_\UB: =2+\max\{ n_\LB-15, 0\}$.  \\
For each link-edge $a_i\in\Et=\EC^\lnk=\{a_1,a_2\}$, 
set 
 $\ell_\LB(a_i):=2+\max\{\lfloor (n_\LB-15)/4\rfloor,0\}$,  
 $\ell_\UB(a_i):=\ell_\LB(a_i)+5$,
 $\bl_\LB(a_i):=0, \bl_\UB(a_i):=3$, 
 $\chch_\LB(a_i):=0, \chch_\UB(a_i):=5$,  
 $\bd_{2,\LB}(a_i):=0$ and $\bd_{2,\UB}(a_i):= \lfloor \ell_\LB(a_i)/3 \rfloor$.\\
To form two benzene rings from the two cycles $C_1$ and $C_2$, set 
   $\Lambda^*(u):=\{{\tt C}\}$, 
 $\bl_\LB(u):=\bl_\UB(u):=\chch_\LB(u):=\chch_\UB(u):=0$, $u\in \VC$,
 $\bd_{2,\LB}(a_i):=\bd_{2,\UB}(a_i):=0,  i\in\{3,5,7,9,11,13\}$,
 $\bd_{2,\LB}(a_i):=\bd_{2,\UB}(a_i):=1,  i\in\{4,6,8,10,12,14\}$.\\
Not to include any triple-bond, set 
 $\bd_{3,\LB}(a):=\bd_{3,\UB}(a):=0, a\in \EC$.
 \\
Set lower bounds
 $\na_\LB$,  $\na^\inte_\LB$,  $\ns^\inte_\LB$,  $\ns^\cnt_\LB$, 
$\ac^\inte_\LB$, $\ac_\LB^\lnk$, $\ec_\LB^\inte$, $\ec_\LB^\lnk$ and  $\ac^\lf_\LB$  to be 0. \\
Set  upper bounds   
 $\na_\UB(\ta):=n^*, \ta\in\{\ttH,\ttC\}$,   
 $\na_\UB(\ta):=5+\max\{ n_\LB-15, 0\}, \ta\in\{\ttO,\ttN\}$,
 $\na_\UB(\ta):=2+\max\{\lfloor (n_\LB-15)/4\rfloor,0\}, 
 \ta\in\Lambda\setminus \{\ttH,\ttC,\ttO,\ttN\}$,  
  $\ns^\cnt_\UB(\mu):=2, \mu\in\Ldg^\inte$, 
 and   $\na^\inte_\UB$,  $\ns^\inte_\UB$, 
$\ac^\inte_\UB$, $\ac_\LB^\lnk$, $\ec_\UB^\inte$,  $\ec_\UB^\lnk$ 
and  $\ac^\lf_\UB$ to be  $n^*$. \\
Set $\mathcal{F}$ to be the set of the 17 chemical rooted trees $\psi_i,i\in[1,17]$
 in  Figure~\ref{fig:fringe-tree-set-b}. 
Set $\mathcal{F}_E :=\mathcal{F}(v) := \mathcal{F}$, $v\in \VC$ and 
$\fc_\LB(\psi):=0, \psi\in \mathcal{F}$,
$\fc_\UB(\psi_i):=12+\max\{ n_\LB-15, 0\}, i\in[1,4]$, 
$\fc_\UB(\psi_i):=8+\max\{\lfloor (n_\LB-15)/2\rfloor,0\}, i\in[5,12]$ and
$\fc_\UB(\psi_i):=5+\max\{\lfloor (n_\LB-15)/4\rfloor,0\}, i\in[13,17], \psi_i\in \mathcal{F}$. 

\item[(c)] $I_{\mathrm{c}}=(\GC,\sint,\sce)$: An instance that represents a relatively simple polymer structure.
The seed graph is illustrated in Figure~\ref{fig:seed-graph-abc}(c), 
where we set  
$\Et=\{a_1, a_2\}$,
$\Eew=\{ a_3, a_4\}$, and $\EC^\lnk=\{ a_1, a_2, a_3 \}$. \\
Set  $\Lambda:= \{ \ttH, \ttC, \ttN, \ttO, \ttCl \}$ 
and  set $\Ldg^\inte$ to be
the set of all possible chemical symbols in $\Lambda\times[1,4]$.\\
Set 
$\Gamma^\inte$ (resp.,  $\Gamma^\lnk$)
to be the set of all the edge-configurations of the interior-edges
(resp.,  the link-edges)
appeared in the dataset. 
Set 
$\Gamma^\inte_\ac$ (resp.,  $\Gamma^\lnk_\ac$) to be
 the set of all the adjacency-configurations of the edge-configurations in 
$\Gamma^\inte$ (resp.,  $\Gamma^\lnk$). \\
We 
set $n_\LB:=15, \nint_\LB:=8, \nlnk_\LB:=2$,
$n^*:=25, \nint_\UB:=20, \nlnk_\UB:=10$.  \\
Set $\bl_\LB(u_i):=\bl_\UB(u_i):=0$,
  $\chch_\LB(u_i):=0$   for each vertex $v_i \in \{u_1, u_2, u_3, u_4\}$, and
  $\chch_\UB(u_1):=1,\chch_\UB(u_2):=0, \chch_\UB(u_3):=4,\chch_\UB(u_4):=2$.\\
For each edge $a_i\in\Et=\{a_1,a_2\}$, 
set 
 $\ell_\LB(a_i):=2$,  
 $\ell_\UB(a_i):=10$, 
 $\bl_\LB(a_i):=0, \bl_\UB(a_i):=3$, 
 $\chch_\LB(a_i):=0, \chch_\UB(a_i):=5$,  
 $\bd_{2,\LB}(a_i):=0$ and $\bd_{2,\UB}(a_i):= 3$.\\ 
Set 
 $\bd_{2,\LB}(a_i):=0$ and $\bd_{2,\UB}(a_3):= 0, \bd_{2,\UB}(a_4):= 1$ for $a_i \in \{a_3, a_4\}$.\\
Not to include any triple-bond, set 
 $\bd_{3,\LB}(a):=\bd_{3,\UB}(a):=0, a\in \EC$.
 \\
Set lower bounds
 $\na_\LB$,  $\na^\inte_\LB$,  $\ns^\inte_\LB$,  $\ns^\cnt_\LB$, 
$\ac^\inte_\LB$, $\ac_\LB^\lnk$, $\ec_\LB^\inte$, $\ec_\LB^\lnk$ and  $\ac^\lf_\LB$  to be 0. \\
Set  upper bounds   
 $\na_\UB(\ta):=n^*, \ta\in\{\ttH,\ttC\}$,   
 $\na_\UB(\ta):=10, \ta\in\{\ttN,\ttO\}$,
 $\na_\UB(\ta):=5, \ta=\ttCl$,
  $\ns^\cnt_\UB(\mu):=2, \mu\in\Ldg^\inte$, 
 and $\ac^\lf_\UB(\nu)$,  $\na^\inte_\UB$,  $\ns^\inte_\UB$, 
$\ac^\inte_\UB$, $\ac_\UB^\lnk$, $\ec_\UB^\inte$,  and  $\ec_\UB^\lnk$ 
 to be  $n^*$. \\
Set $\mathcal{F}$ to be the set of  all the chemical rooted trees that appeared in the dataset.
Set $\mathcal{F}_E :=\mathcal{F}(v) := \mathcal{F}$, $v\in \VC$ and 
$\fc_\LB(\psi):=0, \fc_\UB(\psi):=n^*, \psi\in \mathcal{F}$.

\item[(d1)] $I_{\mathrm{d1}}=(\GC,\sint,\sce)$: An instance that represents a polymer structure with one ring.
The seed graph is illustrated in Figure~\ref{fig:seed-graph-d}(a), 
where we set  
$\Ew=\{a_1, a_2 \}$, 
$\Eew=\{ a_3, a_4, a_5, a_6, a_7, a_8, a_9\}$, and $\EC^\lnk=\{ a_1, a_2, a_9 \}$. \\
Set  $\Lambda:= \{ \ttH, \ttC, \ttN, \ttO, \ttS_{(2)}, \ttCl, \ttF \}$ 
and  set $\Ldg^\inte$ to be
the set of all possible chemical symbols in $\Lambda\times[1,4]$.\\
Set 
$\Gamma^\inte$ (resp.,  $\Gamma^\lnk$)
to be the set of all the edge-configurations of the interior-edges
(resp.,  the link-edges)
appeared in the dataset. 
Set 
$\Gamma^\inte_\ac$ (resp.,  $\Gamma^\lnk_\ac$) to be
 the set of all the adjacency-configurations of the edge-configurations in 
$\Gamma^\inte$ (resp.,  $\Gamma^\lnk$). \\
We 
set $n_\LB:=\nint_\LB:=8, \nlnk_\LB:=2$,
$n^*:=\nint_\UB:=20, \nlnk_\UB:=10$.  \\
Set $\bl_\LB(u_i):=0$  for each vertex $u_i, i\in[1,8]$,
$\bl_\UB(u_i):=0$ for each vertex $u_i, i\in[1,3] \cup [5,8]$, $\bl_\UB(u_4):=1$,
  $\chch_\LB(u_i):=0$ for each vertex $u_i, i\in[1,3] \cup [5,8]$, $\chch_\LB(u_4):=1$,
  $\chch_\UB(u_i):=0$ for $u_i, i\in\{1,2,5,8\}$, $\chch_\UB(u_i):=1$ for $u_i, i\in\{3,6,7\}$, 
   $\chch_\UB(u_4):=3$.\\
For each edge $a_i\in\Ew=\{a_1,a_2\}$, 
set 
 $\ell_\LB(a_i):=1$,  
 $\ell_\UB(a_i):=3$, 
 $\bl_\LB(a_i):=0, \bl_\UB(a_i):=1$, 
 $\chch_\LB(a_i):=0, \chch_\UB(a_i):=3$,  
 $\bd_{2,\LB}(a_i):=0$ and $\bd_{2,\UB}(a_i):= 1$.\\ 
Set 
 $\bd_{2,\LB}(a_i):=\bd_{2,\UB}(a_i):=0$ for $a_i \in \{a_4,a_6,a_8,a_9\}$ and $\bd_{2,\LB}(a_i):=\bd_{2,\UB}(a_i):= 1$ for $a_i \in \{a_3, a_5,a_7\}$.\\
Not to include any triple-bond, set 
 $\bd_{3,\LB}(a):=\bd_{3,\UB}(a):=0, a\in \EC$.
 \\
Set lower bounds
 $\na_\LB$,  $\na^\inte_\LB$,  $\ns^\inte_\LB$,  $\ns^\cnt_\LB$, 
$\ac^\inte_\LB$, $\ac_\LB^\lnk$, $\ec_\LB^\inte$, $\ec_\LB^\lnk$ and  $\ac^\lf_\LB$  to be 0. \\
Set  upper bounds   
 $\na_\UB(\ta):=n^*, \ta\in\{\ttH,\ttC\}$,   
 $\na_\UB(\ta):=8, \ta\in\{\ttN,\ttO\}$,
  $\na_\UB(\ta):=2, \ta=\ttS_{(2)}$,
 $\na_\UB(\ta):=4, \ta\in\{\ttF,\ttCl\}$,
  $\ns^\cnt_\UB(\mu):=2, \mu\in\Ldg^\inte$, 
 and $\ac^\lf_\UB(\nu)$,  $\na^\inte_\UB$,  $\ns^\inte_\UB$, 
$\ac^\inte_\UB$, $\ac_\UB^\lnk$, $\ec_\UB^\inte$,  and  $\ec_\UB^\lnk$ 
 to be  $n^*$. \\
Set $\mathcal{F}$ to be the set of  all the chemical rooted trees that appeared in the dataset.
Set $\mathcal{F}_E :=\mathcal{F}(v) := \mathcal{F}$, $v\in \VC$ and 
$\fc_\LB(\psi):=0, \fc_\UB(\psi):=n^*, \psi\in \mathcal{F}$.

\item[(d2)] $I_{\mathrm{d2}}=(\GC,\sint,\sce)$: An instance that represents a polymer structure with two rings.
The seed graph is illustrated in Figure~\ref{fig:seed-graph-d}(b), 
where we set  
$\Et=\{a_1\}$,
$\Ew=\{a_2, a_3 \}$, 
$\Eew=\{ a_4, a_5, a_6, a_7, a_8, a_9, a_{10}, a_{11}\}$, and $\EC^\lnk=\{ a_2, a_3, a_{11} \}$. \\
Set  $\Lambda:= \{ \ttH, \ttC, \ttN, \ttO, \ttS_{(2)}, \ttCl, \ttF \}$ 
and  set $\Ldg^\inte$ to be
the set of all possible chemical symbols in $\Lambda\times[1,4]$.\\
Set 
$\Gamma^\inte$ (resp.,  $\Gamma^\lnk$)
to be the set of all the edge-configurations of the interior-edges
(resp.,  the link-edges)
appeared in the dataset. 
Set 
$\Gamma^\inte_\ac$ (resp.,  $\Gamma^\lnk_\ac$) to be
 the set of all the adjacency-configurations of the edge-configurations in 
$\Gamma^\inte$ (resp.,  $\Gamma^\lnk$). \\
We 
set $n_\LB:=\nint_\LB:=5, \nlnk_\LB:=2$,
$n^*:=\nint_\UB:=20, \nlnk_\UB:=10$.  \\
Set $\bl_\LB(u_i):=0$  for each vertex $u_i, i\in[1,9]$,
$\bl_\UB(u_i):=0$ for each vertex $u_i, i\in[1,6] \cup \{8\}$, $\bl_\UB(u_7):=\bl_\UB(u_9):=1$,
  $\chch_\LB(u_i):=0$ for each vertex $u_i, i\in[1,9]$, 
  $\chch_\UB(u_i):=0$ for $u_i, i\in [1,5]\cup\{8\}$, $\chch_\UB(u_6):=1$, 
   $\chch_\UB(u_7):=\chch_\UB(u_9):=3$.\\
For each edge $a_i\in\{a_1,a_2,a_3\}$, 
set 
 $\ell_\LB(a_1):=3$,  $\ell_\LB(a_2):=\ell_\LB(a_3):=1$,  
 $\ell_\UB(a_1):=5$,  $\ell_\UB(a_2):=\ell_\UB(a_3):=2$, 
 $\bl_\LB(a_i):=0, \bl_\UB(a_i):=1$, 
 $\chch_\LB(a_i):=0, \chch_\UB(a_i):=1$,  
 $\bd_{2,\LB}(a_1):=1$,  $\bd_{2,\LB}(a_2):=\bd_{2,\LB}(a_3):=0$,  
 $\bd_{2,\UB}(a_1):= 3$, and $\bd_{2,\UB}(a_2):=\bd_{2,\UB}(a_3):=1$.\\ 
Set 
 $\bd_{2,\LB}(a_i):=\bd_{2,\UB}(a_i):=0$ for $a_i \in \{a_4,a_6,a_8,a_{10},a_{11}\}$ and $\bd_{2,\LB}(a_i):=\bd_{2,\UB}(a_i):= 1$ for $a_i \in \{ a_5,a_7,a_9\}$.\\
Not to include any triple-bond, set 
 $\bd_{3,\LB}(a):=\bd_{3,\UB}(a):=0, a\in \EC$.
 \\
Set lower bounds
 $\na_\LB$,  $\na^\inte_\LB$,  $\ns^\inte_\LB$,  $\ns^\cnt_\LB$, 
$\ac^\inte_\LB$, $\ac_\LB^\lnk$, $\ec_\LB^\inte$, $\ec_\LB^\lnk$ and  $\ac^\lf_\LB$  to be 0. \\
Set  upper bounds   
 $\na_\UB(\ta):=n^*, \ta\in\{\ttH,\ttC\}$,   
 $\na_\UB(\ta):=8, \ta\in\{\ttN,\ttO\}$,
  $\na_\UB(\ta):=2, \ta=\ttS_{(2)}$,
 $\na_\UB(\ta):=4, \ta\in\{\ttF,\ttCl\}$,
  $\ns^\cnt_\UB(\mu):=2, \mu\in\Ldg^\inte$, 
 and $\ac^\lf_\UB(\nu)$,  $\na^\inte_\UB$,  $\ns^\inte_\UB$, 
$\ac^\inte_\UB$, $\ac_\UB^\lnk$, $\ec_\UB^\inte$,  and  $\ec_\UB^\lnk$ 
 to be  $n^*$. \\
Set $\mathcal{F}$ to be the set of  all the chemical rooted trees that appeared in the dataset.
Set $\mathcal{F}_E :=\mathcal{F}(v) := \mathcal{F}$, $v\in \VC$ and 
$\fc_\LB(\psi):=0, \fc_\UB(\psi):=n^*, \psi\in \mathcal{F}$.

\item[(d3)] $I_{\mathrm{d3}}=(\GC,\sint,\sce)$: An instance that represents a polymer structure with three rings.
The seed graph is illustrated in Figure~\ref{fig:seed-graph-d}(c), 
where we set  
$\Ew=\{a_1, a_2, a_3 \}$, 
$\Eew=\{ a_i \mid i\in[4,18]\}$ , and $\EC^\lnk=\{ a_2, a_3, a_{18} \}$. \\
Set  $\Lambda:= \{ \ttH, \ttC, \ttN, \ttO, \ttS_{(2)}, \ttCl, \ttF \}$ 
and  set $\Ldg^\inte$ to be
the set of all possible chemical symbols in $\Lambda\times[1,4]$.\\
Set 
$\Gamma^\inte$ (resp.,  $\Gamma^\lnk$)
to be the set of all the edge-configurations of the interior-edges
(resp.,  the link-edges)
appeared in the dataset. 
Set 
$\Gamma^\inte_\ac$ (resp.,  $\Gamma^\lnk_\ac$) to be
 the set of all the adjacency-configurations of the edge-configurations in 
$\Gamma^\inte$ (resp.,  $\Gamma^\lnk$). \\
We 
set $n_\LB:=\nint_\LB:=14, \nlnk_\LB:=2$,
$n^*:=\nint_\UB:=30, \nlnk_\UB:=20$.  \\
Set $\bl_\LB(u_i):=0$  for each vertex $u_i, i\in[1,15]$,
$\bl_\UB(u_i):=0$ for each vertex $u_i, i\in[1,14]$, $\bl_\UB(u_{15}):=1$,
  $\chch_\LB(u_i):=0$ for each vertex $u_i, i\in[1,15]$, 
  $\chch_\UB(u_i):=0$ for $u_i, i\in [1,5]\cup\{8,9,11,14\}$, $\chch_\UB(u_{10}):=\chch_\UB(u_{13}):=1$, 
   $\chch_\UB(u_6):=\chch_\UB(u_7):=\chch_\UB(u_{12}):=2$,   $\chch_\UB(u_{15}):=3$.\\
For each edge $a_i\in\{a_1,a_2,a_3\}$, 
set 
 $\ell_\LB(a_i):=1$, 
 $\ell_\UB(a_i):=2$, 
 $\bl_\LB(a_i):=0$, $\bl_\UB(a_1):=1$, $\bl_\UB(a_2):=\bl_\UB(a_3):=0$, 
 $\chch_\LB(a_i):=0$, $\chch_\UB(a_1):=3$, $\chch_\UB(a_2):=\chch_\UB(a_3):=1$,  
 $\bd_{2,\LB}(a_1):=1$,  $\bd_{2,\LB}(a_2):=\bd_{2,\LB}(a_3):=0$,  
 $\bd_{2,\UB}(a_1):= 3$, and $\bd_{2,\UB}(a_2):=\bd_{2,\UB}(a_3):=1$.\\ 
Set 
 $\bd_{2,\LB}(a_i):=\bd_{2,\UB}(a_i):=0$ for $a_i \in \{a_5,a_7,a_9,a_{10},a_{12},a_{14},a_{16},a_{17},a_{18}\}$ and $\bd_{2,\LB}(a_i):=\bd_{2,\UB}(a_i):= 1$ for $a_i \in \{ a_6,a_8,a_{11},a_{13},a_{15}\}$.\\
Not to include any triple-bond, set 
 $\bd_{3,\LB}(a):=\bd_{3,\UB}(a):=0, a\in \EC$.
 \\
Set lower bounds
 $\na_\LB$,  $\na^\inte_\LB$,  $\ns^\inte_\LB$,  $\ns^\cnt_\LB$, 
$\ac^\inte_\LB$, $\ac_\LB^\lnk$, $\ec_\LB^\inte$, $\ec_\LB^\lnk$ and  $\ac^\lf_\LB$  to be 0. \\
Set  upper bounds   
 $\na_\UB(\ta):=n^*, \ta\in\{\ttH,\ttC\}$,   
 $\na_\UB(\ta):=8, \ta\in\{\ttN,\ttO\}$,
  $\na_\UB(\ta):=2, \ta=\ttS_{(2)}$,
 $\na_\UB(\ta):=4, \ta\in\{\ttF,\ttCl\}$,
  $\ns^\cnt_\UB(\mu):=2, \mu\in\Ldg^\inte$, 
 and $\ac^\lf_\UB(\nu)$,  $\na^\inte_\UB$,  $\ns^\inte_\UB$, 
$\ac^\inte_\UB$, $\ac_\UB^\lnk$, $\ec_\UB^\inte$,  and  $\ec_\UB^\lnk$ 
 to be  $n^*$. \\
Set $\mathcal{F}$ to be the set of  all the chemical rooted trees that appeared in the dataset.
Set $\mathcal{F}_E :=\mathcal{F}(v) := \mathcal{F}$, $v\in \VC$ and 
$\fc_\LB(\psi):=0, \fc_\UB(\psi):=n^*, \psi\in \mathcal{F}$.

 \end{itemize}

\end{document}